\documentclass[a4paper,11pt]{article}

\usepackage{amsfonts}
\usepackage{amsmath}
\usepackage{amssymb}
\usepackage{amssymb}
\usepackage[titletoc,toc,title]{appendix}
\usepackage{authblk}
\usepackage{braket}
\usepackage{cancel}
\usepackage[font={footnotesize,it}]{caption}
\usepackage{cite}
\usepackage{comment}
\usepackage{enumitem}
\usepackage{epsfig}
\usepackage{float}
\usepackage[margin=2.5cm]{geometry}
\usepackage{graphicx}
\usepackage{epstopdf}
\usepackage{hyperref}
\usepackage{cleveref}
\usepackage{physics}
\usepackage{ragged2e}
\usepackage{soul}
\usepackage{subcaption}
\usepackage{xcolor}

\hypersetup{
	citecolor=red,
	colorlinks=true,
	filecolor=red,
	linkcolor=blue,
	linktocpage=true,
	urlcolor=blue
}

{\makeatletter\g@addto@macro\bfseries{\boldmath}\makeatother}
\numberwithin{equation}{section}

\newcommand{\TTbar}{\text{T}\bar{\text{T}}}
\newcommand{\ttbar}{T \overline{T}}
\newcommand{\zbar}{\raisebox{0.2ex}{--}\kern-0.6em Z}

\def\ttbar{\textrm{T}\bar{\textrm{T}}}

\def\td{\textrm{d}}
\def\cM{\mathcal{M}}
\def\cD{\mathcal{D}}
\def\cO{\mathcal{O}}
\def\cL{\mathcal{L}}
\def\cH{\mathcal{H}}
\def\cS{\mathcal{S}}
\def\cE{\mathcal{E}}
\def\cI{\mathcal{I}}
\def\cP{\mathcal{P}}
\def\bZ{\mathbb{Z}}
\def\bP{\mathbb{P}}

\begin{document}

	\title{Entanglement, $\text{T}\bar{\text{T}}$ and rotating black holes}

	\author[]{Debarshi Basu\thanks{\noindent E-mail:~ {\texttt{debarshi.128@gmail.com}}}}

	\author[]{Saikat Biswas\thanks{\noindent E-mail:~ {\texttt{saikatb21@iitk.ac.in}}}}

	\affil[]{
		Department of Physics,\\
		Indian Institute of Technology,\\ 
		Kanpur 208 016, India
	}	
	
	\date{}
	
	\maketitle
	
	\thispagestyle{empty}

	\begin{abstract}
		
		\bigskip
		
		\noindent
	In this work, we investigate the entanglement structure in a $\textrm{T}\bar{\textrm{T}}$-deformed holographic CFT$_2$ with a conserved angular momentum. We utilize conformal perturbation theory to compute the leading order correction to the entanglement entropy and the reflected entropy due to the $\textrm{T}\bar{\textrm{T}}$ deformation in the limit of large central charge. In the dual bulk perspective described by a rotating BTZ black hole with a finite radial cut-off, we compute the holographic entanglement entropy and the entanglement wedge cross-section and observe perfect agreement with our field theoretic computation for small values of the deformation parameter.

	\end{abstract}

	\clearpage
	\tableofcontents
	\clearpage
	\section{Introduction}
The Anti-de Sitter/Conformal Field Theory (AdS/CFT) correspondence \cite{Maldacena:1997re,Gubser:1998bc}, has fundamentally reshaped our understanding of quantum gravity and gauge theories. This duality posits a deep connection between a gravitational theory in an AdS space and a conformal field theory on the boundary of that space. Over the past few decades, AdS/CFT has become a cornerstone of theoretical physics, offering insights into black hole physics, strongly coupled quantum field theories, and the structure of spacetime \cite{VanRaamsdonk:2010pw,Maldacena:2013xja}. In this connection, entanglement entropy has emerged as a crucial tool for probing quantum correlations in the boundary theory and understanding the geometry of the bulk spacetime. The Ryu-Takayanagi formula \cite{Ryu:2006bv,Ryu:2006ef}, which links the entanglement entropy of a boundary region to the area of a minimal surface in the bulk, has provided a profound geometric interpretation of quantum entanglement\footnote{For further developments in this connection, see \cite{Hubeny:2007xt, Lewkowycz:2013nqa, Dong:2016hjy,Casini:2011kv}.}. This has led to significant insights into black hole physics as well as strongly correlated systems. However, while entanglement entropy is well-suited for pure states, it is less effective in capturing the entanglement structure in mixed states. To address these limitations, several entanglement and correlation measures have been introduced in the literature, including entanglement negativity \cite{Vidal:2002zz,Plenio:2005cwa}, reflected entropy \cite{Dutta:2019gen}, entanglement of purification \cite{Takayanagi:2017knl}, balanced partial entanglement \cite{Wen:2021qgx}. These measures offer more nuanced insights into the quantum correlations present in mixed states.
	
On a related note, $\TTbar$ deformation is an intriguing modification of two-dimensional conformal field theories (CFT$_2$) \cite{Zamolodchikov:2004ce,Cavaglia:2016oda,Smirnov:2016lqw}, characterized by a deformation parameter $\mu$. The $\TTbar$ deformation modifies the spectrum, the entanglement structure, and the overall behavior of the theory, making it a powerful tool for exploring the interplay between UV and IR physics and the non-local effects in quantum field theories, they preserve integrability and remain solvable, with exact results available for the energy spectrum, partition function, and S-matrix \cite{McGough:2016lol}. Furthermore, a holographic dual for these $\TTbar$-deformed CFTs was proposed \cite{McGough:2016lol}, characterized by AdS geometries with the asymptotic boundary located at a finite radial distance. Some recent works in this direction include \cite{Asrat:2017tzd,Shyam:2017znq,Kraus:2018xrn,Cottrell:2018skz,Taylor:2018xcy,Hartman:2018tkw,Shyam:2018sro,Caputa:2019pam,Giveon:2017myj,Tian:2023fgf,Dei:2024sct,Du:2024bqk,Tian:2024vln}. Specifically, this AdS geometry mirrors the one dual to the undeformed CFT, except that the cut-off surface is shifted further into the bulk. The proposal was further validated by matching the two-point functions, energy spectrum, and partition function of the deformed CFT with holographic computations, confirming the consistency of the duality. The entanglement structure for various pure and mixed states has been extensively explored in the context of the $\TTbar$	deformations \cite{Chen:2018eqk,Donnelly:2018bef,Lewkowycz:2019xse,Banerjee:2019ewu,Jeong:2019ylz,Murdia:2019fax,He:2019vzf,Asrat:2019end,Basu:2023bov,Basu:2023aqz,Basu:2024bal,Chang:2024voo}. 
	
The holographic characterization of the entanglement entropy in such $\TTbar$-deformed thermal CFT$_2$s with conserved charge has recently been explored in \cite{Banerjee:2024wtl} utilizing the mixed boundary conditions proposal of \cite{Guica:2019nzm}. However, a field theoretic analysis of the same is still elusive. In this article, we bridge this gap by investigating a perturbative replica technique computation of the entanglement entropy and the reflected entropy. In particular, we compute the leading order correction to the entanglement entropy and the reflected entropy for various bipartite states in a $\TTbar$-deformed CFT$_2$ with conserved angular momentum, both from the field theory perspective and their holographic characterization, utilizing the techniques introduced in \cite{Chen:2018eqk,Jeong:2019ylz}. 

Our approach involves a perturbative analysis of the field theory action with respect to the deformation parameter $\mu$, under the assumption that $\mu \ll 1$. Within this perturbative framework, we derive the Rényi version of entanglement and reflected entropies, identifying the leading-order corrections for both measures. In this connection, we introduce a novel approach to map the twisted cylinder to the usual thermal cylinder with a standalone temporal identification, in order to evaluate specific integrals that are crucial for our analysis. This technique is then adapted to the bulk geometry to derive the appropriate metric necessary for implementing the cut-off AdS geometry. Later by connecting the deformation parameter $\mu$ to the new cutoff surface in the dual AdS$_3$ geometry, we compute the length of the Ryu-Takayanagi (RT) surface and the entanglement wedge cross-section (EWCS) in the bulk dual geometry involving a non-extremal rotating BTZ black hole. Remarkably, our findings demonstrate exact agreement between the bulk and field theory results. Expanding our analysis, we also explore the timelike entanglement \cite{Doi:2022iyj,Doi:2023zaf} for various bipartite states in $\TTbar$ deformed CFT$_2$, thereby providing a comprehensive understanding of entanglement in such theories.

The rest of the article is structured as follows. In \cref{sec-2} we begin with an exploration of CFT$_2$ with a conserved charge, providing a brief review of $\TTbar$	deformation, its dual AdS$_3$ geometries, and the holographic interpretation of reflected entropy. In \cref{sec-3}, we compute the entanglement entropy and the reflected entropy for various bipartite state configurations, including two disjoint intervals, two adjacent intervals, and a single interval in a finite-temperature CFT$_2$ with a conserved charge. We also explore the nature of timelike entanglement in these $\TTbar$-deformed theories. In \cref{sec-4}, we extend our analysis to the bulk by computing the Ryu-Takayanagi (RT) surface and the entanglement wedge cross section (EWCS) within the context of cutoff AdS$_3$ geometries. Finally, in \cref{sec-5}, we summarize our findings and present our conclusions, drawing connections between the field theory results and their holographic duals.
	

	\section{Review of earlier literature}\label{sec-2}
	\subsection{Conserved angular momentum in a CFT$_2$}\label{CFT-review}
	In this work, we will be focusing on a CFT$_2$ at a finite temperature $\beta^{-1}$ and a conserved angular momentum $J$. The Euclidean partition function for the CFT$_2$ is given by
	\begin{align}
		\mathcal{Z}(\beta,\Omega)&=\textrm{Tr}\,e^{-\beta \left(H+i\Omega_E J\right)}\notag\\
		&=\textrm{Tr}\exp[-\beta_+L_0-\beta_-\bar{L}_0]\label{partition-sum}\,,
	\end{align}
	where $\Omega_E=-i\Omega$ is the (Euclidean) chemical potential for the ``angular momentum''\footnote{In this study, our primary focus will be on a non-compact spatial dimension, which means that the quantity $J$ is interpreted as linear momentum in this context. However, this non-compact spatial direction may be understood as the limiting case of the circumference of a very large circle. From the bulk dual perspective, we may regard this as a BTZ black string arising as a limit of a BTZ black hole. In this connection, we will continue to refer to $J$ as the angular momentum of the rotating BTZ black hole.} $J$. Note that, in the second equality, we have made use of the following identifications for the Hamiltonian $H$ and the momentum $J$ with the usual Virasoro zero modes:
	\begin{align}
		H=L_0+\bar{L}_0~~,~~J=L_0-\bar{L}_0\,,
	\end{align}
	and defined the (effective) temperatures associated with the left and right moving modes as follows
	\begin{align}
		\beta_{\pm}=\beta\left(1\pm i\Omega_E\right)\label{temperatures}\,.
	\end{align}
	Note that this state is characterized by non-zero expectation values $\left<E\right>$ and $\left<J\right>$ for the energy and momentum:
	\begin{align}
		\left<E\right>=E_++E_--\frac{c}{12}~~,~~\left<J\right>=E_+-E_-\,,\label{energy-momentum-CFT}
	\end{align}
	where $E_{\pm}$ are the eigenvalues of the Virasoro generators $L_0$ and $\bar{L}_0$.
	This thermal CFT$_2$ with a conserved angular momentum is defined on a thermal cylinder $\mathcal{M}$ (with topology $\mathbb{S}^1\times \mathbb{R}$) with a twisted identification, dubbed henceforth as a ``twisted cylinder'', with a flat metric
	\begin{align}
		\textrm{d}s^2_\textrm{CFT}=\textrm{d}\tau^2+\textrm{d}x^2\,.
	\end{align}
	 For a holographic CFT, this state is dual to the planar rotating BTZ black hole (string). The twisted identifications which lead to the partition sum in \cref{partition-sum}, are given as\footnote{The generators of translation along the $x$ and $\tau$ directions are given by $-H$ and $i J$.}
	\begin{align}
		(x,\tau)\sim(x-i\beta\Omega_E,\tau+\beta)\,.\label{twisted-identification-xt}
	\end{align}
	In terms of the complex coordinates $w=x+i\tau\,,\,\bar{w}=x-i\tau$, used to describe this cylinder, such twisted identifications corresponds to different compactifications of the holomorphic (left-moving) and anti-holomorphic (right-moving) sectors:
	\begin{align}
		(w,\bar{w})\sim(w+i\beta_+,\bar{w}+i\beta_-)\,,\label{twisted-identification-w}
	\end{align}
	leading to distinct left and right-moving temperatures given in \cref{temperatures}. For computational purposes, in the following discussions, we conformally map the twisted cylinder $(\omega,\bar{\omega})$ to the complex plane with coordinates $(z,\bar{z})$ utilizing the transformations
	\begin{align}
		w\to z=e^{\frac{2\pi}{\beta_+}w}~~,~~\bar{w}\to \bar{z}=e^{\frac{2\pi}{\beta_-}\bar{w}}\label{plane-to-cylinder-map}\,.
	\end{align}
	Under these maps, the stress-energy tensors transform as follows
	\begin{align}
		T(w)=\left(\frac{2\pi}{\beta_+}z\right)^2T(z)-\frac{\pi^2c}{6\beta_+^2}~~,~~\bar{T}(\bar{w})=\left(\frac{2\pi}{\beta_-}\bar{z}\right)^2\bar{T}(\bar{z})-\frac{\pi^2c}{6\beta_-^2}\,.\label{Schwarzian-transformation}
	\end{align}
	\subsection{$\ttbar$ deformation and holography}\label{sec:TTb-holography}
	In this subsection, we briefly review the salient features of a solvable irrelevant deformation of a CFT$_2$, obtained by turning on a coupling through the determinant of the stress-energy tensor. Such theories, featuring the double-trace operator \cite{Zamolodchikov:2004ce,McGough:2016lol}
	\begin{align}
		\left(\ttbar\right)\equiv\frac{1}{8}\left(T_{ab}T^{ab}-\left(T_a^a\right)^2\right)\,,\label{TTbar-operator}
	\end{align}
	are governed by the following flow equation on flat backgrounds
	\begin{align}
		\frac{\td S_\textrm{QFT}^{(\mu)}}{\td\mu}=-4\pi^2\int \td^2 w \left(\ttbar\right)_\mu\,,
	\end{align}
	where $\mu\geq 0$ is the coupling having the dimensions of length squared, and the notation $\left(\ttbar\right)_\mu$ signifies that the operator is constructed out of the stress-tensor in the deformed theory\footnote{This may be  achieved by replacing $T_{ab}$ in \cref{TTbar-operator} by $T_{ab}^{(\mu)}=\frac{2}{\sqrt{-h}}\frac{\delta S_\textrm{QFT}^{(\mu)}}{\delta h^{ab}}\big|_{h_{ab}=\eta_{ab}}$ in the Lorentzian signature.}.
	The deformed theory forms a trajectory in the $2d$ theory space. The initial condition on this trajectory is given by the seed CFT$_2$:
	\begin{align*}
		S_\textrm{QFT}^{(\mu)}\Big|_{\mu=0}=S_\textrm{CFT}\,.
	\end{align*}
	
	\begin{figure}[ht]
		\centering
		\includegraphics[width=0.25\textwidth]{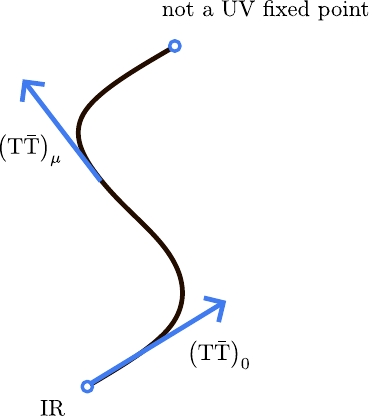}
		\caption{The $\textrm{T}\overline{\textrm{T}}$ deformation is an instantaneous deformation driven by the $\textrm{T}\overline{\textrm{T}}$ operator at each point along the deformation.}
	\end{figure}
	
	 In flat backgrounds, the $\ttbar$ operator satisfies the following factorization property\cite{Zamolodchikov:2004ce}
	\begin{align}
			\left<\ttbar\right>\equiv\frac{1}{8}\left(\left<T_{ab}\right>\langle T^{ab}\rangle-\left<T_a^a\right>^2\right)\,,\label{TTbar-factorization}
	\end{align}
	leading to the trace-flow equation
	\begin{align}
		\langle\Theta\rangle=4\pi^2\mu\left(\langle T\rangle \langle\bar{T}\rangle-\langle\Theta\rangle^2\right)\,,
	\end{align}
	where $\Theta=T_{w\bar{w}}\equiv\frac{1}{4}T^a_a$ is the trace of the stress-energy tensor and 
	\begin{align*}
		T\equiv T_{ww}~~,~~\bar{T}\equiv T_{\bar{w}\bar{w}}\,.
	\end{align*}
	The $\ttbar$ deformation of a CFT$_2$ features an exactly solvable energy spectrum. For example, for the thermal CFT$_2$ with a conserved angular momentum discussed in  subsection \ref{CFT-review}, the energy spectrum is given as \cite{McGough:2016lol,Apolo:2023aho}
	\begin{align}
		E(\mu,\beta)\beta=-\frac{2\pi}{\tilde{\mu}}\left(1-\sqrt{\displaystyle 1+2\tilde{\mu}\langle E\rangle+\tilde{\mu}^2\langle J\rangle^2}\right)~~,~~\tilde{\mu}=\frac{4\pi^3\mu}{\beta^2}\,.\label{energy-spectrum}
	\end{align} 
	From \cref{energy-momentum-CFT}, when $E_+=E_-=0$, we find that the ground state energy becomes ill-defined beyond the Hagedorn temperature $\beta_\mathcal{H}=2\pi\sqrt{\frac{\pi c\mu}{6}}$.
	
	For a holographic CFT$_2$ dual to three dimensional Einstein gravity in asymptotically AdS spacetime, its $\ttbar$ deformation with $\mu>0$\footnote{The bulk dual for $\mu\leq 0$ has recently been proposed in \cite{Apolo:2023vnm,Apolo:2023ckr} through a gluing procedure.} has been proposed to be dual to the same gravitational theory in a compact region of the AdS spacetime by pushing the holographic screen inside the bulk \cite{McGough:2016lol}. In other words, the holographically dual spacetime is described by the same metric, but now satisfying Dirichlet boundary condition at a finite cut-off $r_c$:
	\begin{align}
		\td s^2=\ell_\textrm{AdS}^2\left(\frac{\td r^2}{r^2}+r^2h_{ab}\td x^a\td x^b\right)~~,~~r\leq r_c=\sqrt{\frac{6\ell_\textrm{AdS}^2}{\pi c\mu}}=\frac{\ell_\textrm{AdS}^2}{\epsilon_c}\,,\label{Cutoff-Dictionary}
	\end{align}
	where $c$ is the central charge of the undeformed CFT$_2$ related to the bulk gravitational constant through the Brown-Henneaux relation \cite{Brown:1986nw}
	\begin{align}
		c=\frac{3\ell_\textrm{AdS}}{2G_N}\,,
	\end{align}
	$\ell_\textrm{AdS}$ being the AdS$_3$ radius. The subscript on the UV cut-off $\epsilon_c$ signifies that the dual CFT$_2$ now lives on the cut-off surface at $r=r_c$. The above holographic proposal has passed several tests including a precise correspondence between all thermodynamic quantities, light cone deformations, exact solutions to the RG flow as well as holographic characterization of entanglement and correlation \cite{McGough:2016lol,Chen:2018eqk,Jeong:2019ylz,Basu:2023bov,Basu:2023aqz,Basu:2024bal,Lewkowycz:2019xse}. As a concrete example, the holographic analogue of the energy spectrum (\ref{energy-spectrum}) is given by the ADM energy of a rotating BTZ black hole of mass $M=\langle E\rangle$ and angular momentum $J$ placed inside a box $r\leq r_c$ with Dirichlet boundary conditions $\td s^2|_{r=r_c}=r_c^2\,\td w\td \bar{w}$ \cite{Brown:1994gs, McGough:2016lol}.
	\subsection{Holographic reflected entropy}
	In this subsection, we describe the mixed state entanglement measure termed as the reflected entropy \cite{Dutta:2019gen} and its holographic description in terms of the entanglement wedge cross-section \cite{Takayanagi:2017knl,Nguyen:2017yqw}. We begin by considering a bipartite mixed state described by the density matrix $\rho_{AB}$ with support on the product Hilbert space $\cH_A\otimes\cH_B$ of the individual parties. As discussed in \cite{Dutta:2019gen}, the canonical purification of this mixed state involves the introduction of the CPT conjugate copies of $A$ and $B$, denoted as $A^\star$ and $B^\star$ respectively and subsequently concocting up a pure state $\ket{\sqrt{\rho_{AB}}}$ on the doubled Hilbert space $\cH_A\otimes\cH_B\otimes\cH_{A^\star}\otimes\cH_{B^\star}$. The reflected entropy $S_R(A:B)$ between the two parties may then be defined as the von Neumann entropy of the reduced density matrix $\rho_{AA^\star}=\textrm{Tr}_{BB^\star}\ket{\sqrt{\rho_{AB}}}\bra{\sqrt{\rho_{AB}}}$, namely
	\begin{align}
		S_R(A:B)=S_\textrm{vN}(\rho_{AA^\star})_{\sqrt{\rho_{AB}}}\,.
	\end{align}
	The authors in \cite{Dutta:2019gen} furthermore developed a novel replica technique to compute the reflected entropy for bipartite states in integrable quantum filed theories. In order to construct the state $\sqrt{\rho_{AB}}$ in the replica fashion, one first considers a $m$-fold ($m\in 2\mathbb{Z}^+$) replication of the original manifold and defines the state $\ket{\psi_m}\equiv\big|\rho_{AB}^{m/2}\big>$ as the canonical purification of the product state $\rho_{AB}^m$. Subsequently, yet another replication in the R\'enyi parameter $n$ results in an $m\times n$-sheeted replica manifold $\cM_{n,m}$ with branch cuts present along the copies of the subsystems $A$ and $B$. Subsequently, these branch cuts are sewed under certain elements $g_A$ and $g_B$ of the replica symmetry group $\cS_{nm}$. The R\'enyi reflected entropy, defined as the R\'enyi entropy of the reduced density matrix 
	\begin{align*}
		\rho_{AA^\star}^{(m)}=\textrm{Tr}_{BB^\star}\ket{\psi_m}\bra{\psi_m}\,,
	\end{align*}
	may then be obtained through a properly weighted partition function on this $nm$-sheeted replica manifold as follows
	\begin{align}
		S_n(AA^\star)_{\psi_m}=\frac{1}{1-n}\log\frac{\mathbb{Z}[\cM_{n,m}]}{\left(\mathbb{Z}[\cM_{1,m}]\right)^n}\,.
	\end{align}
	Finally, the reflected entropy is obtained by analytically continuing both the replica parameters to unity,
	\begin{align}
		S_R(A:B)=\lim_{n,m \to 1}S_n(AA^\star)_{\psi_m}\,.
	\end{align}
	
	The authors of \cite{Dutta:2019gen} have also established a holographic dual of the reflected entropy through the minimal cross-section of the entanglement wedge, the codimension one bulk region dual to the density matrix $\rho_{AB}$ \cite{Takayanagi:2017knl,Nguyen:2017yqw}. In particular, the holographic reflected entropy is exactly twice the area of the minimal entanglement wedge cross-section (EWCS). In a holographic theory, the entanglement wedge $\cE_{AB}$ is constructed as the bulk region bounded by the extremal surface(s) $\Gamma_{AB}$ computing the entanglement entropy for the bipartite state $\rho_{AB}$, and the subsystem $A\cup B$. In the context of AdS$_3$/CFT$_2$ correspondence, extremal surfaces are essentially geodesics and a natural framework to compute the EWCS consists in embedding the AdS$_3$ spacetime in $\mathbb{R}^{2,2}$:
	\begin{align}
		\td s^2=\eta_{\mu\nu}\td X^\mu\td X^\nu~~,~~X^\mu X_\mu=-1~~,~~(\mu=1,\cdots,4).\label{AdS-embedding}
	\end{align}
	In this embedding coordinate formalism, the EWCS corresponding to two disjoint subsystems $A=[X_1,X_2]$ and $B=[X_3,X_4]$ may be comprehensively obtained as \cite{Kusuki:2019evw,Kusuki:2019rbk}
	\begin{align}
		E_W(A:B)=\frac{1}{4G_N}\cosh^{-1}\left(\frac{1+\sqrt{u}}{\sqrt{v}}\right)\,,\label{EW-disj-formula}
	\end{align}
	where $u$ and $v$ are defined as
	\begin{align}
		u=\frac{\zeta_{12}\zeta_{34}}{\zeta_{13}\zeta_{24}}~~,~~v=\frac{\zeta_{14}\zeta_{23}}{\zeta_{13}\zeta_{24}}\,,
	\end{align}
	with $\zeta_{ij}=-X_i\cdot X_j$ being the unique (conformal) invariant associated with the pair of points $(X_i,X_j)$. In a similar fashion, for two adjacent subsystems $A=[X_1,X_2]$ and $B=[X_2,X_3]$, the EWCS is given as \cite{Basu:2023jtf}
	\begin{align}
		\frac{1}{4G_N}\cosh^{-1}\left(\sqrt{\frac{2\zeta_{12}\zeta_{23}}{\zeta_{13}}}\right)\,.\label{EW-adj-formula}
	\end{align}
	\section{$\ttbar$ deformed CFT$_2$ with conserved angular momentum}\label{sec-3}
	In this section, we investigate the effect of $\ttbar$ -deformation on the entanglement structure of a thermal CFT$_2$ with angular momentum defined on the twisted cylinder $\mathcal{M}$. We focus on the case with a small deformation parameter $\mu$, such that the QFT action may be perturbatively expanded around the seed CFT,
	\begin{align}
		S^{(\mu)}_\textrm{QFT}=S_\textrm{CFT}+\mu\int_\mathcal{M} \td^2w\left(T\bar{T}-\Theta^2\right)\,, \label{perturbative-action}
	\end{align}
	where (with a slight abuse of notations) $T\,,\,\bar{T}$ and and $\Theta$ denote the stress tensor components in the undeformed theory. Physical quantities, such as the entanglement entropy, may now be obtained through suitable perturbative techniques around the seed theory \cite{Chen:2018eqk,Jeong:2019xdr,Asrat:2020uib,Asrat:2019end,Basu:2023bov,Basu:2023aqz,Basu:2024bal}. In the following, we particularly focus on the entanglement entropy and the reflected entropy for various bipartite states in a thermal CFT$_2$ with angular momentum deformed by coupling with the $\ttbar$ operator and subsequently study their holographic characterization in the dual bulk theory described by a spinning BTZ black hole with Dirichlet boundary conditions placed at a finite cut-off.
	
	\subsection{Entanglement entropy}
\begin{figure}[ht]
		\centering
		\includegraphics[width=0.45\textwidth]{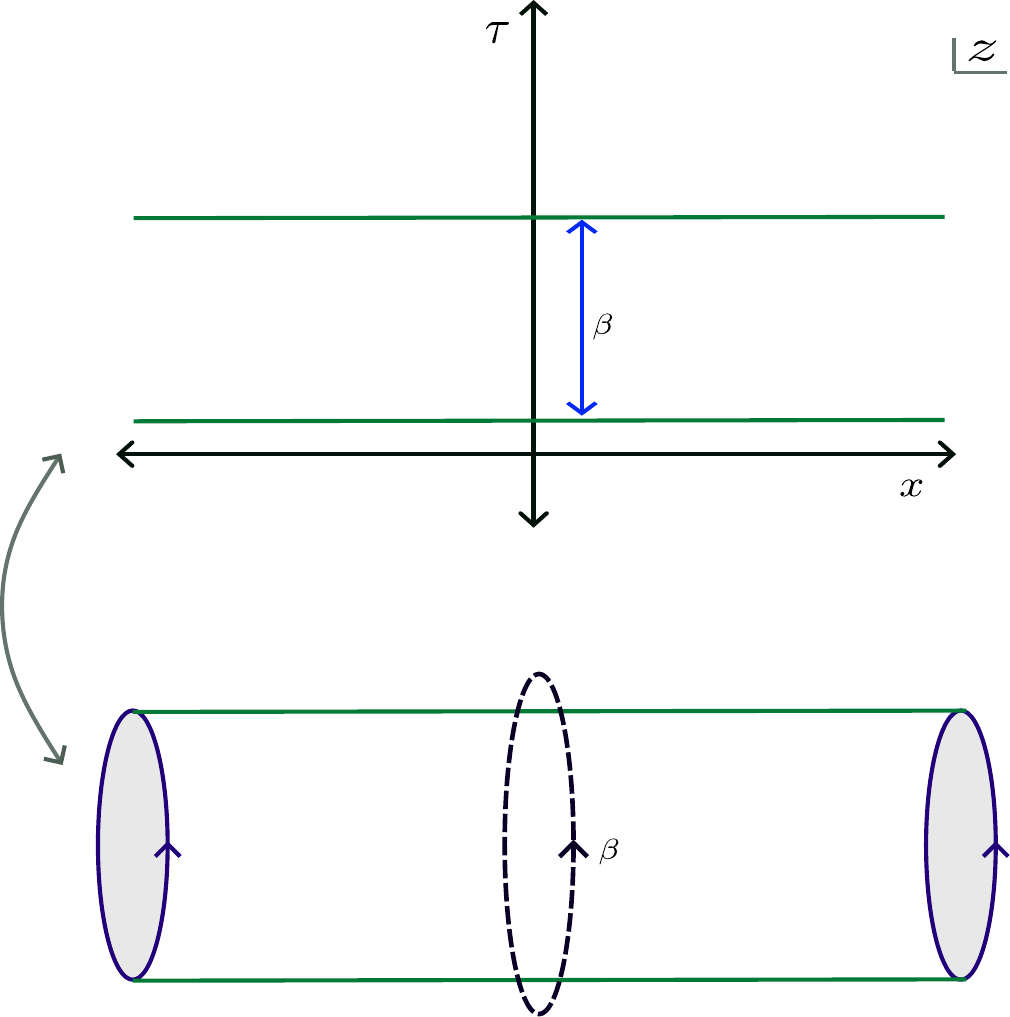}
		\caption{The temporal periodicity of the thermal cylinder is obtained by identifying the two green lines.}
		\label{fig:cylinder}
	\end{figure}

	We begin with the computation of the entanglement entropy of a subsystem $A$ in the thermal CFT$_2$ with angular momentum defined on the twisted cylinder $\cM$. To this end, we recall that the replica technique to compute the $n$-th R\'enyi entropy involves the partition function on an $n$-sheeted replica manifold $\cM_n$ with branch cuts present along copies of $A$ which are sewed cyclically \cite{Calabrese:2004eu,Calabrese:2009qy}:
	\begin{align}
		S_n\left(A\right):=\frac{1}{1-n}\log\frac{\bZ[\cM_n]}{\left(\bZ[\cM]\right)^n}\,.
	\end{align}
	In a $\ttbar$-deformed theory, we have
	\begin{align}
		\bZ[\cM_n]=\int_{\cM_n}\cD \Phi e^{S^{(\mu)}_\textrm{QFT}[\Phi\,;\cM_n]}
	\end{align}
	where $S^{(\mu)}_\textrm{QFT}[\Phi\,;\cM_n]$ denotes the action for the $\ttbar$-deformed theory on the replica manifold $\cM_n$. For a small deformation parameter $\mu\ll 1$, one may expand around the seed CFT$_2$ using \cref{perturbative-action} to obtain the following expression for the replica partition function \cite{Chen:2018eqk,Jeong:2019ylz}
	\begin{align}
		\frac{\bZ[\cM_n]}{(\bZ[\cM])^n}=\left[\frac{\displaystyle\int_{\cM_n}\cD\Phi\, e^{-S_\textrm{CFT}[\Phi]}}{\left(\displaystyle\int_{\cM}\cD\Phi e^{-S_\textrm{CFT}[\Phi]}\right)^n}\right]\left(1-\mu\int_{\cM_n}\langle T\bar{T}\rangle_{\cM_n}+n\mu\int_\cM \langle T\bar{T}\rangle_{\cM}+\cO\left(\mu^2\right)\right)
	\end{align}
	where\footnote{Note that the trace of the stress-energy tensor $\Theta$ has been omitted in these expressions, since any correlation function with a $\Theta$ insertion on a flat manifold vanishes identically \cite{Chen:2018eqk}.} 
	\begin{align}
		\langle T\bar{T}\rangle_{\cM_n}=\frac{\displaystyle\int_{\cM_n}\cD\Phi\,T\bar{T} e^{-S_\textrm{CFT}[\Phi]}}{\displaystyle\int_{\cM_n}\cD\Phi\, e^{-S_\textrm{CFT}[\Phi]}}\label{TTb-expectation-Mn}\,.
	\end{align}
    As a result the leading order correction to the R\'enyi entropy $S_n(A)$ due to the deformation is given by \cite{Chen:2018eqk,Jeong:2019ylz}
	\begin{align}
		\delta S_n(A)=\frac{\mu}{n-1}\left(\int_{\mathcal{M}^n}\langle T\overline{T}\rangle_{\mathcal{M}^n}-n\int_{\mathcal{M}} \langle T\overline{T}\rangle_{\mathcal{M}}\right).\label{Renyi-EE-correction}
	\end{align}
	In the following we specialize to a single interval $A=[(x_1,t_1),(x_2,t_2)]$ on the twisted cylinder $\cM$. To compute the expectation value $\langle T\overline{T}\rangle_{\mathcal{M}^n}$ on the replica manifold, we employ the twist operator formalism developed in \cite{Jeong:2019ylz} to obtain \cite{Calabrese:2004eu,Calabrese:2009qy,Sun:2019ijq}
	\begin{align}
		\int_{\mathcal{M}^n}\langle T\overline{T}\rangle_{\mathcal{M}^n} 
		&=\frac{1}{n}\int_\mathcal{M}\frac{\langle T^{(n)}(w)\overline{T}^{(n)}(\bar{w})\sigma_n(w_1,\bar{w}_1)\bar{\sigma}_n(w_2,\bar{w}_2)\rangle_{\mathcal{M}}}{\langle\sigma_n(w_1,\bar{w}_1)\bar{\sigma}_n(w_2,\bar{w}_2)\rangle_{\mathcal{M}}}
	\end{align}
	where $T^{(n)}(w)\,,\bar{T}^{(n)}(\bar{w})$ are the total stress energy tensors of $n$ replica sheets, $\sigma_n$ and $\bar{\sigma}_n$ are the twist operators inserted at the endpoint of the interval which employ twisted boundary conditions at their locations to cyclically sew copies of the original manifold \cite{Calabrese:2004eu,Calabrese:2009qy}. These twist operators are conformal primaries with dimensions
	\begin{align}
		h_{\sigma_n} =\bar{h}_{\sigma_n}=\frac{c}{24}\left(n-\frac{1}{n}\right).
	\end{align}
	In order to compute the correlation functions on the twisted cylinder $\mathcal{M}$, we employ the conformal map \cref{plane-to-cylinder-map}. The resulting correlation function with stress tensor insertions may be obtained utilizing the conformal Ward identities \cite{Jeong:2019ylz} and the conformal anomaly in \cref{Schwarzian-transformation}. 
    Finally we obtain
	\begin{align}
		n\,\langle T\overline{T}\rangle_{\mathcal{M}^n}
		=&\frac{1}{\langle\sigma_n(z_1, \bar{z}_1)\bar{\sigma}_n(z_2, \bar{z}_2)\rangle_{\mathcal{C}}}\left[-\frac{\pi^2nc}{6{\beta_+}^2}+\left(\frac{2\pi}{\beta_+}z\right)^2\sum_{j=1}^{2}\left(\frac{h_{\sigma_n}}{(z-z_j)^2}+\frac{1}{z-z_j}\partial_{z_j}\right)\right]\notag\\
		&\times\left[-\frac{\pi^2nc}{6{\beta_-}^2}+\left(\frac{2\pi}{\beta_-}\bar{z}\right)^2\sum_{j=1}^{2}\left(\frac{\bar{h}_{\sigma_n}}{(\bar{z}-\bar{z}_j)^2}+\frac{1}{\bar{z}-\bar{z}_j}\partial_{\bar{z}_j}\right)\right]
		\langle\sigma_n(z_1, \bar{z}_1)\bar{\sigma}_n(z_2, \bar{z}_2)\rangle_{\mathcal{C}},
		\label{TTb-expectation}
	\end{align}
	The two-point twist correlator is given by
	\begin{align}
		\langle\sigma_n(z_1, \bar{z}_1)\bar{\sigma}_n(z_2, \bar{z}_2)\rangle_{\mathcal{C}} &= \frac{c_n}{(z_1-z_2)^{h_{\sigma_{n}}+\bar{h}_{\sigma_{n}}}(\bar{z}_1-\bar{z}_2)^{h_{\sigma_{n}}+\bar{h}_{\sigma_{n}}}}
	\end{align}
	Substituting the above expression in \cref{TTb-expectation}, and subsequently utilizing \cref{Renyi-EE-correction}, we may now obtain the leading order correction to the entanglement entropy by employing the replica limit $n\to 1$ as follows
	\begin{align}
			\delta S(A) 
			=&-\mu \left(\frac{c}{12}\right)^2\frac{8\pi^4}{(\beta_+ \beta_-)^2}\int_{\mathcal{M}}\dd^2 w\left(\frac{z^2(z_1-z_2)^2}{(z-z_1)^2(z-z_2)^2}+\frac{\bar{z}^2(\bar{z}_1-\bar{z}_2)^2}{(\bar{z}-\bar{z}_1)^2(\bar{z}-\bar{z}_2)^2}\right) \notag\\
				=&-\mu \left(\frac{c}{12}\right)^2\frac{8\pi^4}{(\beta_+ \beta_-)^2}\int_{0}^{i\Omega_E}\td x\int_{0}^{\beta}\td\tau \Bigg[\frac{ e^{\frac{4 \pi  (x+i \tau )}{\beta_+}} \left(z_1-z_2\right)^2}{ \left(e^{\frac{2 \pi  (x+i \tau )}{\beta_+}}-z_1\right)^2 \left(e^{\frac{2 \pi  (x+i \tau )}{\beta_+}}-z_2\right)^2}\notag\\&\qquad\qquad\qquad\qquad\qquad\qquad\qquad\qquad+\frac{ e^{\frac{4 \pi  (x-i \tau )}{\beta_-}} \left(\bar{z}_1-\bar{z}_2\right)^2}{ \left(e^{\frac{2 \pi  (x-i \tau )}{\beta_-}}-\bar{z}_1\right)^2 \left(e^{\frac{2 \pi  (x-i \tau )}{\beta_-}}-\bar{z}_2\right)^2}\Bigg]\,.\label{delta-S-integral}
			\end{align} 
				\begin{figure}[ht]
				\begin{center}
					\includegraphics[width=0.6\textwidth]{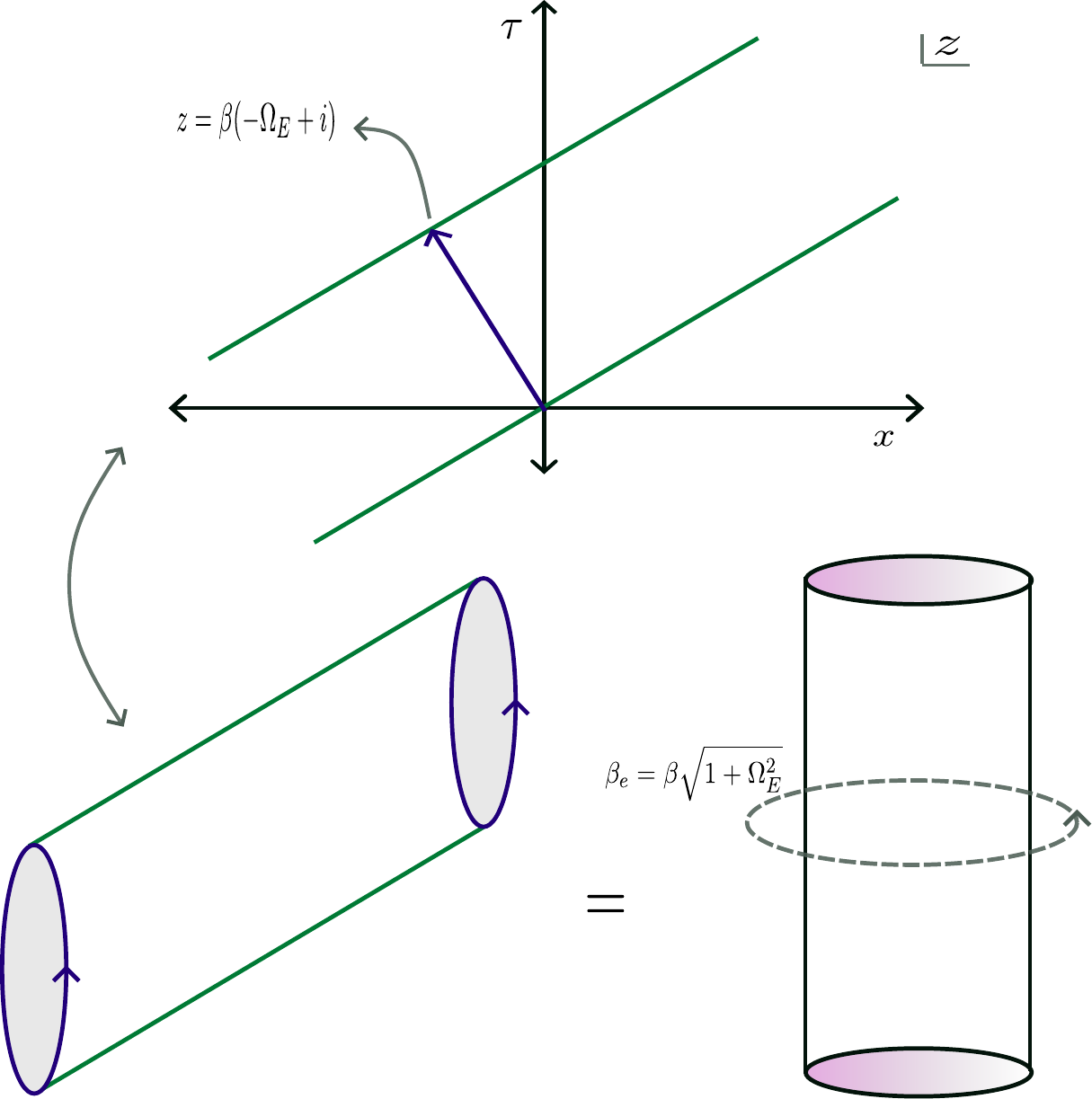}
					\caption{Structure of the twisted cylinder $\cM$. The two green lines are identified which leads to the twisted identifications in \cref{twisted-identification-xt}.}
					\label{twisted cylinder}
				\end{center}
			\end{figure}
	Similar definite integrals were evaluated in \cite{Chen:2018eqk,Jeong:2019ylz} through careful consideration of various logarithmic branch cuts. However, in the present context, the evaluation of the above definite integrals requires a bit more machinery and a careful investigation of the twisted cylinder manifold.	To proceed, we note that the twisted identifications in \cref{twisted-identification-xt,twisted-identification-w} correspond to identifying $\tau=\Omega_E\,x$ with $\tau=\Omega_E\,x+\beta(1+\Omega_E^2)$ on the $(\tau,x)$ plane, as depicted in \cref{twisted cylinder}. This clearly leads to the domain of integration as in \cref{delta-S-integral}. It is evident that, if we consider a rotation by an angle $\theta_0=\tan^{-1}(\Omega_E)$ and redefine the $x$ and $\tau$ coordinates as
	\begin{align}
		\begin{pmatrix}
			x^\prime \\ \tau^\prime
		\end{pmatrix}
		=\frac{1}{\sqrt{1+\Omega_E^2}} \begin{pmatrix}
			1 & \Omega_E\\
			-\Omega_E  & 1
		\end{pmatrix}
		\begin{pmatrix}
			x \\ \tau
		\end{pmatrix}\label{Bichi-map}
	\end{align}
	then the compactifications in the new directions correspond to the standalone thermal identification:
	\begin{align}
		\tau^\prime\sim \tau^\prime+\beta\sqrt{1+\Omega_E^2}\,,\label{effective-temp}
	\end{align}
	as depicted in \cref{fig:cylinder}.
	In terms of the complex coordinates $(w^\prime,\bar{w}^\prime)$, the earlier twisted identifications transform to standard thermal identifications:
	\begin{align}
		w^\prime\sim w^\prime+i\beta\sqrt{1+\Omega_E^2}~~,~~\bar{w}^\prime\sim \bar{w}^\prime-i\beta\sqrt{1+\Omega_E^2}\,.
	\end{align}
	In other words, we now have a standard thermal cylinder on which lives a CFT$_2$ at a finite temperature $\beta_e:=\beta\sqrt{1+\Omega_E^2}$ devoid of any angular momentum. A few comments are in order:
		\begin{itemize}
		\item First of all, the coordinate transformation in \cref{Bichi-map} is essentially a rotation : $w\to e^{i\theta_0}w$ with $\theta_0=\tan^{-1}(\Omega_E)$, which is a global conformal transformation (see \cref{appB} for more details). hence no information about the state is lost, only transformed to the redefinition of the temperature.
		\item The bulk dual of this coordinate transformation (the Banados map) does not change the bulk metric. Hence the underlying physics remains unaltered.
	\end{itemize}
	
	The definite integrals in \cref{delta-S-integral} reduces to the following in the rotated frame $(x^\prime,\tau^\prime)$:
	\begin{align}
		\mathcal{I}=J\int_{-\infty}^\infty \dd x^\prime \int_{0}^{\beta_e}\dd \tau^\prime\Bigg[&\frac{e^{\frac{4 \pi  (x^\prime+i \tau^\prime )}{\beta_e}} \left(z^\prime_1-z^\prime_2\right)^2}{ \left(e^{\frac{2 \pi  (x^\prime+i \tau^\prime )}{\beta_e}}-z^\prime_1\right)^2 \left(e^{\frac{2 \pi  (x^\prime+i \tau^\prime )}{\beta_e}}-z^\prime_2\right)^2}+\textrm{h.c}\Bigg]
	\end{align} 
	where $J=1$ is the Jacobian of the coordinate transformation and $z^\prime_k=e^{\frac{2\pi(x^\prime_k+i \tau^\prime_k )}{\beta_e}}$.
	The above integral was evaluated in \cite{Chen:2018eqk,Jeong:2019ylz}, with the result\footnote{Note that the argument of the logarithm contains the absolute value since we consider the entanglement entropy of a boosted interval as opposed to the spatial interval considered in \cite{Jeong:2019ylz}.}:
	\begin{align}
		\mathcal{I}&=\frac{\beta_e^2}{2\pi}\left(\frac{z^\prime_1+z^\prime_2}{z^\prime_1-z^\prime_2}\log\left|\frac{z^\prime_1}{z^\prime_2}\right|+\frac{\bar{z}^\prime_1+\bar{z}^\prime_2}{\bar{z}^\prime_1-\bar{z}^\prime_2}\log\left|\frac{\bar{z}^\prime_1}{\bar{z}^\prime_2}\right|\right)\notag\\
		&=\beta_e(x^\prime_2-x^\prime_1)\frac{ \left(e^{2 \pi  \left(\frac{x_1-i \tau _1}{\beta_-}+\frac{x_1+i \tau _1}{\beta_+}\right)}-e^{2 \pi  \left(\frac{x_2-i \tau _2}{\beta_-}+\frac{x_2+i \tau _2}{\beta_+}\right)}\right)}{ \left(e^{\frac{2 \pi  \left(x_1-i \tau _1\right)}{\beta_-}}-e^{\frac{2 \pi  \left(x_2-i \tau _2\right)}{\beta_-}}\right) \left(e^{\frac{2 \pi  \left(x_1+i \tau _1\right)}{\beta_+}}-e^{\frac{2 \pi  \left(x_2+i \tau _2\right)}{\beta_+}}\right)}
	\end{align}
	Transforming back to the original twisted cylinder and subsequently considering the Wick rotation $\tau=-it\,,\Omega_E=-i\Omega$, we obtain the following expression for the leading order correction to the entanglement entropy
	\begin{align}
		\delta S(A)=-\mu\frac{\pi^4c^2}{18(\beta_+ \beta_-)^{3/2}}\frac{(x_{21}-\Omega\,t_{21})}{\sqrt{1-\Omega^2}}\left[\coth \left(\frac{\pi  (x_{21}+ t_{21}) }{\beta_+}\right)+\coth \left(\frac{\pi  (x_{21}-t_{21}) }{\beta_- }\right)\right]\label{EE-corrections}
	\end{align}
	As a consistency check, we find that for a vanishing angular speed, our result reduces to that of a thermal CFT$_2$ with $\TTbar$ deformation \cite{Jeong:2019ylz}. For a small angular speed $\Omega\ll 1$ (in the bulk, this corresponds to a very slowly spinning black hole), $\delta S(A)$ becomes
	\begin{align}
		\delta S(A)&=-\frac{\mu \pi^4c^2}{18\beta^3}\Bigg((x_{21}-\Omega\,t_{21})\left[\coth \left(\frac{\pi  (x_{21}+ t_{21}) }{\beta}\right)+\coth \left(\frac{\pi  (x_{21}-t_{21}) }{\beta}\right)\right]\notag\\
		&-\frac{\pi  x_{21} \Omega}{\beta}\left[\left(x_{21}-t_{21}\right) \csch^2\left(\frac{\pi  \left(t_{21}-x_{21}\right)}{\beta }\right)+\left(t_{21}+x_{21}\right) \csch^2\left(\frac{\pi  \left(t_{21}+x_{21}\right)}{\beta }\right)\right]\Bigg)
	\end{align}
	
\begin{figure}[h!]
		\centering
		\includegraphics[width=0.45\textwidth]{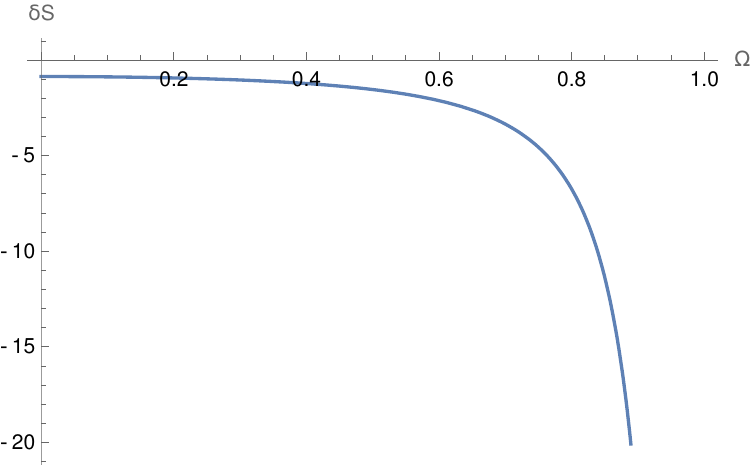}
		\hspace{.3cm}
		\includegraphics[width=0.45\textwidth]{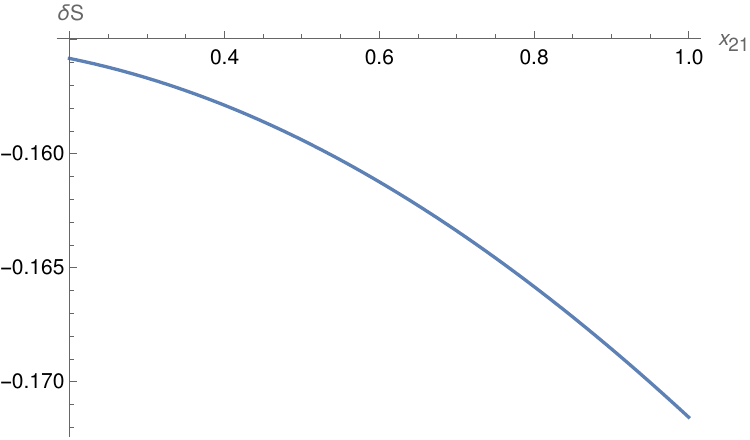}
		\caption{Leading order correction for purely spacelike interval, in the left pannel $\abs{x_{21}}=20$ and in the right pannel $\Omega=0.5$\,. We have set $\beta=2\pi$,\, $\mu=1$  and $c=1$.}
		\label{fig:plots-spacelike-EE}
	\end{figure}

\begin{figure}[h!]
		\centering
		\includegraphics[width=0.45\textwidth]{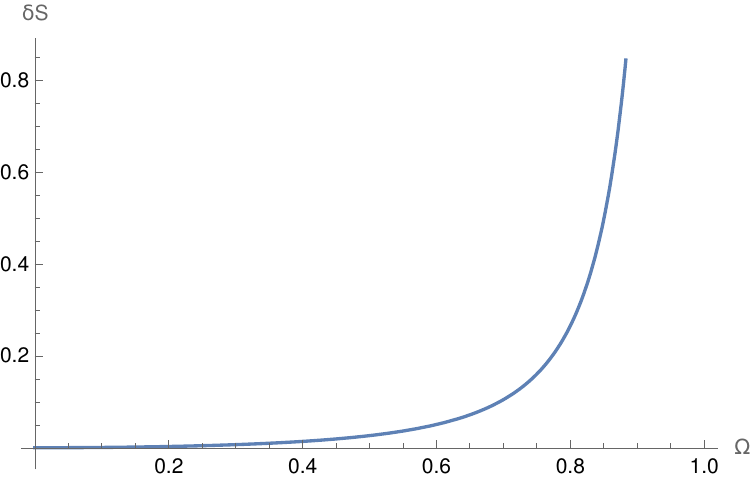}
		\includegraphics[width=0.45\textwidth]{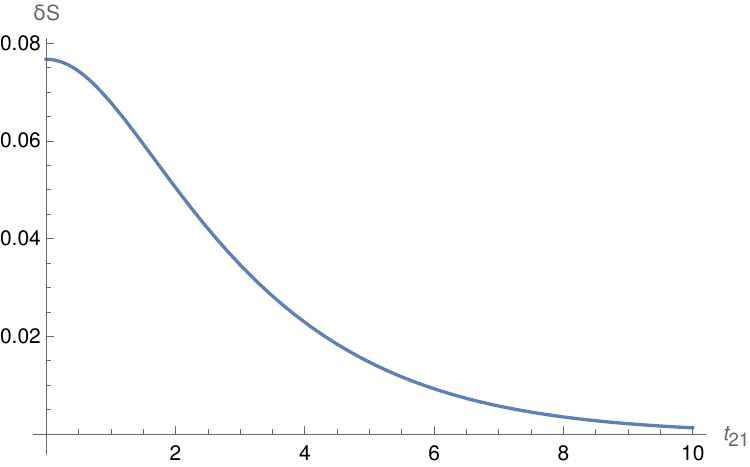}
		\caption{Leading order correction for purely timelike interval, in the left pannel $\abs{t_{21}}=2$ and in the right pannel $\Omega=0.6$\,. We have set $\beta=2 \pi$, $\mu=1$ and $c=1$.}
		\label{fig:plots-timelike-EE}
	\end{figure}

	In the following, we analyze the behavior of (correction to) the entanglement entropy for spacelike and timelike subsystems separately.
	\paragraph{Purely spacelike interval:} We first consider a purely spatial interval $A=[(x_1,t),(x_2,t)]$. In this case, the leading order correction to the entanglement entropy is given by
	\begin{align}
		\delta S(A)=-\frac{\mu\pi^4 c^2x_{21}}{18\beta^3(1-\Omega^2)^2}\left[\coth\left(\frac{\pi x_{21}}{\beta_+}\right)+\coth\left(\frac{\pi x_{21}}{\beta_-}\right)\right]
	\end{align} 
	which is always negative as the quantity inside the square braces never vanishes (recall that $\Omega<1)$. In \cref{fig:plots-spacelike-EE}, we have plotted $\delta S(A)$ with respect to the subsystem size (right panel) and the angular speed $\Omega$ (left panel).
	\paragraph{Purely timelike interval:} Next we consider a purely timelike interval $A=[(x,t_1),(x,t_2)]$, for which $\delta S(A)$ is given as
	\begin{align}
		\delta S^{(\textrm{T})}(A)=\frac{\mu\pi^4c^2\Omega\, t_{21}}{18\beta^3(1-\Omega^2)^2}\frac{\sinh\left[\frac{2\pi t_{21} \Omega}{\beta(1-\Omega^2)}\right]}{\sinh\left(\frac{\pi t_{21}}{\beta_+}\right)\sinh\left(\frac{\pi t_{21}}{\beta_-}\right)}\,.\label{deltaS-tiemlike}
	\end{align}
	Interestingly, the leading correction to the timelike entanglement is non-vanishing for our thermal cylinder due to the presence of angular momentum. Furthermore, contrary to the case of spacelike interval, we find that $\delta S(A)$ is always positive\footnote{However, for a very large timelike interval $\delta S(A)$ becomes vanishingly small.}. To our knowledge, this is the first instance of a positive (leading) correction to the entanglement entropy due to $\ttbar$-deformation \cite{Chen:2018eqk,Jeong:2019xdr}. We will have more to say about this behavior in subsection \ref{HEE}. In \cref{fig:plots-timelike-EE}, we have plotted $\delta S(A)$ with respect to the subsystem size (right panel) and the angular speed $\Omega$ (left panel).

\begin{figure}[h!]
    	\centering
    	\includegraphics[width=0.495\textwidth]{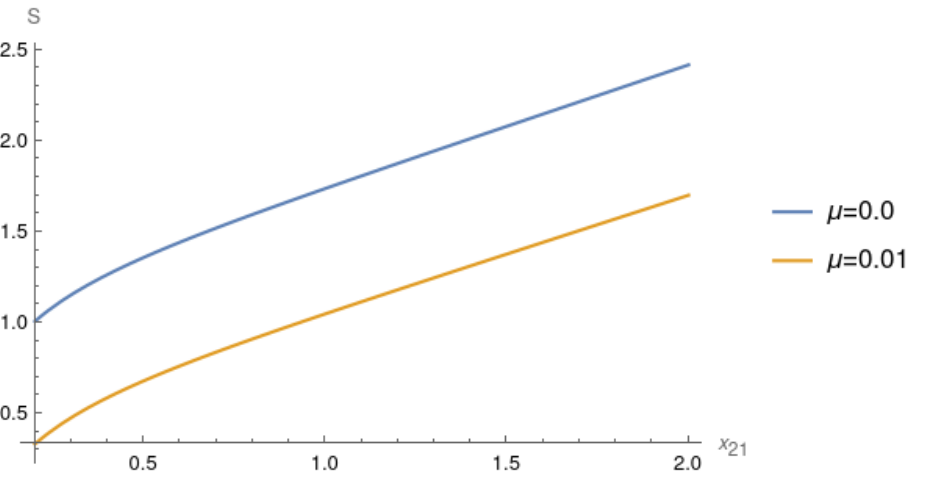}
    	\includegraphics[width=0.495\textwidth]{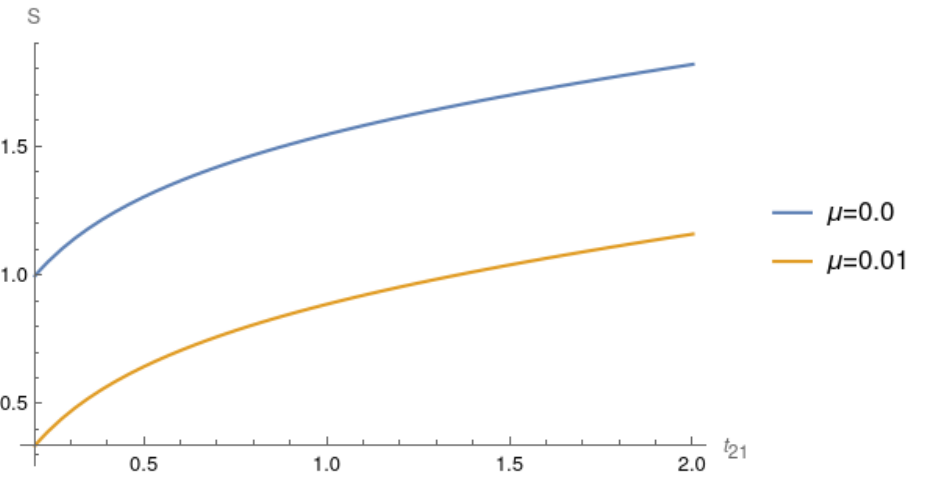}
    	\caption{Comparison of the variation of entanglement entropy with subsystem size, for an undeformed CFT$_2$ and that with $\ttbar$ deformation: left- spacelike interval, right-timelike interval. We have set $\Omega=0.1$,\,$\beta=2 \pi$ and $c=1$.}
    	\label{fig:plots-difference}
    \end{figure}

    Although the leading order corrections for a purely timelike interval remains positive, we have verified that the total entanglement entropy for the undeformed CFT$_2$ is always greater than that with a $\ttbar$ deformation (cf. \cref{fig:plots-difference}). 

	\subsection{Reflected entropy}
	In this subsection, we compute the reflected entropy for various bipartite states in a $\TTbar$-deformed CFT$_2$ with conserved angular momentum, defined on the twisted cylinder $\cM$. A replica technique to compute the reflected entropy in such $\TTbar$-deformed theories was developed in \cite{Asrat:2020uib,Basu:2024bal}, which we briefly recapitulate below. The partition function computing the R\'enyi reflected entropy on the $nm$-sheeted replica manifold $\cM_{n,m}$ may be written as follows 
    \begin{equation}
    	\mathbb{Z}[\cM_{n,m}]= \int_{\cM_{n,m}} \mathcal{D}\Phi \,e^{-S^{(\mu)}_\text{QFT}[\Phi;\cM_{n,m}]},
    \end{equation}
    with $S^{(\mu)}_\text{QFT}[\Phi;\cM_{n,m}]$ being the action of the $\TTbar$-deformed theory on the replica manifold $\mathcal{M}_{nm}$. As earlier, we consider an infinitesimal deformation around the seed CFT$_2$, with $\mu\ll 1$. Then the (normalized) replica partition function may be expanded around the seed theory to obtain
    \begin{align}
    	\frac{\bZ[\cM_{n,m}]}{(\bZ[\cM_{1,m}])^n}=\left[\frac{\displaystyle\int_{\cM_{n,m}}\cD\Phi\, e^{-S_\textrm{CFT}[\Phi]}}{\left(\displaystyle\int_{\cM_{1,m}}\cD\Phi e^{-S_\textrm{CFT}[\Phi]}\right)^n}\right]\left(1-\mu\int_{\cM_{n,m}}\langle T\bar{T}\rangle_{\cM_{n,m}}+n\mu\int_{\cM_{1,m}} \langle T\bar{T}\rangle_{\cM_{1,m}}+\cO\left(\mu^2\right)\right)\,,
    \end{align}
    with $\langle T\bar{T}\rangle_{\cM_{n,m}}$ bearing an expression similar to \cref{TTb-expectation-Mn}. Hence, the leading order correction to the R\'enyi reflected entropy may be written as follows \cite{Asrat:2020uib,Basu:2024bal}
    \begin{align}
    	\delta S_n (A A^\star)_{\psi_m}= \frac{\mu}{n-1} \left(\int_{\mathcal{M}_{n,m}}\langle T \overline{T} \rangle_{\mathcal{M}_{n,m}} - n \int_{\mathcal{M}_{1,m}} \langle T \overline{T} \rangle_{\mathcal{M}_{1,m}} \right)\label{SR-correction}\,.
    \end{align}
    Subsequently, performing the analytic continuation of the replica parameters $n,m\to 1$, the leading correction to the reflected entropy is obtained.
    
    The expectation value of the $\TTbar$ operator on the replica manifold for a generic bipartite state may be obtained by utilizing the twist operator formalism similar to case of the entanglement entropy as follows \cite{Asrat:2020uib}
    \begin{equation}\label{TT_exp}
    	\begin{aligned}
    		\int_{\mathcal{M}_{n,m}} \langle T \overline{T} \rangle_{\mathcal{M}_{n,m}}= \frac{1}{nm}\int_\mathcal{M}\frac{\langle T^{(nm)} (w) \bar T^{(nm)} (\bar w) \Pi_i \sigma_i  (w_i,\bar w_i) \rangle_\mathcal{M}}{\langle \Pi_i \sigma_i  (w_i,\bar w_i) \rangle_\mathcal{M}} \,,
    	\end{aligned}
    \end{equation}
    where T$^{(nm)}$ is the total energy tensor on the $nm$-replicated manifold and $\sigma_i$ are appropriate twist operators inserted at the endpoints of the subsystems $A$ and $B$. Subsequently, the correlation function with $T$ and $\bar T$ insertions may be evaluated by employing the conformal Ward identities. Similarly the expectations value for $\TTbar$ can be obtained for the replica manifold $\mathcal{M}_m$.
    \subsubsection{Two disjoint intervals}
    We begin with the case of two boosted disjoint intervals $A=[(x_1,\tau_1),(x_2,\tau_2)]$ and $B=[(x_3,\tau_3),(x_4,\tau_4)]$. Upon utilizing the conformal map (\ref{plane-to-cylinder-map}) to the complex plane, the relevant four-point twist correlators have the following large central charge structure \cite{Fitzpatrick:2014vua,Dutta:2019gen} 
    \begin{align}
    	\log\langle\sigma\rangle&\equiv\log\langle\sigma_{g^{}_A}(z_1)\sigma_{g^{-1}_A}(z_2)\sigma_{g^{}_B}(z_3)\sigma_{g^{-1}_B}(z_4)\rangle\notag\\&\approx -2h_{g^{}_A}\log(1-\eta)-2\bar h_{g^{}_A}\log(1-\bar\eta)+h_{g^{-1}_Ag^{}_B}\log\left(\frac{1+\sqrt{\eta}}{1-\sqrt{\eta}}\right)+\bar h_{g^{-1}_Ag^{}_B}\log\left(\frac{1+\sqrt{\bar\eta}}{1-\sqrt{\bar\eta}}\right)\label{4-point}
    \end{align}
    where $\eta=\frac{(z_1-z_2)(z_3-z_4)}{(z_1-z_3)(z_2-z_4)}$ is the conformal cross-ratio, $h_{g^{-1}_Ag^{}_B}$ is the conformal dimension corresponding to the dominant (composite) operator $\sigma_{g^{-1}_A g^{}_B}$ exchanged between $\sigma_{g^{-1}_A}$ and $\sigma_{g^{}_B}$. The four-point twist correlator on the $m$-sheeted replica manifold may be readily obtained from the above expression as
    \begin{align}
    	\log\langle\sigma_m\rangle&\equiv\log\langle\sigma_{g^{}_m}(z_1)\sigma_{g^{-1}_m}(z_2)\sigma_{g^{}_m}(z_3)\sigma_{g^{-1}_m}(z_4)\rangle=\lim_{n \to 1}\log\langle\sigma\rangle\,.
    \end{align}
    The conformal dimensions of these twist operators are given by \cite{Dutta:2019gen}
    \begin{align}\label{C-dim}
    	h_{g^{}_A}=h_{g^{}_B}=n h_{g^{}_m}=\frac{nc}{24}\left(m-\frac{1}{m}\right)~~,~~h_{g^{-1}_Ag^{}_B}=\frac{c}{12}\left(n-\frac{1}{n}\right)\,.
    \end{align}
    Now utilizing \cref{TT_exp,4-point} and the conformal Ward identities, we may obtain the leading order correction to the R\'enyi reflected entropy from \cref{SR-correction} as follows \cite{Basu:2024bal}
    \begin{equation} \label{Snm-disj}
    	\begin{aligned}
    		\delta S_n \left(A A^\star \right)_{\psi_m} = \frac{\mu}{n-1}\int_\mathcal{M} &\Bigg[-\frac{2 \pi^4 c}{3 \beta_+^2 \beta_-^2} \Bigg(z^2 \sum_{i=1}^{4} \left( \frac{{h_{g^{}_i}}-n\, {h_{g^{}_m}}}{(z-{z_i})^2} + \frac{\partial_{z_i}({\log \langle \sigma \rangle}-{n \, \log \langle \sigma_m \rangle})}{z-{z_i}} \right)\\
    		&\qquad \qquad \quad +{\bar{z}}^2 \sum_{i=1}^{4} \left( \frac{{\bar{h}_{g^{}_i}}-n\, {\bar{h}_{g^{}_m}}}{(\bar{z}-{\bar{z}_i})^2} + \frac{\partial_{\bar{z}_i}({\log \langle \sigma \rangle}-{n \, \log \langle \sigma_m \rangle})}{\bar{z}-{\bar{z}_i}} \right) \Bigg)\\
    		&+\frac{16 \pi^2}{\beta_+ \beta_- \, m} \Bigg(\frac{z^2 {\bar{z}}^2}{n} \sum_{i,j=1}^{4} \left( \frac{{h_{g^{}_i}}}{(z-{z_i})^2} + \frac{\partial_{z_i}{\log \langle \sigma \rangle}}{z-{z_i}} \right) \left( \frac{{\bar{h}_{g^{}_j}}}{(\bar{z}-{\bar{z}_j})^2} + \frac{\partial_{\bar{z}_j}{\log \langle \sigma \rangle}}{\bar{z}-{\bar{z}_j}} \right)\\
    		& -n \, z^2 {\bar{z}}^2 \sum_{i,j=1}^{4} \left( \frac{{h_{g^{}_m}}}{(z-{z_i})^2} + \frac{\partial_{z_i}{\log \langle \sigma_m \rangle}}{z-{z_i}} \right) \left( \frac{{\bar{h}_{g^{}_m}}}{(\bar{z}-{\bar{z}_j})^2} + \frac{\partial_{\bar{z}_j}{\log \langle \sigma_m \rangle}}{\bar{z}-{\bar{z}_j}}\right)\Bigg)\Bigg].
    	\end{aligned}
    \end{equation} 
    Finally, taking the replica limit $n,m\to 1$ we may obtain the leading correction to the reflected entropy for the present configuration as follows
    \begin{align}
    	\delta S_R \left(A:B\right)= - \frac{\pi^4 c^2 \mu}{9 \beta_+^2\beta_-^2}\int_{0}^{i\Omega_E}\td x\int_{0}^{\beta}\td\tau\left( \frac{z^2 \sqrt{(z_1-z_2) (z_1-z_3) (z_2-z_4) (z_3-z_4)}}{(z-z_1) (z-z_2) (z-z_3) (z-z_4)}+ \text{h.c.} \right)  \,,
    \end{align}
    where, $z = e^{\frac{2\pi(x+i\tau)}{\beta_+}}\,,\,\bar{z} = e^{\frac{2 \pi (x + i \tau)}{\beta_-}}$. In order to compute the above definite integral, we again employ the rotated frame $(x^\prime,\tau^\prime)$ given in \cref{Bichi-map}. As a result, the definite integral under consideration reduces to (recall that the Jacobian of transformation is unity)
    \begin{align*}
    	\cI_\textrm{disj}= \int_{-\infty}^\infty \dd x^\prime \int_{0}^{\beta_e}\dd \tau^\prime \left[ \frac{e^{\frac{4\pi(x^\prime + i \tau^\prime)}{\beta_e}} \sqrt{(z^\prime_1-z^\prime_2) (z^\prime_1-z^\prime_3) (z^\prime_2-z^\prime_4) (z^\prime_3-z^\prime_4)}}{\left(e^{\frac{2 \pi (x^\prime + i \tau^\prime)}{\beta_e}}-z^\prime_1\right) \left(e^{\frac{2 \pi (x^\prime + i \tau^\prime)}{\beta_e}}-z^\prime_2\right)\left(e^{\frac{2 \pi (x^\prime + i \tau^\prime)}{\beta_e}}-z^\prime_3\right)\left(e^{\frac{2 \pi (x^\prime + i \tau^\prime)}{\beta_e}}-z^\prime_4\right)}+ \text{c.c.} \right]  \,,
    \end{align*}
    As earlier, we may evaluate this integral utilizing the tools developed in \cite{Chen:2018eqk,Jeong:2019ylz} with a careful consideration of various branch cuts, resulting in
    \begin{align}
    	\cI_\textrm{disj}&=\frac{\beta_e^2}{4\pi}\sqrt{\frac{(z_1-z_2)(z_3-z_4)}{(z_1-z_3)(z_2-z_4)}}\left(\frac{z_2+z_1}{z_2-z_1}\log\left|\frac{z^\prime_2}{z^\prime_1}\right|+\frac{z_4+z_3}{z_4-z_3}\log\left|\frac{z^\prime_4}{z^\prime_3}\right|-\frac{z_4+z_1}{z_4-z_1}\log\left|\frac{z^\prime_4}{z^\prime_1}\right|-\frac{z_3+z_2}{z_3-z_2}\log\left|\frac{z^\prime_3}{z^\prime_2}\right|\right)\notag\\
    	&+\frac{\beta_e^2}{4\pi}\sqrt{\frac{(\bar z_1-\bar z_2)(\bar z_3-\bar z_4)}{(\bar z_1-\bar z_3)(\bar z_2-\bar z_4)}}\left(\frac{\bar z_2+\bar z_1}{\bar z_2-\bar z_1}\log\left|\frac{\bar z^\prime_2}{\bar z^\prime_1}\right|+\frac{\bar z_4+\bar z_3}{\bar z_4-\bar z_3}\log\left|\frac{\bar z^\prime_4}{\bar z^\prime_3}\right|-\frac{\bar z_4+\bar z_1}{\bar z_4-\bar z_1}\log\left|\frac{\bar z^\prime_4}{\bar z^\prime_1}\right|-\frac{\bar z_3+\bar z_2}{\bar z_3-\bar z_2}\log\left|\frac{\bar z^\prime_3}{\bar z^\prime_2}\right|\right)\label{deltaSR-disj}
    \end{align}
    Transforming back to the original unprimed coordinates and subsequently performing the Wick rotation $\tau=-i\,t\,,\Omega_E=-i\,\Omega$, the leading order correction to the reflected entropy due to the $\TTbar$-deformation may be written as
    \begin{align}
    	\delta S_R(A:B)=&-\frac{\pi ^4 \beta  c^2 \mu }{18 \beta_-^2 \beta_+^2}\sqrt{\eta}\left(\cP_{12}+\cP_{34}-\cP_{14}-\cP_{23}\right)\notag\\
    	&-\frac{\pi ^4 \beta  c^2 \mu }{18 \beta_-^2 \beta_+^2}\sqrt{\bar\eta}\left(\bar\cP_{12}+\bar\cP_{34}-\bar\cP_{14}-\bar\cP_{23}\right)\,,\label{SR-disj-corrections}
    \end{align}
    where we have defined the functions
    \begin{align}
    	\cP_{ij}=(x_{ij}-t_{ij}\Omega)\coth\left(\frac{\pi(x_{ij}+t_{ij})}{\beta_+}\right)~~,~~\bar\cP_{ij}=(x_{ij}-t_{ij}\Omega)\coth\left(\frac{\pi(x_{ij}-t_{ij})}{\beta_-}\right)\label{P-functions}
    \end{align}
    Furthermore, in \cref{deltaSR-disj}, $\eta,\bar\eta$ denote the finite temperature cross-ratios:
    \begin{align}
    	\eta=\frac{\sinh \left(\frac{\pi  \left(x_{21}+t_{21}\right)}{\beta_+}\right) \sinh \left(\frac{\pi  \left(x_{43}+t_{43}\right)}{\beta_+}\right)}{\sinh \left(\frac{\pi  \left(x_{31}+t_{31}\right)}{\beta_+}\right) \sinh \left(\frac{\pi  \left(x_{42}+t_{42}\right)}{\beta_+}\right)}~~,~~\bar\eta=\frac{\sinh \left(\frac{\pi  \left(x_{21}-t_{21}\right)}{\beta_-}\right) \sinh \left(\frac{\pi  \left(x_{43}-t_{43}\right)}{\beta_-}\right)}{\sinh \left(\frac{\pi  \left(x_{31}-t_{31}\right)}{\beta_-}\right) \sinh \left(\frac{\pi  \left(x_{42}-t_{42}\right)}{\beta_-}\right)}\label{cross-ratios}
    \end{align}
    The leading correction to the reflected entropy between two boosted disjoint intervals in a $\TTbar$-deformed thermal CFT$_2$ defined on a cylinder with temporal compactification may be reproduced from the above expression simply by setting $\beta_+=\beta_-$ or $\Omega=0$. A direct comparison to the results reported in \cite{Basu:2024bal} serves as a strong consistency check of the above analysis.
    \subsubsection*{Timelike entanglement}
    We now consider the case of two purely timelike intervals $A=[(x,t_1),(x,t_2)]$ and $B=[(x_,t_3),(x,t_4)]$. Once again, due to the presence of conserved angular momentum, we find a non-trivial leading correction to the reflected entropy:
    \begin{align}
    	\delta S_R(A:B)&=\frac{\mu\Omega\pi ^4 \beta  c^2\sqrt{\eta_t}}{18 \beta_-^2 \beta_+^2}\left[t_{12}\coth\left(\frac{\pi t_{12}}{\beta_+}\right)+t_{34}\coth\left(\frac{\pi t_{34}}{\beta_+}\right)-t_{14}\coth\left(\frac{\pi t_{14}}{\beta_+}\right)-t_{23}\coth\left(\frac{\pi t_{23}}{\beta_+}\right)\right]\notag\\
    	&+\frac{\mu\Omega\pi ^4 \beta  c^2\sqrt{\bar\eta_t}}{18 \beta_-^2 \beta_+^2}\left[t_{12}\coth\left(\frac{\pi t_{12}}{\beta_-}\right)+t_{34}\coth\left(\frac{\pi t_{34}}{\beta_-}\right)-t_{14}\coth\left(\frac{\pi t_{14}}{\beta_-}\right)-t_{23}\coth\left(\frac{\pi t_{23}}{\beta_-}\right)\right]\,,
    \end{align}
    with
    \begin{align}
    	\eta_t=\frac{\sinh \left(\frac{\pi  t_{21}}{\beta_+}\right) \sinh \left(\frac{\pi t_{43}}{\beta_+}\right)}{\sinh \left(\frac{\pi t_{31}}{\beta_+}\right) \sinh \left(\frac{\pi t_{42}}{\beta_+}\right)}~~,~~\bar\eta_t=\frac{\sinh \left(\frac{\pi  t_{21}}{\beta_-}\right) \sinh \left(\frac{\pi t_{43}}{\beta_-}\right)}{\sinh \left(\frac{\pi t_{31}}{\beta_-}\right) \sinh \left(\frac{\pi t_{42}}{\beta_-}\right)}\,.\label{Timelike-cross-ratios}
    \end{align}
    Evidently, for a vanishing angular speed $\Omega$, the leading order correction vanishes identically conforming to earlier literature \cite{Basu:2024bal}. However, contrary to the correction to timelike entanglement entropy, we have found that the leading correction to the reflected entropy for timelike subsystems is not always positive. 
    \subsubsection{Two adjacent intervals}
    In this subsection, we focus on the correlation between two adjacent Lorentz boosted intervals $A=[(x_1,\tau_1),(x_2,\tau_2)]$ and $B=[(x_2,\tau_2),(x_3,\tau_3)]$. The relevant correlation functions for the twist operators on the replica manifolds $\cM_{n,m}$ and $\cM_{1,m}$ are given, respectively, by \cite{Dutta:2019gen}
    \begin{align}
    	\langle \sigma \rangle & \equiv \langle \sigma_{g^{}_A} (z_1,\bar{z}_1) \sigma_{g^{-1}_A g^{}_B} (z_2,\bar{z}_2) \sigma_{g^{-1}_B} (z_3,\bar{z}_3) \rangle \label{log-sigma-adj}\notag\\
    	&=(2m)^{-4h_{g_A^{-1}g^{}_B}}|z_1-z_2|^{-4h_{g^{}_A}}|z_2-z_3|^{-4h_{g^{}_A}}|z_1-z_3|^{4h_{g^{}_A}-4h_{g_A^{-1}g^{}_B}}\\
    	\langle \sigma_m \rangle & \equiv \langle \sigma_{g^{}_m} (z_1,\bar{z}_1) \sigma_{g_m^{-1}} (z_3,\bar{z}_3) \rangle=|z_1-z_2|^{-2h_{g^{}_m}}\,,\label{log-sigma-m-adj}
    \end{align}
    where the conformal dimensions are given in \cref{C-dim}. Now utilizing the above expression in \cref{TT_exp}, the expectation value of the $\TTbar$ operator on the replica manifolds may be readily computed through the conformal Ward identities and the conformal map (\ref{plane-to-cylinder-map}) to the complex plane.
    The leading order correction to the R\'enyi reflected entropy for the two adjacent intervals may now be computed from \cref{SR-correction} In this case using \cref{TT_exp} as follows \cite{Basu:2024bal},
    \begin{equation}\label{Snm-adj}
    	\begin{aligned}
    		\delta S_n \left(A A^\star \right)_{\psi_m} = \frac{\mu }{n-1} &\int_\mathcal{M}\Bigg[-\frac{2 \pi^4 c}{3 \beta_+^2 \beta_-^2} \Bigg(\sum_{i=1}^{3} \left( z^2  \left( \frac{{h_{g_i}}}{(z-{z_i})^2} + \frac{\partial_{z_i}{\log \langle \sigma \rangle}}{z-{z_i}} \right) + {\bar{z}}^2  \left( \frac{{\bar{h}_{g_i}}}{(\bar{z}-{\bar{z}_i})^2} + \frac{\partial_{\bar{z}_i}{\log \langle \sigma \rangle}}{\bar{z}-{\bar{z}_i}} \right) \right)\\
    		& -\sum_{i=1,3} \left( n \, z^2 \left( \frac{{h_{g_m}}}{(z-{z_i})^2} + \frac{\partial_{z_i}{\log \langle \sigma_m \rangle}}{z-{z_i}} \right) -n \, {\bar{z}}^2 \left( \frac{{\bar{h}_{g_m}}}{(\bar{z}-{\bar{z}_i})^2} + \frac{\partial_{\bar{z}_i}{\log \langle \sigma_m \rangle}}{\bar{z}-{\bar{z}_i}}\right)\right)\Bigg)\\
    		&+\frac{16 \pi^2}{\beta_+ \beta_- \, m} \Bigg(\frac{z^2 {\bar{z}}^2}{n} \sum_{i,j=1}^{3} \left( \frac{{h_{g_i}}}{(z-{z_i})^2} + \frac{\partial_{z_i}{\log \langle \sigma \rangle}}{z-{z_i}} \right) \left( \frac{{\bar{h}_{g_j}}}{(\bar{z}-{\bar{z}_j})^2} + \frac{\partial_{\bar{z}_j}{\log \langle \sigma \rangle}}{\bar{z}-{\bar{z}_j}} \right)\\
    		& -n \, z^2 {\bar{z}}^2 \sum_{i,j=1,3} \left( \frac{{h_{g_m}}}{(z-{z_i})^2} + \frac{\partial_{z_i}{\log \langle \sigma_m \rangle}}{z-{z_i}} \right) \left( \frac{{\bar{h}_{g_m}}}{(\bar{z}-{\bar{z}_j})^2} + \frac{\partial_{\bar{z}_j}{\log \langle \sigma_m \rangle}}{\bar{z}-{\bar{z}_j}}\right)\Bigg)\Bigg]\,,
    	\end{aligned}
    \end{equation}
    Finally, taking the replica limit $n,m\to 1$, we may obtain the leading order correction to the reflected entropy from \cref{Snm-adj} in terms of the following integral on the twisted cylinder,
    \begin{equation} \label{Snm-adj-int}
    	\begin{aligned}
    		\delta S_R(A:B) = - \frac{\pi^4 c^2 \mu}{9 \beta_+^2 \beta_-^2}\int_\mathcal{M} \text{d}^2 w \left( \frac{z^2 (z_2 - z_1) (z_2 - z_3)}{(z - z_1) (z - z_2)^2 (z - z_3)} + \text{c.c.} \right)\,.
    	\end{aligned}
    \end{equation}
    In order to simplify the computation of the above integral on the twsited cylinder, we once again employ the rotated frame given in \cref{Bichi-map} to obtain
    \begin{align}
    	\cI_\textrm{adj}=\int_{-\infty}^\infty \dd x^\prime \int_{0}^{\beta_e}\dd \tau^\prime \left( \frac{e^{\frac{4\pi(x^\prime + i \tau^\prime)}{\beta_e}} (z_2^\prime - z^\prime_1) (z^\prime_2 - z^\prime_3)}{\left(e^{\frac{2\pi(x^\prime + i \tau^\prime)}{\beta_e}} - z^\prime_1\right) \left(e^{\frac{2\pi(x^\prime + i \tau^\prime)}{\beta_e}}- z^\prime_2\right)^2 \left(e^{\frac{2\pi(x^\prime + i \tau^\prime)}{\beta_e}} - z^\prime_3\right)} + \text{c.c.} \right)\,.
    \end{align}
    We may now evaluate the above definite integral on the regular cylinder by utilizing the techniques developed in \cite{Chen:2018eqk,Jeong:2019ylz}. After carefully addressing the branch cuts that arise during the integration, the result we obtain for the integration is given as
    \begin{align}
    	\mathcal{I}_\textrm{adj}=&\frac{\beta_e}{2 \pi} \left( \frac{z_1}{z_1-z_2}+\frac{z_1}{z_3-z_1}+ \frac{\bar{z}_1}{\bar{z}_1-\bar{z}_2}+\frac{\bar{z}_1}{\bar{z}_3-\bar{z}_1}\right)\log\abs{\frac{z_2^\prime}{z_1^\prime}}\notag \\\ &-\frac{\beta_e}{2 \pi} \left(\frac{z_1}{z_1-z_3}+\frac{z_2}{z_3-z_2}+\frac{\bar{z}_1}{\bar{z}_1-\bar{z}_3}+\frac{\bar{z}_2}{\bar{z}_3-\bar{z}_2}\right)\log\abs{\frac{z_3^\prime}{z_2^\prime}}\,.
    \end{align}
     Transforming back to the original unprimed coordinates and subsequently performing the Wick rotation $\tau=-i\,t\,,\Omega_E=-i\,\Omega$, the leading order correction to the reflected entropy due to the $\TTbar$-deformation may be written as
    \begin{align}
    	\delta S_R\left(A:B\right)=&-\frac{\pi ^4 \beta  c^2 \mu }{18 \beta_-^2 \beta_+^2}\left(x_{21}-t_{21} \Omega \right)\left( \coth \left(\frac{\pi  \left(x_{21}+t_{21}\right)}{\beta_+}\right)+ \coth \left(\frac{\pi  \left(x_{21}-t_{21}\right)}{\beta_-}\right)\right)\notag \\ &-\frac{\pi ^4 \beta  c^2 \mu }{18 \beta_-^2 \beta_+^2}\left(x_{32}-t_{32} \Omega \right)\left( \coth \left(\frac{\pi  \left(x_{32}+t_{32}\right)}{\beta_+}\right)+ \coth \left(\frac{\pi  \left(x_{32}-t_{32}\right)}{\beta_-}\right)\right) \notag \\
    	&+\frac{\pi ^4 \beta  c^2 \mu }{18 \beta_-^2 \beta_+^2}\left(x_{31}-t_{31} \Omega \right)\left( \coth \left(\frac{\pi  \left(x_{31}+t_{31}\right)}{\beta_+}\right)+ \coth \left(\frac{\pi  \left(x_{31}-t_{31}\right)}{\beta_-}\right)\right) \,.\label{SR-adj-correction}
    \end{align}
    The leading correction to the reflected entropy between two boosted adjacent intervals in a $\TTbar$-deformed thermal CFT$_2$ defined on a cylinder with temporal compactification may be reproduced from the above expression simply by setting $\beta_+=\beta_-$ or $\Omega=0$. A direct comparison to the results reported in \cite{Basu:2024bal} serves as a strong consistency check of the above analysis.
     \subsubsection*{Timelike entanglement}
     We now consider two purely timelike intervals $A=[(x,t_1),(x,t_2)]$ and $B=[(x_,t_2),(x,t_3)]$ in the presence of conserved angular momentum. Following the above analysis for boosted spacelike intervals, we may compute the correction to the reflected entropy arising from the $\ttbar$
     deformation, which is expressed as follows,
      \begin{align}
     	\delta S_R^{\left(T\right)}=\frac{\pi ^4 \beta  c^2 \mu \Omega}{18 \beta_-^2 \beta_+^2}\Bigg[t_{21}  \left( \coth \left(\frac{\pi  t_{21}}{\beta_+}\right)-\coth \left(\frac{\pi  t_{21}}{\beta_-}\right)\right)&+t_{32}\left( \coth \left(\frac{\pi  t_{32}}{\beta_+}\right)-\coth \left(\frac{\pi  t_{32}}{\beta_-}\right)\right)\notag\\&-t_{31}\left( \coth \left(\frac{\pi  t_{31}}{\beta_+}\right)-\coth \left(\frac{\pi  t_{31}}{\beta_-}\right)\right)\Bigg]\,.\label{SR-adj-correction-timelike}
     \end{align}
	 Clearly the above expression vanishes identically for zero angular momentum which aligns with the result obtained earlier \cite{Basu:2024bal}.
    	\subsubsection{A single interval}
    	\label{sec:Single-EW}
    	Finally, we consider a Lorentz boosted single spacelike interval $A \equiv [(0,0) , (\ell,t)]$ in a CFT$_2$ with conserved charge on the twisted cylinder $\cM$, deformed by the $\TTbar$ operator. A naive guess for the relevant correlator is a two point function of twist operators inserted at the endpoints of the subsystem $A$. However, as discussed in \cite{Basu:2022nds,Afrasiar:2022wzn} in the context of an undeformed CFT$_2$, this requires tracing out an infinite branch cut\footnote{Such investigation in \cite{Basu:2022nds} was inspired by a similar pathology discussed in \cite{Calabrese:2014yza} in the context of the entanglement negativity.} present along the complement $A^c$. A careful investigation of the replica manifold reveals that such a pathology stems from a non-trivial gluing of the replica sheets for the present configuration and hence requires the introduction of auxiliary subsystems \cite{Basu:2022nds}. For the case of a deformed CFT$_2$ where the replica manifold is essentially modified by the insertion of stress energy tensor components, one may utilize a similar construction as in \cite{Basu:2022nds} and introduce two auxiliary subsystems $B_1 \equiv [(-L,-T) , (0,0)]$ and $B_2 \equiv [(\ell,t) , (L,T)]$, adjacent to the interval $A$ on the either sides. The leading order correction to the reflected entropy may then be obtained with this modified setup, taking the bipartite limit $B_1\cup B_2\equiv B\to A^c \left(L\to \infty \right)$ as the end of calculations.
    
    For this configuration where the interval $A$ is sandwiched between $B_1$ and $B_2$, the relevant four point correlation function of twist operators on the replica manifold $\cM_{n,m}$ is given as \cite{Basu:2022nds,Afrasiar:2022wzn} 
    \begin{equation}\label{four-point-single}
    	\begin{aligned} 
    		\langle \sigma_{g^{}_B} (z_1,\bar{z}_1) \sigma_{g^{-1}_B g^{}_A} (z_2,\bar{z}_2) \sigma_{g^{-1}_A g^{}_B} (z_3,\bar{z}_3) \sigma_{g^{-1}_B} (z_4,\bar{z}_4) \rangle =&\frac{1}{z_{14}^{2 h_{g^{}_B}} z_{23}^{2 h_{g^{-1}_A g^{}_B}}} \frac{\mathcal{F}_{mn}(\eta)}{\eta^{h_{g^{-1}_A g^{}_B}}}\\
    		\times & \frac{1}{\bar z_{14}^{2 \bar {h}_{g^{}_B}} \bar z_{23}^{2 \bar h_{g^{-1}_A g^{}_B}}} \frac{\mathcal{\bar F}_{mn}(\bar \eta)}{\bar \eta^{\bar h_{g^{-1}_A g^{}_B}}}\,, 
    	\end{aligned}
    \end{equation}
    where, $\eta=\frac{(z_1-z_2)(z_3-z_4)}{(z_1-z_3)(z_2-z_4)}$ and $\bar{\eta}=\frac{(\bar z_1-\bar z_2)(\bar z_3-\bar z_4)}{(\bar z_1-\bar z_3)(\bar z_2-\bar z_4)}$ are the cross ratios. The functions $\mathcal{F}_{mn}(\eta)$ and $\mathcal{\bar F}_{mn}(\bar \eta)$ are non-universal and depend on the full operator context of the theory. These have the following limiting values \cite{Basu:2022nds},
    \begin{align}
    	\lim_{\eta\to1}\mathcal{F}_{mn}(\eta)=	\lim_{\bar{\eta}\to1}\mathcal{\bar F}_{mn}(\bar \eta)=1\ , \quad	\mathcal{F}_{mn}(0)=	\mathcal{\bar F}_{mn}(0)=C_{mn}\ ,
    \end{align}
    $C_{mn}$ being a non-universal OPE coefficient. Now utilizing 
    \cref{SR-correction,TT_exp,plane-to-cylinder-map}, the leading order correction to the R\'enyi reflected entropy between $A$ and $B$ may be found as follows,
    \begin{equation}\label{Snm-sing-beta}
    	\begin{aligned}
    		\delta S_n \left(A A^\star \right)_{\psi_m} = \frac{\mu}{n-1}& \int_\mathcal{M}\Bigg[-\frac{2 \pi^4 c}{3 \beta_+^2\beta_-^2} \Bigg(z^2 \sum_{i=1}^{4} \left( \frac{{h_{g^{}_i}}}{(z-{z_i})^2} + \frac{\partial_{z_i}{\log \langle \sigma \rangle}}{z-{z_i}} \right) + {\bar{z}}^2 \sum_{i=1}^{4} \left( \frac{{\bar{h}_{g^{}_i}}}{(\bar{z}-{\bar{z}_i})^2} + \frac{\partial_{\bar{z}_i}{\log \langle \sigma \rangle}}{\bar{z}-{\bar{z}_i}} \right)\\
    		&-n \, z^2 \sum_{i=1,4} \left( \frac{{h_{g^{}_m}}}{(z-{z_i})^2} + \frac{\partial_{z_i}{\log \langle \sigma_m \rangle}}{z-{z_i}} \right) -n \, {\bar{z}}^2 \sum_{i=1,4} \left( \frac{{\bar{h}_{g^{}_m}}}{(\bar{z}-{\bar{z}_i})^2} + \frac{\partial_{\bar{z}_i}{\log \langle \sigma_m \rangle}}{\bar{z}-{\bar{z}_i}}\right) \Bigg)\\
    		&+\frac{16 \pi^2}{\beta_+\beta_- \, m} \Bigg(\frac{z^2 {\bar{z}}^2}{n} \sum_{i,j=1}^{4} \left( \frac{{h_{g^{}_i}}}{(z-{z_i})^2} + \frac{\partial_{z_i}{\log \langle \sigma \rangle}}{z-{z_i}} \right) \left( \frac{{\bar{h}_{g^{}_j}}}{(\bar{z}-{\bar{z}_j})^2} + \frac{\partial_{\bar{z}_j}{\log \langle \sigma \rangle}}{\bar{z}-{\bar{z}_j}} \right)\\
    		&-n \, z^2 {\bar{z}}^2 \sum_{i,j=1,4} \left( \frac{{h_{g^{}_m}}}{(z-{z_i})^2} + \frac{\partial_{z_i}{\log \langle \sigma_m \rangle}}{z-{z_i}} \right) \left( \frac{{\bar{h}_{g^{}_m}}}{(\bar{z}-{\bar{z}_j})^2} + \frac{\partial_{\bar{z}_j}{\log \langle \sigma_m \rangle}}{\bar{z}-{\bar{z}_j}}\right)\Bigg)\Bigg]\,,
    	\end{aligned}
    \end{equation}
    where,
    \begin{align} 
    	\log \langle \sigma \rangle \equiv& \log \langle \sigma_{g^{}_B} (z_1,\bar{z}_1) \sigma_{g^{-1}_B g^{}_A} (z_2,\bar{z}_2) \sigma_{g^{-1}_A g^{}_B} (z_3,\bar{z}_3) \sigma_{g^{-1}_B} (z_4,\bar{z}_4) \rangle \label{log-sigma-sing}\\
    	\log \langle \sigma_m \rangle \equiv &\log \langle \sigma_{g_m} (z_1,\bar{z}_1) \sigma_{g_m^{-1}} (z_4,\bar{z}_4) \rangle = - h_{g_m} \log ( z_4 - z_1 ) - \bar{h}_{g_m} \log ( \bar{z}_4 - \bar{z}_1 )\,. \label{log-sigma-m-sing}
    \end{align}
    In the replica limit $n,m\to1$, the above expression may be simplified to obtain the following result for the leading order correction to the reflected entropy between $A$ and $B$:
     \begin{align}
     	\delta S_R(A:B) = -\frac{\pi^4 c^2 \mu}{9 \beta_+^2 \beta_-^2} \int_\mathcal{M} \text{d}^2 w \left(H\left(z,z_1,z_2,z_3,z_4\right) + H\left(\bar{z},\bar{z}_1,\bar{z}_2,\bar{z}_3,\bar{z}_4\right) \right)\,,
     \end{align}
     where the integrand $H$ has the  following form,
     \begin{align}
      H=&z^2\left(\frac{1}{z-z_2}-\frac{1}{z-z_3}\right)\left(-\frac{1}{z-z_1}+\frac{1}{z-z_2}-\frac{1}{z-z_3}+\frac{1}{z-z_4}\right)\notag\\
      &-\frac{z^2\eta f^\prime(\eta)}{z_{13}z_{24}(z-z_1)(z-z_2)(z-z_3)(z-z_4)}\,,
     \end{align}
      with, $z = e^{\frac{2\pi(x+i\tau)}{\beta_+}}\,,\,\bar{z} = e^{\frac{2 \pi (x - i \tau)}{\beta_-}}$ and the functions $f$ and $\bar{f}$ are defined as\footnote{Note that the non-universal functions $\mathcal{F}_{mn},\bar{\mathcal{F}}_{mn}$ are expected to be sub-leading in the large central charge limit \cite{Basu:2022nds}. However, to simplify expressions, we have extracted factors of $\frac{c}{6}$ and the resulting auxiliary functions scale as $f,\bar{f}\sim \cO\left(\frac{1}{c}\right)$.},
      \begin{align}
      	\frac{c}{6}\log f\left(\eta\right)=\lim_{m,n\to1}\frac{1}{1-n}\log[\mathcal{F}_{mn}\left(\eta\right)]\ ,\quad \frac{c}{6}\log\bar{f}\left(\bar{\eta}\right)=\lim_{m,n\to1}\frac{1}{1-n}\log[\bar{\mathcal{F}}_{mn}\left(\bar{\eta}\right)]\ .
      \end{align}
      To evaluate the definite integral, we once again apply the rotated frame $(x^\prime,\tau^\prime)$ in (\ref{Bichi-map}), which reduces the integral into a more manageable form:
      \begin{align}
     	\mathcal{I}_\textrm{sing}=\int_\mathcal{M} \text{d}^2 w\left(H\left(z,z_1,z_2,z_3,z_4\right) + \text{c.c} \right)= \int_{-\infty}^\infty \dd x^\prime \int_{0}^{\beta_e}\dd \tau^\prime\left[H\left( e^{\frac{2\pi(x^\prime+i\tau^\prime)}{\beta_e}},z_1^\prime,z_2^\prime,z_3^\prime,z_4^\prime\right) + \text{c.c} \right]\,,
      \end{align}
      where $z_i^\prime=e^{\frac{2\pi(x^\prime_i+i\tau^\prime_i)}{\beta_e}}$. Now we can compute the integral using the machinery developed in \cite{Chen:2018eqk,Jeong:2019ylz}, leading to the result
      \begin{align}
      		\mathcal{I}_\textrm{sing}=&\frac{\beta_e}{2 \pi}\Bigg[\left(-\frac{z_1}{z_2-z_1}-\frac{z_1}{z_1-z_3}\right)\log\abs{\frac{z_2^\prime}{z_1^\prime}}+\left(\frac{z_1}{z_1-z_3}+ \frac{2z_2}{z_3-z_2}+\frac{z_2}{z_2-z_4}\right)\log\abs{\frac{z_3^\prime}{z_2^\prime}} \notag \\ &+\left(\frac{z_2}{z_2-z_4}+\frac{z_3}{z_4-z_3}\right)\log\abs{\frac{z_4^\prime}{z_3^\prime}}+\Bigg(\frac{z_1(z_2-z_3)}{(z_2-z_1)(z_3-z_1)}\log|z^\prime_1|+\frac{z_2(z_4-z_1)}{(z_2-z_1)(z_4-z_2)}\log|z^\prime_2|\notag\\
      		&+\frac{z_3(z_1-z_4)}{(z_3-z_1)(z_4-z_3)}\log|z^\prime_3|+\frac{z_4(z_3-z_2)}{(z_4-z_2)(z_4-z_3)}\log|z^\prime_4|\Bigg)\eta f^\prime(\eta)+ \textrm{c.c}\Bigg]
      \end{align}
      After performing the integration, we apply the inverse map of (\ref{Bichi-map}) to return to the original coordinate system. Next, we implement the Wick rotation and finally take the bipartite limit $L\to \infty$ to obtain the first-order correction to the reflected entropy for a boosted spacelike interval as follows,
    \begin{align}
    	\delta S_R(A:A^c)=&-\mu\frac{\pi^4\beta c^2}{9\beta _+^2\beta_-^2}(\ell-t \Omega)\left[\coth \left(\frac{\pi  (\ell+t)}{\beta_+}\right)+\coth \left(\frac{\pi  (\ell-t)}{\beta_-}\right)-2\right] \notag \\&+\mu\frac{\pi^4\beta c^2}{9\beta _+^2\beta_-^2\sqrt{1-\Omega^2}}(\ell-t \Omega)\left[e^{-\frac{2\pi\left(\ell+t\right)}{\beta_+}}f^\prime\left(e^{-\frac{2\pi\left(\ell+t\right)}{\beta_+}}\right)+e^{-\frac{2\pi\left(\ell-t\right)}{\beta_-}}\bar{f}^\prime\left(e^{-\frac{2\pi\left(\ell-t\right)}{\beta_-}}\right)\right]\ .\label{SR-sing-correction}
    \end{align} 
    The leading correction to the reflected entropy for a single Lorentz boosted intervals in a $\TTbar$-deformed thermal CFT$_2$ defined on a cylinder with temporal compactification may be reproduced from the above expression simply by setting $\beta_+=\beta_-$ or $\Omega=0$, which may be checked against the results reported in \cite{Basu:2024bal}.
     \subsubsection*{Timelike entanglement}
     We now consider a boosted, purely timelike interval $A=\left[\left(0,0\right),\left(0,t\right)\right]$ in a thermal CFT$_2$ in the presence of a conserved angular momentum. In this timelike scenario, the time direction is compactified, so no auxiliary intervals appear. The relevant correlation function in this case therefore simplifies to a two-point function. Therefore, the correction term to the reflected entropy in a CFT$_2$ with a conserved charge can be computed, similar to the case of entanglement entropy, as follows:
     \begin{align}
     	\delta S_R^{\textrm{T}}\left(A:A^c\right)=&\mu\frac{\pi^4\beta c^2}{9\beta _+^2\beta_-^2}t \Omega\left[\coth \left(\frac{\pi  t}{\beta_+}\right)-\coth \left(\frac{\pi  t}{\beta_-}\right)\right]\,.
     \end{align}
 	Note that when the angular momentum vanishes, the leading order correction to the reflected entropy becomes zero, aligning with the known result \cite{Basu:2024bal} for a single timelike interval.
	\section{Cut-off AdS holography}\label{sec-4}
	In this section, we investigate the holographic characterization of entanglement and correlation in the $\TTbar$-deformed CFT$_2$ with conserved angular momentum utilizing the cut-off proposal of \cite{McGough:2016lol} discussed briefly in subsection \cref{sec:TTb-holography}. The holographic dual of the $\TTbar$-deformed CFT$_2$ defined on the twisted cylinder is described by a non-extremal rotating BTZ black hole geometry in the bulk with line element (we have set the AdS radius to unity, $\ell_\textrm{AdS}=1$)
	\begin{align}
		\dd s^2&=-\frac{\left(r^2-r_+^2\right)\left(r^2-r_-^2\right)}{r^2}\dd t^2+\frac{r^2}{\left(r^2-r_+^2\right)\left(r^2-r_-^2\right)}\dd r^2
		+r^2\left(\dd x-\frac{r_+ r_-}{r^2}\dd t \right)^2\notag\\
		&=\frac{r^2}{\left(r^2-r_+^2\right)\left(r^2-r_-^2\right)}\dd r^2-\frac{(r^2-r_+^2)}{(r_+^2-r_-^2)}\left(r_+\dd t-r_-\dd x\right)^2+\frac{(r^2-r_-^2)}{(r_+^2-r_-^2)}\left(r_+\dd x-r_-\dd t\right)^2\,.\label{rotating-BTZ-metric}
	\end{align}
	In the above metric, $r_{\pm}$ denote the outer and inner horizons respectively and the spatial and temporal directions are compactified as follows
	\begin{align}
		x\sim x-\beta\Omega~~,~~t\sim t+i\beta\,,
	\end{align}
	where the angular speed $\Omega$ and temperature $\beta$ of the black hole are related to the horizon radii as follows\footnote{The ADM mass and angular momentum of the rotating black hole may also be written in terms of the horizon radii (cf. \cref{appB} for more details),
		\begin{align}
			M=r_+^2+r_-^2~~,~~J=2r_+r_-\,.
	\end{align}}
	\begin{align}
		\beta=\frac{2\pi r_+}{r_+^2-r_-^2}~~,~~\Omega=\frac{r_-}{r_+}\,.\label{Temperature-AngSpeed}
	\end{align}
	One may define the effective left and right moving temperatures form the above relations:
	\begin{align}
		\beta_{\pm}=\beta(1\pm\Omega)=\frac{2\pi}{r_+\pm r_-}\,.
	\end{align}
	
	According to the holographic correspondence discussed in \cite{McGough:2016lol}, the dual field theory is located at the asymptotic boundary $r=r_c$ (cf. \cref{Cutoff-Dictionary}) and the metric on this holographic screen is given by
	\begin{align}
		\dd s^2=-\frac{(r_c^2-r_+^2)}{(r_+^2-r_-^2)}\left(r_+\dd t-r_-\dd x\right)^2+\frac{(r_c^2-r_-^2)}{(r_+^2-r_-^2)}\left(r_+\dd x-r_-\dd t\right)^2
	\end{align}
	Interestingly, the above metric may be recast in a conformally flat form by performing the conformal map 
	\begin{align}
		x^\prime=\frac{1}{\sqrt{r_+^2-r_-^2}}\left(r_+x-r_-t\right)~~,~~t^\prime=\frac{1}{\sqrt{r_+^2-r_-^2}}\left(r_+t-r_-x\right)\,.	\label{bulk-rotation}
	\end{align} 
	Utilizing \cref{Temperature-AngSpeed}, it is evident that this is nothing but the rotation of the twisted cylinder given in (\ref{Bichi-map}). In these coordinates, the metric on the conformal boundary is given by
	\begin{align}
		\dd s^2\sim -\left(\dd t^\prime\right)^2+\frac{r_c^2-r_-^2}{r_c^2-r_+^2}\left(\td \tilde{x}^\prime\right)^2\equiv -\left(\dd t^\prime\right)^2+\left(\td x^\prime\right)^2\,,
	\end{align}
	where, we have defined the conformal coordinate
	\begin{align}
		x^\prime=\tilde{x}^\prime\left(\frac{r_c^2-r_-^2}{r_c^2-r_+^2}\right)^{-1/2}\,.\label{conformal-coordinate}
	\end{align}
	From the periodicity of the timelike coordinate $t^\prime$, it is evident that the boundary theory has the effective inverse temperature\footnote{In fact, the coordinate transformations in \cref{bulk-rotation} together with the redefinition of the holographic coordinate $r\to R=\sqrt{r^2-r_-^2}$, the metric in \cref{rotating-BTZ-metric} may be recast into the form of a non-rotating BTZ black hole
		\begin{align}
			\dd s^2=\frac{\dd R^2}{R^2-R_h^2}-(R^2-R_h^2)(\dd t^\prime)^2+R^2(\dd x^\prime)^2\,,
		\end{align}
		with horizon at $R_h=\sqrt{r_+^2-r_-^2}$. This in turn, implies that the kinematics of a rotating BTZ black hole with Hawing temperature $\beta^{-1}$ is equivalent to that of a non-rotating BTZ black hole with an effective inverse temperature $\beta_e=\frac{2\pi}{\sqrt{r_+^2-r_-^2}}=\beta\sqrt{1-\Omega^2}$, where in the second equality, we have utilized (\ref{Temperature-AngSpeed}).} $\beta_e=\beta\sqrt{1-\Omega^2}\,$ conforming to our finding in \cref{effective-temp}.
	
	The rotating BTZ black hole may be embedded in $\mathbb{R}^{2,2}$ according to \cref{AdS-embedding} utilizing the following coordinate transformations
	\begin{align}
		&X_1=\sqrt{\frac{r^2-r_+^2}{r_+^2-r_-^2}}\cosh\left(r_+x-r_-t\right)\,,\notag\\
		&X_2=\sqrt{\frac{r^2-r_-^2}{r_+^2-r_-^2}}\sinh\left(r_+t-r_-x\right)\,,\notag\\
		&X_3=\sqrt{\frac{r^2-r_+^2}{r_+^2-r_-^2}}\sinh\left(r_+x-r_-t\right)\,,\notag\\
		&X_4=\sqrt{\frac{r^2-r_-^2}{r_+^2-r_-^2}}\cosh\left(r_+t-r_-x\right)\,.\label{Embedding-coordinates}
	\end{align}
	In the following, we will utilize the above embedding coordinates to compute lengths of various extremal curves (geodesics) in the cut-off rotating BTZ geometry.
	\subsection{Holographic entanglement entropy}\label{HEE}
	In the following, we compute the holographic entanglement entropy utilizing the Ryu-Takayanagi prescription \cite{Ryu:2006bv,Ryu:2006ef} extended to a BTZ black hole spacetime with a finite radial cut-off. Consider the boosted subsystem $A=[(x_1,t_1),(x_2,t_2)]$ on the twisted cylinder. The length of the geodesic connecting these endpoints may be obtained utilizing the embedding coordinates in \cref{Embedding-coordinates} as follows
	\begin{align}
		\cL=\cosh^{-1}\left(\zeta_{12}\right)=\cosh^{-1}\left(-X_1\cdot X_2\right)\,,
	\end{align}
	where $\zeta_{12}$ is the unique conformal invariant associated with the endpoints $X_1^\mu$ and $x_2^\mu$ of our subsystem, written in the embedding coordinates. Utilizing \cref{Embedding-coordinates,conformal-coordinate} this invariant may be computed as
	\begin{align}
		\zeta_{12}&=\left(\frac{r_c^2-r_+^2}{r_+^2-r_-^2}\right)\cosh\left[\sqrt{r_+^2-r_-^2}\left(t^\prime_2-t^\prime_1\right)\right]-\left(\frac{r_c^2-r_-^2}{r_+^2-r_-^2}\right)\cosh\left[\sqrt{\frac{(r_c^2-r_+^2)(r_+^2-r_-^2)}{(r_c^2-r_-^2)}}|x^\prime_{2}-x^\prime_1|\right]\notag\\
		&=\left(\frac{r_c^2-r_+^2}{r_+^2-r_-^2}\right)\cosh\left(r_-x_{21}-r_+t_{21}\right)-\left(\frac{r_c^2-r_-^2}{r_+^2-r_-^2}\right)\cosh\left[\sqrt{\frac{r_c^2-r_+^2}{r_c^2-r_-^2}}\left(r_+x_{21}-r_-t_{21}\right)\right]\,,\label{zeta12}
	\end{align}
	where, in the second equality, we have inverted the coordinate transformation (\ref{bulk-rotation}). Now using the Ryu-Takayanagi prescription and the holographic dictionary in \cref{Cutoff-Dictionary}, we may obtain the holographic entanglement entropy for the single interval under consideration as follows
	\begin{align}
		S(A)=\frac{1}{4G_N}&\cosh^{-1}\Bigg[\frac{1}{\hat{\mu}(1-\Omega^2)}\Bigg(\left(\hat{\mu}-\frac{\beta^2}{4\pi^2} \left(1-\Omega ^2\right)^2\right) \cosh \left[\frac{2 \pi  (t_{21}-x_{21}\Omega)}{\beta  \left(1-\Omega ^2\right)}\right]\notag\\
		&-\left(\Omega^2\hat{\mu}-\frac{\beta^2}{4\pi^2} \left(1-\Omega ^2\right)^2\right)\cosh \left[\frac{2 \pi  (x_{21}-t_{21}\Omega)}{\beta  \left(1-\Omega ^2\right)}\sqrt{\frac{4\pi^2\hat{\mu}-\beta^2\left(1-\Omega ^2\right)^2}{4\pi^2\Omega^2\hat{\mu}-\beta^2\left(1-\Omega ^2\right)^2}}\right]\Bigg)\Bigg]\label{HEE-holo}
	\end{align}
	where we have defined a re-scaled deformation parameter $\hat{\mu}=\frac{\pi c\mu}{6}$. Note that the above expression is non-perturbative in the deformation parameter $\hat{\mu}$. We have verified that the entanglement entropy in \cref{HEE-holo} violates the strong sub-additivity property which indicates that the field theory is non-local. We have also computed the $C$-function which decreases monotonically with the subsystem size which set the scale for the holographic renormalization group flow, thereby validating Zamolodchikov's $C$-theorem \cite{Zamolodchikov:1986gt}. We refrain from quoting the analytic expressions here as they do not provide significant insight.
	
	In order to compare our holographic result in \cref{HEE-holo}, we now consider the limit of a small deformation parameter, which corresponds to a large cut-off radius $r_c$. The leading order correction to the entanglement entropy due to the deformation is then given as
	\begin{align}
		S(A)&=\frac{c}{6}\log\left[\frac{\beta_+\beta_-}{\pi^2\epsilon_c^2}\sinh\left(\frac{\pi(x_{21}+t_{21})}{\beta_+}\right)\sinh\left(\frac{\pi(x_{21}-t_{21})}{\beta_-}\right)\right]\notag\\
		&-\frac{2\pi^2\hat{\mu}}{\beta^3\left(1-\Omega ^2\right)^2}\pi(x_{21}-t_{21}\Omega)\left[\coth\left(\frac{\pi(x_{21}+t_{21})}{\beta_+}\right)+\coth\left(\frac{\pi(x_{21}-t_{21})}{\beta_-}\right)\right]\notag\\
		&-\frac{2\pi^2\hat{\mu}}{\beta^2\left(1-\Omega ^2\right)^2}\left[\left(1+\Omega ^2\right)-(1-\Omega^2)\coth\left(\frac{\pi(x_{21}+t_{21})}{\beta_+}\right)\coth\left(\frac{\pi(x_{21}-t_{21})}{\beta_-}\right)\right]\,,\label{HEE-corrections}
	\end{align}
	where $\epsilon_c=\sqrt{\displaystyle\hat{\mu}}$ is the UV cut-off in the dual field theory.
    Since the bulk rotating BTZ black hole is dual to the deformed CFT$_2$ in the limit of high temperatures, we may further consider the limit $\beta\ll |x_{21}|,t_{21}$. Dropping the third term, the above expression is easily seen to match with our perturbative field theoretic computations in \cref{EE-corrections}.

    \subsubsection*{Timelike entanglement}
    We now consider the case of a purely timelike interval $A=[(x,0),(x,t)]$. As described in \cite{Doi:2023zaf}, for a real Euclidean theory, the corresponding Lorentzian metric becomes complex valued under the analytic continuation of the inner horizon $r_-\to ir_-$ and consequently, the entanglement entropy in \cref{HEE-holo} also becomes complex-valued and should be interpreted as a pseudo entropy \cite{Nakata:2020luh}. This is manifestly seen from the perturbative expansion in \cref{HEE-corrections}, as $|x_{21}|\to 0$ the argument of the logarithmic term becomes negative and hence procures a imaginary part,
    \begin{align}
    	S^{(\textrm{T})}(A)=&\frac{c}{6}\log\left[\frac{\beta_+\beta_-}{\pi^2\epsilon_c^2}\sinh\left(\frac{\pi t}{\beta_+}\right)\sinh\left(\frac{\pi t}{\beta_-}\right)\right]+\frac{c}{6}\pi i\notag\\
    	&+\frac{2\pi^3\hat{\mu}\Omega\,t}{\beta^3\left(1-\Omega ^2\right)^2}\left[\coth\left(\frac{\pi t}{\beta_+}\right)+\coth\left(\frac{\pi t}{\beta_-}\right)\right]\,.\label{Timelike-HEE}
    \end{align} 
    In the above expression, the first two terms correspond to the timelike entanglement entropy for an undeformed CFT$_2$ on the twisted cylinder \cite{Doi:2022iyj,Doi:2023zaf} having an UV cut-off $\epsilon_c$, while the rest of the expression corresponds to effects of the $\TTbar$-deformation as the holographic screen is pushed inside the bulk. Utilizing the holographic dictionary (\ref{Cutoff-Dictionary}), the above expression may be straightforwardly seen to match half of the reflected entropy obtained in \cref{deltaS-tiemlike}. Furthermore, as discussed earlier, the leading order correction due to the deformation remains real-valued and positive for the timelike subsystem. This seems to reflect the spurious fact that the spacelike part of the corresponding RT surface gets larger despite the inward displacement of the holographic screen, which may be ameliorated by the re-scaling of the UV cut-off of the dual field theory as seen from \cref{Cutoff-Dictionary}. 
	\subsubsection{Bounds on $\mu $ and Hagedorn behavior}
	Now we investigate the reality of the holographic entanglement entropy in \cref{HEE-holo}. In particular, the reality puts a sharp upper bound on the (re-scaled) deformation parameter:
	\begin{align}
		\hat{\mu}<\hat{\mu}_\textrm{max}=\frac{\beta^2(1-\Omega^2)^2}{4\pi^2\Omega^2}\,.
	\end{align}
	Note that, at $\hat{\mu}=\hat{\mu}_\textrm{max}$ the entanglement entropy diverges. Furthermore, there exists a universal zero of the holographic entanglement entropy at the value
	\begin{align}
		\hat{\mu}_0=\frac{\beta^2}{4\pi^2}(1-\Omega^2)^2\,.\label{Hagedorn-rotating}
	\end{align}
	Evidently, at $\hat{\mu}=\hat{\mu}_0$ the spacelike geodesic computing the HEE becomes null. Furthermore, we have checked numerically that beyond $\hat{\mu}_0$ the HEE becomes complex valued for a certain window, depending upon the size of the subsystem under consideration.
	
\begin{figure}[h!]
	 	\centering
	 	\includegraphics[width=0.65\textwidth]{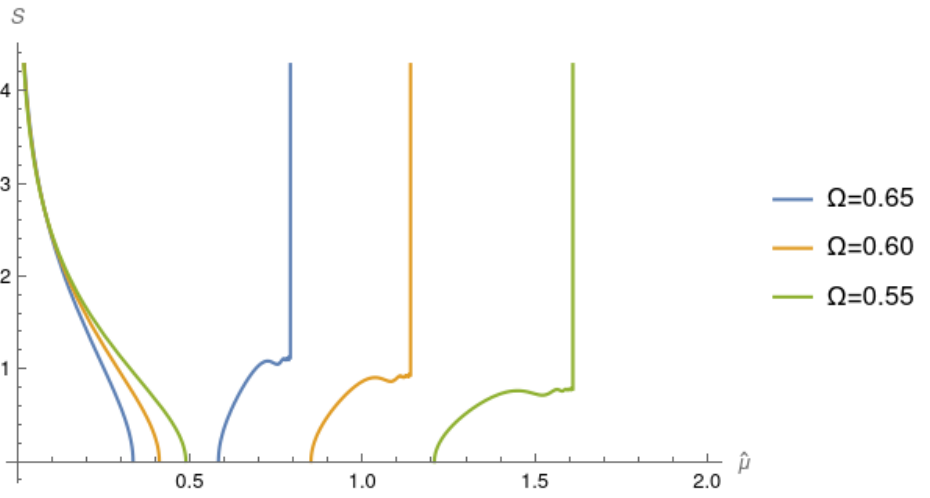}
	 	\includegraphics[width=0.65\textwidth]{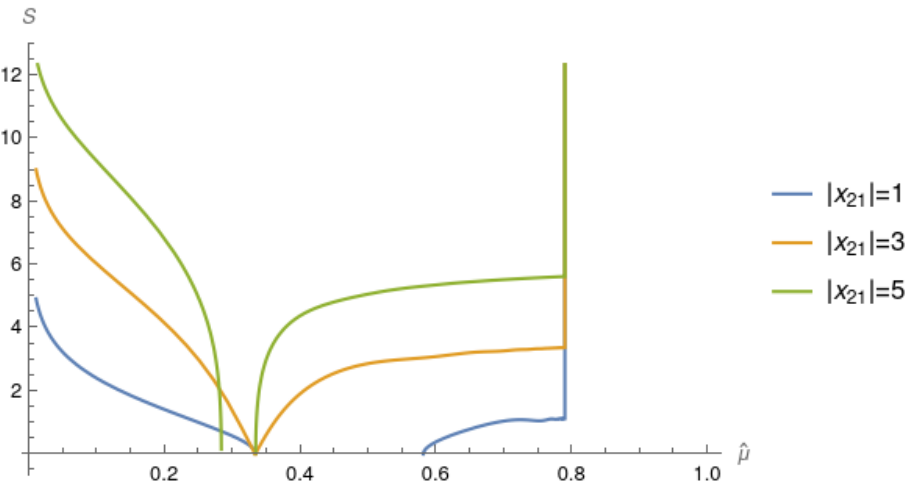}
	 	\caption{Plot of holographic entanglement entropy for a single interval on a constant time slice with respect to the strength of deformation $\hat{\mu}$. Top panel: $|x_{21}|=1\,,\,\Omega=0.65,0.6,0.55$; bottom panel: $\Omega=0.65\,,\,|x_{21}|=1,3,5$. We have set $\beta=2\pi$ and $4G_N=1$.}
	 	\label{fig:HEE-plots}
	 \end{figure}

	For purely spacelike subsystems on a constant time slice, we have plotted the entanglement entropy against the strength of deformation in \cref{fig:HEE-plots}. Note that for certain subsystem sizes, the HEE exhibits a gap in its spectrum, where its value becomes complex. However, there exists a critical size when the gap closes up. Furthermore, the zeros of the HEE change branches below and above this critical size. It will be interesting to further investigate the physical origin of such phenomena.
	 
\begin{figure}[h!]
		\centering
		\includegraphics[width=0.65\textwidth]{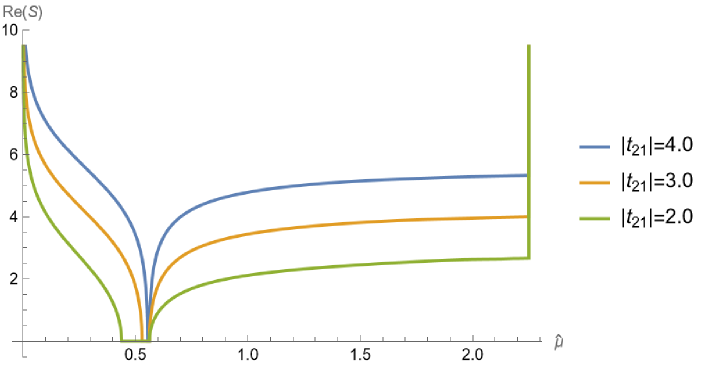}
		\caption{Variation of holographic entanglement entropy for a timelike interval with respect to the strength of deformation $\hat{\mu}$. We have set $\mu=0.5\,,\,\beta=2\pi$ and $4G_N=1$.}
		\label{fig:HEE-plots-timelike1}
	\end{figure}

\begin{figure}[h!]
		\centering
		\includegraphics[width=0.65\textwidth]{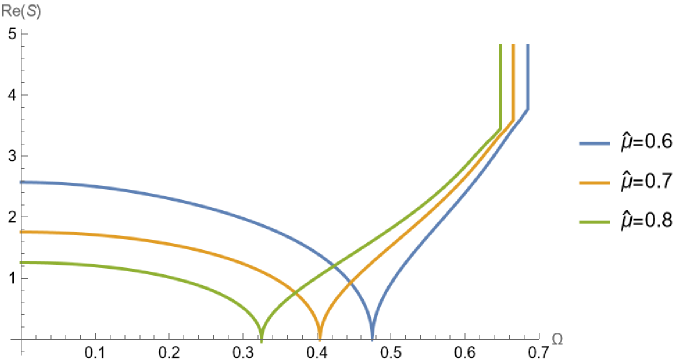}
		\caption{Variation of holographic entanglement entropy for a timelike interval with respect to the angular speed $\Omega$; we have set $|t_{21}|=2.0\,,\,\beta=2\pi\,,\,4G_N=1$.}
		\label{fig:HEE-plots-timelike2}
	\end{figure}

	On the other hand, for timelike subsystems, even without the $\TTbar$-deformation, the entanglement entropy is complex-valued as described in \cite{Doi:2022iyj,Doi:2023zaf}, which may be interpreted as a pseudo entropy. With $\TTbar$-deformation, the timelike entanglement entropy also captures the essence of the spectrum being complex-valued. In \cref{fig:HEE-plots-timelike1,fig:HEE-plots-timelike2}, we have plotted the real part of the timelike entanglement entropy for various sizes of the subsystem with respect to the spin $\Omega$ of the black hole as well as the deformation parameter $\hat{\mu}$. Similar to the case of spacelike intervals, we found that for different sizes of the timelike subsystem, the plot of the HEE with respect to $\hat{\mu}$ exhibits a band-gap. Once again, for a critical size this gap closes up.

	It is also instructive to analyze the reality of the holographic entanglement entropy for a deformed thermal CFT$_2$ with zero angular momentum dual to a non-rotating BTZ black hole with a finite radial cut-off. Setting the angular speed $\Omega=0$ in \cref{HEE-holo}, we may obtain the holographic entanglement entropy for a boosted interval in a $\TTbar$-deformed thermal CFT$_2$ dual to a non-rotating BTZ black hole as follows
	\begin{align}
		S(A)=\frac{1}{4G_N}\cosh^{-1}\left[\left(1-\frac{\beta^2}{4\pi^2\hat{\mu}}\right)\cosh\left(\frac{2\pi t_{21}}{\beta}\right)+\frac{\beta^2}{4\pi^2\hat{\mu}}\cosh\left(\frac{2\pi x_{21}}{\beta}\sqrt{1-\frac{4\pi^2\hat{\mu}}{\beta^2}}\right)\right]\,,
	\end{align}
	which may be checked against the non-perturbative holographic results in \cite{Chen:2018eqk,Jeong:2019ylz} for the case of a single interval on a constant time slice. Similar to the rotating case, we found that the upper bound to the strength of deformation is given by
	\begin{align}
		\hat{\mu}_0=\frac{\beta^2}{4\pi^2}\,.
	\end{align}
	\begin{figure}[h!]
		\centering
		\includegraphics[width=0.5\textwidth]{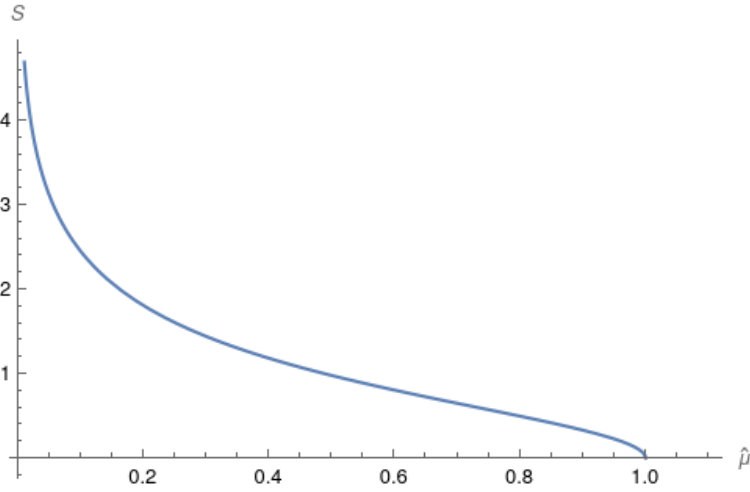}
		\caption{Plot of HEE against $\hat{\mu}$ for a non-rotating BTZ black hole. We have set $\beta=2\pi,|x_{21}|=1$.}
	\end{figure}
	However, contrary to the rotating case, at $\hat{\mu}_0$ the HEE vanishes and beyond $\hat{\mu}_0$ it becomes complex valued. Interestingly, this value of $\hat{\mu}$ corresponds to the Hagedorn temperature where the ground state energy is ill-defined\footnote{Note that this is contrary to the findings in \cite{Banerjee:2024wtl}, where the authors employ the mixed boundary condition proposal of \cite{Guica:2019nzm} to obtain divergent behavior of entanglement at the Hagedorn temperature.}. In this connection, (\ref{Hagedorn-rotating}) may be regarded as the spinning analogue of the Hagedorn behavior.
	

	\subsection{Entanglement wedge cross-section}
	In this section, we compute the bulk EWCS for various bipartite mixed states in the rotating BTZ geometry with a finite radial cut-off dual to the $\TTbar$-deformed CFT$_2$ with conserved angular momentum. We expand our holographic results about the seed theory for a small deformation parameter and find agreement with our field theoretic computations in the large central charge limit.
	\subsubsection{Two disjoint intervals}
\begin{figure}[ht]
    	\centering
    	\includegraphics[width=0.65\textwidth]{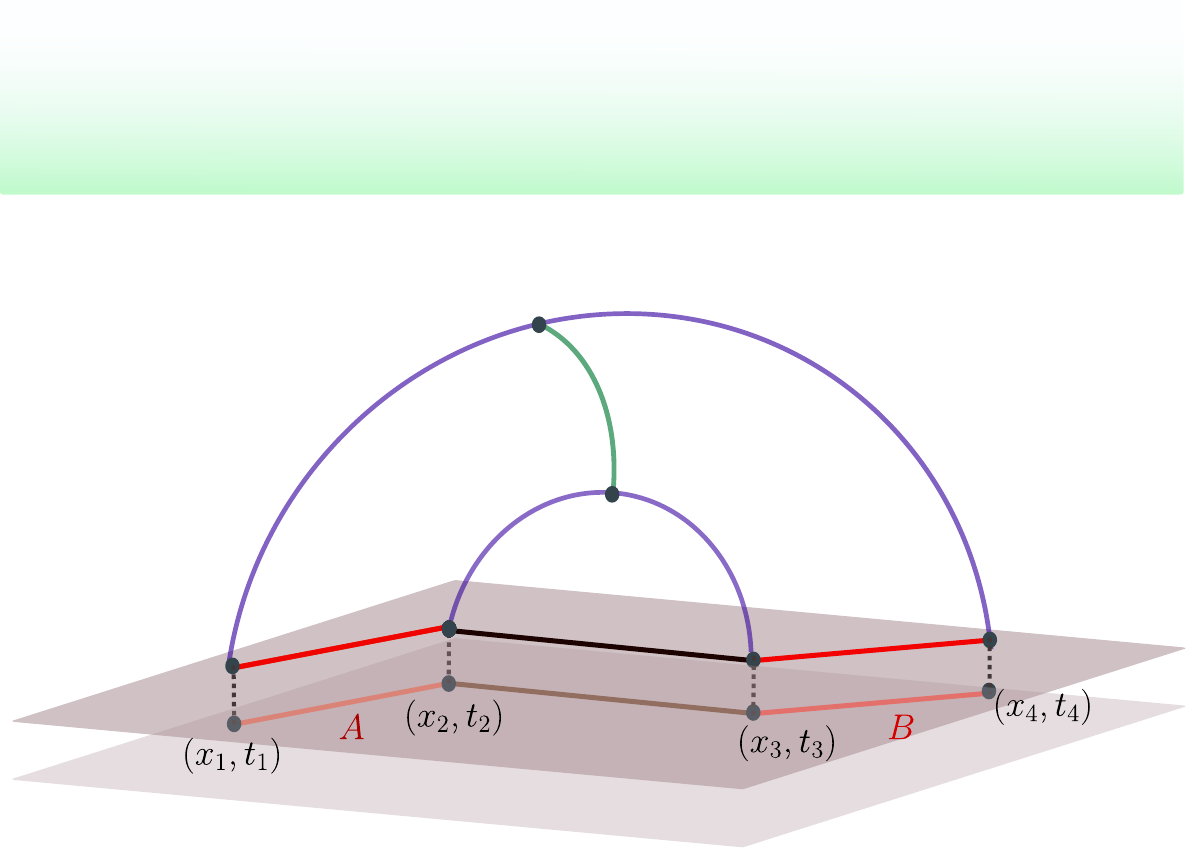}
    	\caption{Schematics of two boosted disjoint intervals $A=[(x_1,t_1),x_2,t_2]$ and $B=[(x_3,t_3),(x_4,t_4)]$ in the cutoff rotating BTZ geometry. The two shaded planes represent the holographic screens before and after the deformation. The interior of the black hole is shaded green. Figure modified from \cite{Basu:2024bal}.}
    	\label{fig:EW-disj}
    \end{figure}

	We begin with the case of two disjoint boosted intervals $A=[(x_1,t_1),(x_2,t_2)]$ on the twisted cylinder, as depicted in \cref{fig:EW-disj}.  We consider the case of a connected entanglement wedge. The green curve depicts the corresponding EWCS. We may obtain the area of the EWCS corresponding to the mixed state $\rho_{AB}$ utilizing the embedding coordinate formalism. The invariant dot product $\zeta_{ij}$ may be obtained similar to \cref{zeta12} and subsequently the EWCS may be obtained by utilizing \cref{EW-disj-formula}. 
	
	We now consider the limit of a small deformation parameter $\hat{\mu}=\frac{1}{r_c^2}$ as well as the high temperature limit $\beta\ll |x_{ij}|,t_{ij}$ to obtain the leading order correction to the EWCS due to the deformation as follows
	\begin{align}
		E_W\left(A:B\right)=&\frac{1}{4G_N}\cosh^{-1}\left[\frac{1+\sqrt{\eta\bar{\eta}}}{\sqrt{\left(1-\eta\right)\left(1-\bar{\eta}\right)}}\right]\notag\\&-\frac{\beta \pi^3 \sqrt{\eta\bar{\eta}}}{2 G_N r_c^2\beta_+^2 \beta_-^2\left(\sqrt{\eta}+\sqrt{\bar{\eta}}\right)}\left[\bP_{21}+\bP_{43}-\bP_{32}-\bP_{41}\right]\notag \\&-\frac{\beta \pi^3}{2 G_N r_c^2\beta_+^2 \beta_-^2\left(\sqrt{\eta}+\sqrt{\bar{\eta}}\right)}\left[\bP_{31}+\bP_{42}-\bP_{32}-\bP_{41}\right]\label{EW-disj}
	\end{align}
	where we have defined 
    \begin{align}
	   \bP_{ij}=\cP_{ij}+\bar\cP_{ij}=\left(x_{ij}-\Omega t_{ij}\right)\left[\coth \left(\frac{\pi  \left(x_{ij}+t_{ij}\right)}{\beta_+}\right)+\coth \left(\frac{\pi  \left(x_{ij}-t_{ij}\right)}{\beta_-}\right)\right]\label{bP-ij}
    \end{align}
    and $\eta,\bar{\eta}$ are the finite temperature cross-ratios given in \cref{cross-ratios}. The first term in \cref{EW-disj} corresponds to the EWCS corresponding to two disjoint boosted intervals in an undeformed CFT$_2$ located at the asymptotic boundary $r\to\infty$ of the bulk spacetime. On the other hand, the rest of the terms may be seen to match identically with the field theoretic results in \cref{SR-disj-corrections} upon utilizing the holographic dictionary in \cref{Cutoff-Dictionary} and the Brown-Henneaux relation in AdS$_3$/CFT$_2$ \cite{Brown:1986nw}. 
    \subsubsection*{Timelike entanglement}
    We now consider the correlation between two purely timelike subsystems $A=[(x,t_1),(x,t_2)]$ and $B=[(x,t_3),(x,t_4)]$ having a timelike separation between them. However, a precise geometric construction of a timelike entanglement wedge in the spirit of \cite{Doi:2022iyj,Doi:2023zaf} remains elusive. We proceed along the lines of \cite{Basu:2024bal} and consider a straightforward analytic continuation of the result (\ref{EW-disj}). 
    \begin{align}
    	E_W^{(\textrm{T})}(A:B)=&\frac{1}{4G_N}\cosh^{-1}\left(\frac{1+\eta_t}{1-\eta_t}\right)+\frac{1}{4G_N}\cosh^{-1}\left(\frac{1+\bar\eta_t}{1-\bar\eta_t}\right)\notag\\
    	&+\frac{\beta \pi^3\sqrt{\eta_t}}{2 G_N r_c^2\beta_+^2 \beta_-^2}\left[t_{12}\coth\left(\frac{\pi t_{12}}{\beta_+}\right)+t_{34}\coth\left(\frac{\pi t_{34}}{\beta_+}\right)-t_{14}\coth\left(\frac{\pi t_{14}}{\beta_+}\right)-t_{23}\coth\left(\frac{\pi t_{23}}{\beta_+}\right)\right]\notag\\
    	&+\frac{\beta \pi^3\sqrt{\bar\eta_t}}{2 G_N r_c^2\beta_+^2 \beta_-^2}\left[t_{12}\coth\left(\frac{\pi t_{12}}{\beta_-}\right)+t_{34}\coth\left(\frac{\pi t_{34}}{\beta_-}\right)-t_{14}\coth\left(\frac{\pi t_{14}}{\beta_-}\right)-t_{23}\coth\left(\frac{\pi t_{23}}{\beta_-}\right)\right]\,,
    \end{align}
    where the cross-ratios $\eta_t,\bar\eta_t$ are defined in \cref{Timelike-cross-ratios}.
    Note that, contrary to the non-rotating case discussed in \cite{Asrat:2017tzd,Basu:2024bal}, the timelike EWCS for the present configuration gets a non-trivial contribution due to the $\TTbar$-deformation. Furthermore, the timelike EWCS is manifestly real indicating no timelike segment in the geometric construction.
	\subsubsection{Two adjacent intervals}\label{sec-EW-adj}
\begin{figure}[h!]
	   	\centering
	   	\includegraphics[width=0.65\textwidth]{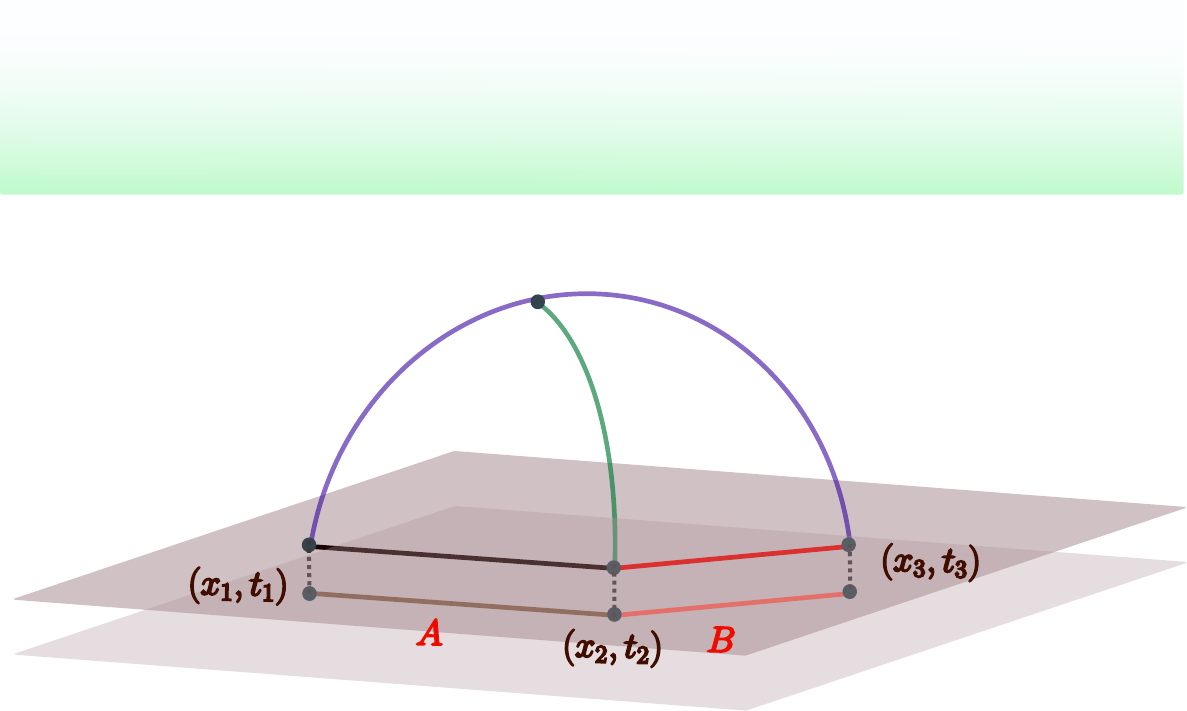}
	   	\caption{Schematics of two boosted adjacent intervals in the cutoff rotating BTZ geometry. Figure modified form \cite{Basu:2024bal}.}
	   	\label{fig:EW-adj}
	   \end{figure}

	Next, we examine the case of two adjacent intervals, $A=\left[\left(x_1,t_1\right),\left(x_2,t_2\right)\right]$ and $B=\left[\left(x_2,t_2\right),\left(x_3,t_3\right)\right]$, on the twisted cylinder. The cutoff rotating BTZ geometry is illustrated in \cref{fig:EW-adj}. The entanglement wedge corresponding to the reduced density matrix $\rho_{AB}$ is always connected and the EWCS is depicted as the green curve. By utilizing \cref{EW-adj-formula} and the embedding coordinates for the rotating BTZ geometry provided in \cref{Embedding-coordinates}, the EWCS between the two adjacent subsystems can be computed in a straightforward manner as earlier. To make connection with the field theoretic computations, we again make the assumption of a small deformation parameter $\hat{\mu}$ as well as consider the high temperature limit to obtain the leading order expression of the EWCS as follows
	   	\begin{align}
	   	E_W\left(A:B\right)&=\frac{1}{8G_N}\log \left[\frac{2\beta_+}{\pi\epsilon_c}\frac{\sinh \left(\frac{\pi  \left(x_{21}+t_{21}\right)}{\beta_+}\right) \sinh \left(\frac{\pi  \left(x_{32}+t_{32}\right)}{\beta_+}\right) }{ \sinh \left(\frac{\pi  \left(x_{31}+t_{31}\right)}{\beta_+}\right) }\right]\notag \\&+ \frac{1}{8G_N}\log \left[\frac{2\beta_-}{\pi\epsilon_c}\frac{\sinh \left(\frac{\pi  \left(x_{21}-t_{21}\right)}{\beta_-}\right) \sinh \left(\frac{\pi  \left(x_{32}-t_{32}\right)}{\beta_-}\right)}{\sinh \left(\frac{\pi  \left(x_{31}-t_{31}\right)}{\beta_-} \right)}\right]\notag \\ &-\frac{\pi ^3 \beta}{4 G_N  r_c^2 \left(\beta_- \beta_+\right){}^2}\left(\bP_{21}+\bP_{32}-\bP_{31}\right)\,,\label{EW-adj-corr}
	   \end{align}
	   where $\bP_{ij}$ is defined in \cref{bP-ij}, and we have utilized the holographic dictionary (\ref{Cutoff-Dictionary}) to incorporate the UV cut-off $\epsilon_c$ of the dual field theory. In the above expression, the logarithmic terms may be identified as the area of the EWCS corresponding to two adjacent subsystems in the undeformed CFT$_2$ with an UV cut-off $\epsilon_c$. On the other hand, the subleading terms proportional to $r_c^{-2}$ correspond to the leading correction due to the $\TTbar$-deformation which may be checked against the corrections to the reflected entropy for the present configuration obtained in \cref{SR-adj-correction} through conformal perturbation theory. 
       \subsubsection*{Timelike entanglement}
       For two purely timelike adjacent intervals $A=\left[\left(x,t_1\right),\left(x,t_2\right)\right]$ and $B=\left[\left(x,t_2\right),\left(x,t_3\right)\right]$, the bulk entanglement wedge may be identified as the region of spacetime bounded by two spacelike and one timelike geodesic segments computing the entanglement entropy of $A\cup B$ \cite{Doi:2022iyj,Doi:2023zaf}. We may compute the minimal EWCS by analytically continuing the result (\ref{EW-adj-corr}) for boosted spacelike intervals as follows
       \begin{align}
       	E_W^{(\textrm{T})}(A:B)=&\frac{1}{8G_N}\log \left[\frac{4\beta_+\beta_-}{\pi^2\epsilon_c^c}\frac{\sinh \left(\frac{\pi t_{21}}{\beta_+}\right) \sinh \left(\frac{\pi t_{32}}{\beta_+}\right) \sinh \left(\frac{\pi t_{21}}{\beta_-}\right) \sinh \left(\frac{\pi t_{32}}{\beta_-}\right)}{ \sinh \left(\frac{\pi  t_{31}}{\beta_+}\right)\sinh \left(\frac{\pi t_{31}}{\beta_-} \right) }\right]+\frac{i\pi}{4G_N}\notag \\ &+\frac{\pi ^3 \beta\Omega}{4 G_N  r_c^2 \left(\beta_- \beta_+\right){}^2}\Bigg[t_{21}  \left( \coth \left(\frac{\pi  t_{21}}{\beta_+}\right)-\coth \left(\frac{\pi  t_{21}}{\beta_-}\right)\right)\notag \\ &\hspace{3.2cm}+t_{32}\left( \coth \left(\frac{\pi  t_{32}}{\beta_+}\right)-\coth \left(\frac{\pi  t_{32}}{\beta_-}\right)\right)\notag\\&\hspace{3.2cm}-t_{31}\left( \coth \left(\frac{\pi  t_{31}}{\beta_+}\right)-\coth \left(\frac{\pi  t_{31}}{\beta_-}\right)\right)\Bigg]\,.
       \end{align}
       Once again, in the above expression, the leading order term correspond to the timelike EWCS corresponding to an undeformed CFT$_2$ with cutoff $\epsilon_c$. On the other hand, the subleading terms are easily seen to match with the corresponding correction to the reflected entropy computed through conformal perturbation theory in \cref{SR-adj-correction-timelike}. Furthermore, the presence of an imaginary part signifies a timelike geodesic segment in the bulk construction for the minimal EWCS.
	\subsubsection{A single interval}
\begin{figure}[ht]
		\centering
		\includegraphics[width=0.65\textwidth]{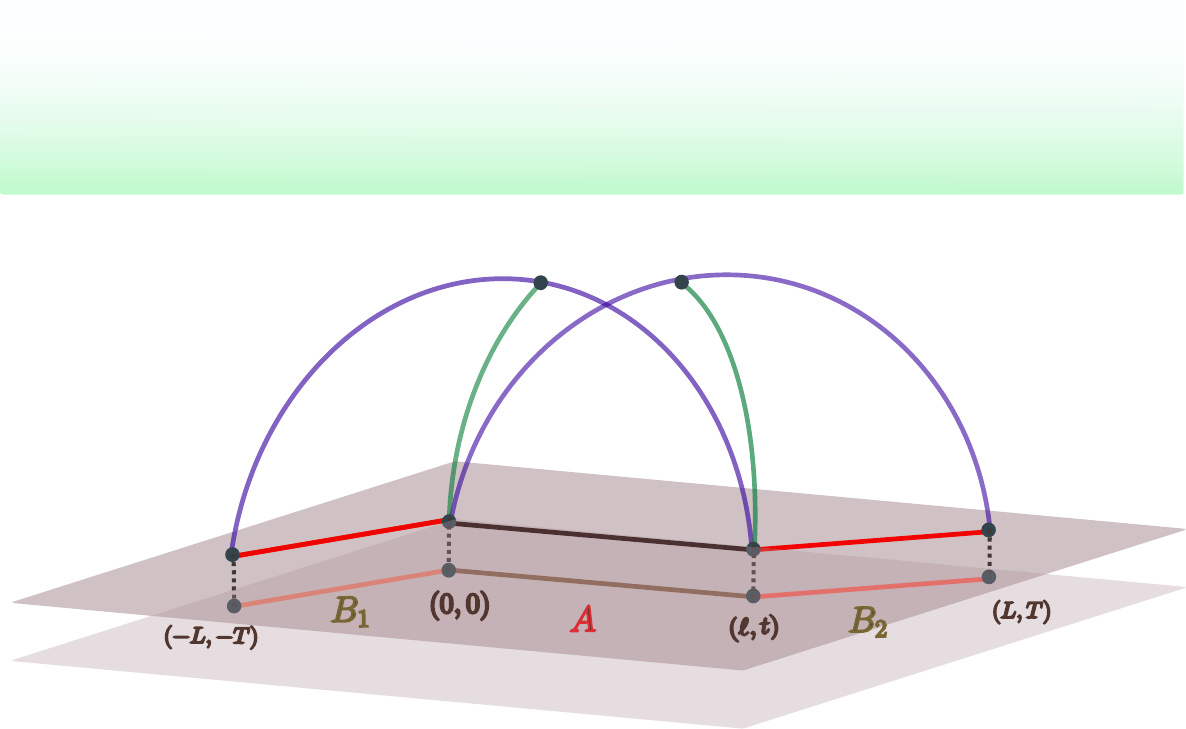}
		\caption{Schematics for the upper bound to the EWCS corresponding to a single interval in the cutoff rotating BTZ geometry. Figure modified from \cite{Basu:2024bal}.}
		\label{fig:EW-single}
	\end{figure}

	Finally, we consider a single boosted subsystem $A=\left[\left(0,0\right),\left(\ell,t\right)\right]$ in the $T\bar{T}$ deformed thermal CFT$_2$ with a conserved charge defined on the twisted cylinder $\cM$. As described in subsection \ref{sec:Single-EW}, the computation of the reflected entropy for for the present configuration requires the introduction of two large auxiliary intervals $B_1 \equiv [(-L,-T) , (0,0)]$ and $B_2 \equiv [(\ell,t) , (L,T)]$, both adjacent to $A$ as shown in \cref{fig:EW-single}. Following this, the bulk codimension one entanglement wedge dual to the reduced density matrix $\rho_{AB}$ with $B\equiv B_1\cup B_2$ may be constructed utilizing the RT surface corresponding to $A\cup B$. We may then determine an upper bound of the EWCS between $A$ and $B$ using the following expression \cite{Basu:2022nds,Basu:2021awn},
	\begin{align}
		\tilde{E}_W\left(A:B\right)=E_W\left(A:B_1\right)+E_W\left(A:B_2\right)
	\end{align}
	The upper bound of the EWCS for the given Lorentz boosted interval in the thermal CFT$_2$ with conserved charge may then be obtain by applying the bipartite limit $B\to A^c$, namely by sending the auxiliary intervals to infinity, $L\to\infty$.
	In this scenario, as $A$ and $B_i$ are adjacent, we may utilize the result (\ref{EW-adj-corr}) from the previous subsection, to obtain
	\begin{align}
		E_W\left(A:B_1\right)=&\frac{1}{8G_N}\log \left[\frac{4\beta_+\beta_-}{\pi^2\epsilon_c^2}\frac{\sinh\left(\frac{\pi(\ell+t)}{\beta_+}\right)\sinh\left(\frac{\pi(\ell-t)}{\beta_-}\right)\sinh\left(\frac{\pi(L+T)}{\beta_+}\right)\sinh\left(\frac{\pi(L-T)}{\beta_-}\right)}{\sinh\left(\frac{\pi(L+\ell+T+t)}{\beta_+}\right)\sinh\left(\frac{\pi(L+\ell-T-t)}{\beta_-}\right)}\right]\notag \\
		&-\frac{\pi ^3 \beta \left(L-T \Omega \right) \left(\coth \left(\frac{\pi  \left(L-T\right)}{\beta_-}\right)+\coth \left(\frac{\pi  \left(L+T\right)}{\beta_+}\right)\right)}{4 G_N r_{c}^2 \left(\beta_- \beta_+\right){}^2}\notag \\ &-\frac{\pi ^3 \beta \left(\ell-t \Omega \right) \left(\coth \left(\frac{\pi  \left(\ell-t\right)}{\beta_-}\right)+\coth \left(\frac{\pi  \left(t+\ell\right)}{\beta_+}\right)\right)}{4 G_N r_{c}^2 \left(\beta_- \beta_+\right){}^2}\notag \\ &+\frac{\pi ^3 \beta \left(\ell+L-(t+T) \Omega \right) \left(\coth \left(\frac{\pi  \left(\ell+L-t-T\right)}{\beta_-}\right)+\coth \left(\frac{\pi  \left(t+T+\ell+L\right)}{\beta_+}\right)\right)}{4 G_N r_{c}^2 \left(\beta_- \beta_+\right){}^2}
	\end{align}
	and
	\begin{align}
		E_W\left(A:B_2\right)=&\frac{1}{8G_N}\log \left[\frac{4\beta_+\beta_-}{\pi^2\epsilon_c^2}\frac{\sinh\left(\frac{\pi(\ell+t)}{\beta_+}\right)\sinh\left(\frac{\pi(\ell-t)}{\beta_-}\right)\sinh\left(\frac{\pi(L-\ell+T-t)}{\beta_+}\right)\sinh\left(\frac{\pi(L-\ell-T+t)}{\beta_-}\right)}{\sinh\left(\frac{\pi(L+T)}{\beta_+}\right)\sinh\left(\frac{\pi(L-T)}{\beta_-}\right)}\right]\notag \\
		& -\frac{\pi ^3 \beta \left(\ell-t \Omega \right) \left(\coth \left(\frac{\pi  \left(\ell-t\right)}{\beta_-}\right)+\coth \left(\frac{\pi  \left(t+\ell\right)}{\beta_+}\right)\right)}{4 G_N r_{c}^2 \left(\beta_- \beta_+\right){}^2}\notag \\ &-\frac{\pi ^3 \beta \left(L-\ell-(T-t) \Omega \right) \left(\coth \left(\frac{\pi  \left(L-\ell-T+t\right)}{\beta_-}\right)+\coth \left(\frac{\pi  \left(L-\ell+T-t\right)}{\beta_+}\right)\right)}{4 G_N r_{c}^2 \left(\beta_- \beta_+\right){}^2}\notag \\ &+\frac{\pi ^3 \beta\left(L-T \Omega \right) \left(\coth \left(\frac{\pi  \left(L-T\right)}{\beta_-}\right)+\coth \left(\frac{\pi  \left(T+L\right)}{\beta_+}\right)\right)}{4 G_N r_{c}^2\left(\beta_- \beta_+\right){}^2}\,.
	\end{align}
	The upper bound of the EWCS for the single boosted subsystem with angular momentum can now be obtained by applying the bipartite limit $L\to \infty$ as follows
	\begin{align}
		\tilde{E}_{W}\left(A:A^c\right)&=\frac{1}{4G_N}\log\left[\frac{4\beta_+\beta_-}{\pi^2\epsilon_c^2}\sinh\left(\frac{\pi(\ell+t)}{\beta_+}\right)\sinh\left(\frac{\pi(\ell-t)}{\beta_-}\right)\right]-\frac{\pi(\ell-\Omega\,t)}{2G_N\beta(1-\Omega^2)}\notag\\
		&-\frac{\pi^3\beta}{2G_N r_c^2\beta _+^2\beta_-^2}(\ell- \Omega\,t)\left[\coth \left(\frac{\pi  (\ell+t)}{\beta_+}\right)+\coth \left(\frac{\pi  (\ell-t)}{\beta_-}\right)-2\right]\,.\label{EW-single-full}
	\end{align}
	As earlier, the first two terms in the above expression correspond to the (upper bound to the) EWCS corresponding to a single interval in the undeformed CFT$_2$ with a modified cut-off $\epsilon_c$ \cite{Basu:2022nds}, whereas the subleading terms proportional to $r_c^{-2}$ correspond to the corrections due to the $\TTbar$-deformation. utilizing the holographic dictionary (\ref{Cutoff-Dictionary}) and the Brown-Henneaux relation, the subleading terms may be identified with the corresponding corrections to the reflected entropy obtained from conformal perturbation theory in \cref{SR-sing-correction}. Note that the non-universal functions $f,\bar{f}$ are expected to be sub-dominant in the large central charge limit and hence do not appear in our holographic computations.
	
	Interestingly, the above expression for the EWCS corresponding to a single interval in the $\TTbar$-deformed CFT$_2$ on the twisted cylinder $\cM$ may be recast in the instructive form
	\begin{align}
		\tilde{E}_{W}\left(A:A^c\right)=S(A)-S^\textrm{Th}(A)+\frac{1}{2G_N}\log 2\,,
	\end{align}
	where $S(A)$ is the corresponding entanglement entropy and $S^\textrm{Th}(A)$ represents the thermal entropy
	\begin{align}
		S^\textrm{Th}(A)=\frac{\pi(\ell-\Omega\,t)}{2G_N\beta(1-\Omega^2)}\left(1-\frac{\pi^2}{r_c^2\beta_+\beta_-}\right)=\frac{\pi c(\ell-\Omega\,t)}{3\beta(1-\Omega^2)}\left(1-\frac{2\pi^2\hat{\mu}}{\beta^2(1-\Omega^2)}\right)\,.
	\end{align}
	The above expression indicates that the thermal entropy gets non-trivial correction due to the $\TTbar$ deformation. This may be interpreted in terms of the wrapping of the corresponding HRT surface around the black hole horizon. As the holographic screen is pushed inside the bulk, the portion of the HRT surface wrapped around the black hole horizon decreases.
	\subsubsection*{Timelike entanglement}
	We now consider a purely timelike interval $A=[(x,0),(x,t)]$ on the twisted cylinder with a compactified time direction. Hence, contrary to the spacelike subsystem considered above, in this case no auxiliary subsystems are needed to remove the pathology of tracing out an infinite branch cut. Furthermore, the entanglement wedge for the present configuration may be identified as the bulk region bounded by the collection of extremal surfaces computing the entanglement entropy of $A$, as demonstrated in \cite{Doi:2022iyj,Doi:2023zaf}. As a result, the minimal EWCS reduces to the HRT surface homologous to the subsystem under consideration. The length of such extremal surface has already been computed in subsection \ref{HEE} with the result (\ref{Timelike-HEE}). Upon utilizing the holographic dictionary in \cref{Cutoff-Dictionary} and the Brown-Henneaux relation in AdS$_3$/CFT$_2$, the EWCS matches identically with half of the reflected entropy for the present configuration.
	\subsection{Holographic mutual information and Markov gap}
	In this subsection, we report the leading order corrections to the holographic Markov gap \cite{Hayden:2021gno} due to the $\TTbar$ deformation. The Markov gap, defined as the difference between the reflected entropy and holographic mutual information, is related to the fidelity of a specific Markov recovery process concerning the canonical purification of a given mixed state.
	Furthermore, in \cite{Hayden:2021gno}, it was posited that in holographic theories, the Markov gap is bounded in terms of the number of non-trivial boundaries of the EWCS as follows
	\begin{align}
		h(A:B)=S_R(A:B)-I(A:B)\geq \ell_\textrm{AdS}\frac{\log 2}{2G_N}\times (\textrm{boundaries of EWCS})+\cO\left(\frac{1}{G_N}\right)\,.\label{Hayden-inequality}
	\end{align}
	In the following, we observe that deforming a CFT$_2$ with the $\TTbar$ operator does not always lead to finite corrections to the Markov gap.
	\subsection{Disjoint intervals}
	For the case of two disjoint boosted intervals with a connected entanglement wedge, the holographic mutual information may readily be obtained from \cref{HEE-corrections} as follows:
	\begin{align}
		I(A:B)&=S(A)+S(B)-S(A\cup B)\notag\\
		&=\frac{1}{4G_N}\log\left[\frac{\eta\bar\eta}{(1-\eta)(1-\bar\eta)}\right]-\mu\frac{\pi^4\beta c^2}{18\left(\beta_+\beta_-\right)^2}\left(\bP_{12}+\bP_{34}-\bP_{14}-\bP_{23}\right)\,,
	\end{align} 
	where, we have defined $\bP_{ij}=\cP_{ij}+\bar\cP_{ij}$, and the functions $\cP\,,\,\bar\cP$ are defined in \cref{P-functions}. Utilizing \cref{SR-disj-corrections,EW-disj}, we may now compute the leading order correction to the holographic Markov gap as follows
	\begin{align}
		h(A:B)&=S_R(A:B)-I(A:B)\notag\\
		=&\frac{c}{6}\log\left[\frac{(1+\sqrt{\eta})^2(1+\sqrt{\bar{\eta}})^2}{\eta\bar\eta}\right]\notag\\
		&-\mu\frac{\pi^4\beta c^2}{18\left(\beta_+\beta_-\right)^2}\left(\sqrt{\eta}-1\right)\left(\cP_{12}+\cP_{34}-\cP_{14}-\cP_{23}\right)\notag\\
		&-\mu\frac{\pi^4\beta c^2}{18\left(\beta_+\beta_-\right)^2}\left(\sqrt{\bar\eta}-1\right)\left(\bar\cP_{12}+\bar\cP_{34}-\bar\cP_{14}-\bar\cP_{23}\right)\,.\label{MarkovGap-Disj}
	\end{align}
	Note that the inequality (\ref{Hayden-inequality}) is obeyed even after deforming the theory with the $\TTbar$ operator (recall that the cross-ratios lie within $0\leq \eta,\bar\eta\leq 1$).
	The corresponding correction to the Markov gap for tow disjoint boosted intervals in the absence of a conserved angular momentum may be obtained by setting $\beta_+=\beta_-$. Furthermore, we observe that for purely timelike subsystems on a spatial slice, the correction to the Markov gap is always positive indicating a better Markov recovery compared to the undeformed CFT$_2$. It would be interesting to investigate the physical origin of such behavior along the lines of \cite{Hayden:2021gno}.
	\subsection{Adjacent intervals}
	Next we consider the case of two adjacent boosted intervals, discussed in subsection \ref{sec-EW-adj}. As earlier, the holographic mutual information may be obtained from \cref{HEE-corrections} as follows
	\begin{align}
		I(A:B)=&\frac{c}{6}\log\left[\frac{\beta_+\beta_-}{\pi^2\epsilon_c^2}\frac{\sinh\left(\frac{\pi(x_{21}+t_{21})}{\beta_+}\right)\sinh\left(\frac{\pi(x_{32}+t_{32})}{\beta_+}\right)\sinh\left(\frac{\pi(x_{21}-t_{21})}{\beta_-}\right)\sinh\left(\frac{\pi(x_{32}-t_{32})}{\beta_-}\right)}{\sinh\left(\frac{\pi(x_{31}+t_{31})}{\beta_+}\right)\sinh\left(\frac{\pi(x_{31}-t_{31})}{\beta_-}\right)}\right]\notag\\
		&-\frac{\pi ^4 \beta  c^2 \mu }{18 \beta_-^2 \beta_+^2}\left(\bP_{12}+\bP_{23}-\bP_{13}\right)
	\end{align}
	Utilizing \cref{SR-adj-correction,EW-adj-corr}, the holographic Markov gap for the present configuration may now be obtained as
	\begin{align}
		h(A:B)=\frac{c}{3}\log 2\,.
	\end{align}
	Interestingly, in this case the holographic Markov gap is given by a constant irrespective of the strength of the deformation. This is also evident from \cref{MarkovGap-Disj} as the adjacent limit is approached as $\eta,\bar\eta\to 1$.
	\subsection{single interval}
	Finally, we investigate the leading order correction to the Markov gap for a single interval on the twisted cylinder. We utilize the monogamy of holographic mutual information \cite{Hayden:2011ag} to obtain
	\begin{align}
		I(A:B_1B_2)\geq& I(A:B_1)+I(A:B_2)\notag\\
		=&\frac{c}{6}\log\left[\frac{\beta_+^2\beta_-^2}{\pi^4\epsilon_c^4}\frac{\sinh^2\left(\frac{\pi(\ell+t)}{\beta_+}\right)\sinh^2\left(\frac{\pi(\ell-t)}{\beta_-}\right)\sinh\left(\frac{\pi(L-\ell+T-t)}{\beta_+}\right)\sinh\left(\frac{\pi(L-\ell-T+t)}{\beta_-}\right)}{\sinh\left(\frac{\pi(L+\ell+T+t)}{\beta_+}\right)\sinh\left(\frac{\pi(L+\ell-T-t)}{\beta_-}\right)}\right]\notag\\
		&-\mu\frac{\pi^4c^2\beta}{9\beta_+^2\beta_-^2}(\ell-t\Omega )\Big[\coth\left(\frac{\pi(\ell+t)}{\beta_+}\right)+\coth\left(\frac{\pi(\ell-t)}{\beta_-}\right)\Big]\notag\\&-\mu\frac{\pi^4c^2\beta}{18\beta_+^2\beta_-^2}\left(L+\ell-\Omega(T+t)\right)\Big[\coth\left(\frac{\pi(L+\ell+T+t)}{\beta_+}\right)+\coth\left(\frac{\pi(L+\ell-T-t)}{\beta_-}\right)\Big]\notag\\
		&+\mu\frac{\pi^4c^2\beta}{18\beta_+^2\beta_-^2}\left(L-\ell-\Omega(T-t)\right)\Big[\coth\left(\frac{\pi(L-\ell+T-t)}{\beta_+}\right)+\coth\left(\frac{\pi(L-\ell-T+t)}{\beta_-}\right)\Big]
	\end{align}
	Subsequently, employing the bipartite limit $L\to\infty$, we obtain the following lower bound to the mutual information between the single interval and its complement:
	\begin{align}
		I(A:A^c)\geq &\frac{c}{3}\log\left[\frac{\beta_+\beta_-}{\pi^2\epsilon_c^2}\sinh\left(\frac{\pi(\ell+t)}{\beta_+}\right)\sinh\left(\frac{\pi(\ell-t)}{\beta_-}\right)\right]-\frac{2\pi c}{3\beta(1-\Omega^2)}(\ell-\Omega\,t)\notag\\
		&-\mu\frac{\pi^4\beta c^2}{9\beta _+^2\beta_-^2}(\ell- \Omega\,t)\left[\coth \left(\frac{\pi  (\ell+t)}{\beta_+}\right)+\coth \left(\frac{\pi  (\ell-t)}{\beta_-}\right)-2\right]
	\end{align}
	Now, from \cref{EW-single-full,HEE-corrections}, we may obtain an upper bound to the holographic Markov gap for the present configuration as follows:
	\begin{align}
		h(A:B)\leq \frac{c}{3}\log 4\,.
	\end{align}
	Note that the inequality (\ref{Hayden-inequality}) is saturated owing to the two boundaries of the (upper bound to the) EWCS. Once again, the Markov gap is independent of the deformation parameter $\mu$.
	
	\section{Summary and discussion}\label{sec-5}
	To summarize, in this article, we have investigated the mixed state entanglement and correlation structure in a $\TTbar$-deformed thermal CFT$_2$ with a conserved angular momentum defined on a twisted cylinder with complex identifications. In particular, we have investigated the entanglement entropy and the reflected entropy in such deformed CFT$_2$ with a particular focus on Lorentz boosted subsystems. We have extended the conformal perturbation theory developed in earlier literature to the case with a conserved charge through a coordinate rotation that effective changes the complex period of the twisted cylinder. In particular, we obtained a CFT$_2$ defined on a cylinder with temporal identification whose bulk dual geometry is given by a non-rotating black hole with an effective temperature which encapsulates the essence of rotation. Utilizing this rotated frame, we obtain the leading order correction to the entanglement entropy and the reflected entropy for various bipartite mixed states in the $\TTbar$-deformed theory with conserved angular momentum. We have also investigated the timelike entanglement in such field theories, and found that the leading corrections for timelike subsystems remain positive, in contrast to the cases with boosted spacelike subsystems.
	
	The holographic dual of the $\TTbar$-deformed theory defined on the twisted cylinder is described by a rotating BTZ black hole with a finite radial cut-off. We have obtained the holographic reflected entropy through the Ryu-Takayanagi prescription by obtaining the induced metric on the cut-off surface through the coordinate rotation discussed above. Interestingly, the holographic entanglement entropy is complex valued and imposing its reality puts an universal upper bound to the strength of the deformation. Furthermore, there also exists an universal zero of the entanglement entropy which might be related to the Hagedorn behavior of the theory. Furthermore, we have found that $\TTbar$-deformed CFT$_2$ have a complex entanglement spectrum and violates the boosted strong sub-additivity property of entanglement entropy, thereby exhibiting non-locality. We have also obtained the minimal cross-section of the entanglement wedge, holographically dual to the reflected entropy, in the cut-off BTZ geometry. Interestingly, to the leading order in the deformation parameter, all our holographic results match identically with earlier field theoretic computations from conformal perturbation theory.

    In this study, we have restricted ourselves to finite temperature systems. however, the entanglement properties of a $\TTbar$-deformed CFT$_2$ with conserved angular momentum at zero temperature may be obtained by taking appropriate limits of our finite temperature results. The bulk dual geometry is described by an extremal rotating BTZ black hole with a finite radial cut-off. Due to the presence of spin, observables in such theories depend on an effective Frolov-Thorne temperature $\beta_{\textrm{FT}}^{-1}$. For example, the leading order correction to the entanglement entropy of a Lorentz boosted interval $A=[(0,0),(x,t)]$ may be obtained from \cref{EE-corrections} by taking the limit $\beta\to\infty$ and subsequently setting $\Omega=1$,
	\begin{align*}
		-\mu\frac{\pi^3c^2}{36\beta_{\textrm{FT}}^2}\left(\frac{x-t}{x+t}\right)\,.
	\end{align*}
	The above result is supported by the extremal limit of the holographic computations for the non-extremal black hole, performed in subsection \ref{HEE}. However, a concrete field theoretic interpretation of the above result is still illusive, and the known perturbative analysis in \cite{Chen:2018eqk,Jeong:2019ylz} seems to indicate a vanishing contribution from the $\TTbar$ operator insertion. Moreover, a straightforward holographic computation of the HRT surface utilizing the extremal BTZ metric does not seem to align with the above result. We hope to return to this issue in the near future.
	
	There are several other open issues related to our study. It will be interesting to study other mixed state entanglement and correlation measures such as the entanglement negativity \cite{Vidal:2002zz,Plenio:2005cwa}, the balanced partial entanglement \cite{Wen:2021qgx} and the entanglement of purification \cite{Takayanagi:2017knl}, in the context of $\TTbar$-deformed CFT$_2$ with conserved charges holographically dual to rotating black holes with finite radial cut-offs. Interestingly, in a recent article \cite{Deng:2023pjs}, the authors have investigated the $\TTbar$ deformation of holographic boundary conformal field theories extending the island formulation of the black hole information loss paradigm to $\TTbar$-deformed baths coupled to semi-classical gravity. The mixed state entanglement structure in Hawking radiation in such a non-conformal bath may be studied through the island framework for the reflected entropy introduced in \cite{Chandrasekaran:2020qtn,Li:2020ceg}. It will also be interesting to explore different holographic descriptions of $\TTbar$-deformed CFTs in this context, such as the mixed boundary condition proposal of \cite{Guica:2019nzm}. Perhaps a more profound understanding of the non-local properties of the $\TTbar$-deformed CFT$_2$ requires non-perturbative field theoretic computations of the entanglement properties, maybe along the lines of \cite{Lewkowycz:2019xse,Chang:2024voo}. We leave these interesting open issues for future investigations.
	\section*{Acknowledgment}
	The authors would like to express their sincere gratitude to Mrinmoy Biswas, Apratim Kaviraj and Diptarka Das for their valuable discussions and insights that greatly contributed to the development of this work. SB acknowledges the financial support provided by the Council of Scientific and Industrial Research (CSIR) under the grant number 09/0092(12686)/2021-EMR-I.
	\appendix
	\section{On rotating BTZ}\label{appB}
	In this appendix, we recollect some details on the (Euclidean) rotating BTZ black hole with complex periods. The Euclidean rotating BTZ metric in the Schwarzschild coordinates is given by
	\begin{align}
		\td s^2=\frac{(r^2-r_+^2)(r^2+\tilde{r}_-^2)}{r^2}\td\tau^2+\frac{r^2}{(r^2-r_+^2)(r^2+\tilde{r}_-^2)}\td r^2+r^2\left(\td\phi+\frac{r_+\tilde{r}_-}{r^2}\td \tau^2\right)\,,\label{EuclideanBTZ}
	\end{align}
	where $\tilde{r}_-=i r_-$ and the complex period 
	\begin{align*}
		(x,\tau)\sim(x-i\beta\Omega_E,\tau+\beta)
	\end{align*}
	is related to the inner and outer horizon radii through \cref{Temperature-AngSpeed}. We may recast \cref{EuclideanBTZ} in the Fefferman-Graham form
	\begin{align}
		\td s^2=\frac{\td\rho^2}{4\rho^2}+\frac{1}{\rho}\left(\td w+\rho\bar{L}_0(\bar{w})\td\bar{w}\right)\left(\td \bar{w}+\rho L_0(w)\td w\right)\,,\label{Flat-FG-expansion}
	\end{align}
	by choosing 
	\begin{align}
		L_0=\frac{1}{4}(r_++i\tilde{r}_-)^2=\frac{\pi^2}{\beta_+^2}~~,~~\bar{L}_0=\frac{1}{4}(r_+-i\tilde{r}_-)^2=\frac{\pi^2}{\beta_-^2}\,,
	\end{align}
	and redefining the radial coordinate as
	\begin{align}
		r^2=\frac{1}{\rho}+L_0+\bar{L}_0+\rho L_0\bar{L}_0\,.
	\end{align}
	Note that, any general asymptotically AdS$_3$ solution may be written in the Fefferman-Graham expansion
	\begin{align}
		\td s^2=\frac{\td\rho^2}{4\rho^2}+g_{ab}\td x^a\td x^b~~,~~g_{ab}=\frac{1}{\rho}g_{ab}^{(0)}+g_{ab}^{(2)}+\rho g_{ab}^{(4)}\,,
	\end{align}
	where the expansion coefficients $g_{ab}^{(2)}$ and $g_{ab}^{(4)}$ may be algebraically obtained from the seed metric $g_{ab}^{(0)}$ by Employing the Einstein equations. Incidentally, the form (\ref{Flat-FG-expansion}) is an example of the Fefferman-Graham expansion with a flat seed metric $g_{ab}^{(0)}$. In particular, $g_{ab}^{(2)}$ determines the seed expectation value of the CFT stress tensor as \cite{Skenderis:1999nb,deHaro:2000vlm,Balasubramanian:1999re}
	\begin{align}
		g_{ab}^{(2)}=8\pi G_N \left(T_{ab}^{(0)}- \textrm{Tr}[T^{(0)}]g_{ab}^{(0)}\right)\,.
	\end{align}
	From the above holographic stress tensor, one may obtain the holographic mass and angular momentum as
	\begin{align}
		&M=-\int_{0}^{2\pi}\td x \, T_{\tau\tau}^{(0)}=\frac{c}{6}(L_0+\bar{L}_0)=\frac{c}{3}\left(r_+^2-\tilde{r}_-^2\right)\notag\\
		&\tilde{J}=-\int_{0}^{2\pi}\td x \, T_{\tau x}^{(0)}=-i\frac{c}{6}(L_0-\bar{L}_0)=\frac{2c}{3}r_+\tilde{r}_-\,.
	\end{align}
	We may obtain a generic torus background by making a coordinate rotation
	\begin{align}
		w\to e^{i\theta}w=w^\prime\,,
	\end{align}
	or, equivalently
	\begin{align}
			\begin{pmatrix}
				x^\prime \\ \tau^\prime
			\end{pmatrix}
			=\begin{pmatrix}
				\cos\theta & -\sin\theta\\
				\sin\theta  & \cos\theta
			\end{pmatrix}
			\begin{pmatrix}
				x \\ \tau
			\end{pmatrix}
	\end{align}
	The periods are transformed accordingly as
	\begin{align}
		\begin{pmatrix}
			\beta^\prime \\ -\beta^\prime\Omega_E^\prime
		\end{pmatrix}
		=\begin{pmatrix}
			\cos\theta & -\sin\theta\\
			\sin\theta  & \cos\theta
		\end{pmatrix}
		\begin{pmatrix}
			\beta \\ -\beta\Omega_E
		\end{pmatrix}
	\end{align}
	In particular, we may obtain an equivalent description of the rotating BTZ black hole in terms of a non-rotating one with the choice $\theta=\tan^{-1}\left(\Omega_E\right)$, leading to
	\begin{align}
		\beta^\prime=\beta\sqrt{1+\Omega_E^2}~~,~~\Omega_E^\prime=0\,.
	\end{align}
	Interestingly, this is nothing but the coordinate rotation used in \cref{Bichi-map} to investigate the entanglement properties in $\TTbar$-deformed CFT$_2$s with a conserved angular momentum.
	\bibliographystyle{utphys}
	\bibliography{reference}

\providecommand{\href}[2]{#2}\begingroup\raggedright\begin{thebibliography}{10}

\bibitem{Maldacena:1997re}
J.~M. Maldacena, ``{The Large N limit of superconformal field theories and
  supergravity},'' \href{http://dx.doi.org/10.4310/ATMP.1998.v2.n2.a1}{{\em
  Adv. Theor. Math. Phys.} {\bfseries 2} (1998) 231--252},
  \href{http://arxiv.org/abs/hep-th/9711200}{{\ttfamily arXiv:hep-th/9711200}}.

\bibitem{Gubser:1998bc}
S.~S. Gubser, I.~R. Klebanov, and A.~M. Polyakov, ``{Gauge theory correlators
  from noncritical string theory},''
  \href{http://dx.doi.org/10.1016/S0370-2693(98)00377-3}{{\em Phys. Lett. B}
  {\bfseries 428} (1998) 105--114},
  \href{http://arxiv.org/abs/hep-th/9802109}{{\ttfamily arXiv:hep-th/9802109}}.

\bibitem{VanRaamsdonk:2010pw}
M.~Van~Raamsdonk, ``{Building up spacetime with quantum entanglement},''
  \href{http://dx.doi.org/10.1142/S0218271810018529}{{\em Gen. Rel. Grav.}
  {\bfseries 42} (2010) 2323--2329},
  \href{http://arxiv.org/abs/1005.3035}{{\ttfamily arXiv:1005.3035 [hep-th]}}.

\bibitem{Maldacena:2013xja}
J.~Maldacena and L.~Susskind, ``{Cool horizons for entangled black holes},''
  \href{http://dx.doi.org/10.1002/prop.201300020}{{\em Fortsch. Phys.}
  {\bfseries 61} (2013) 781--811},
  \href{http://arxiv.org/abs/1306.0533}{{\ttfamily arXiv:1306.0533 [hep-th]}}.

\bibitem{Ryu:2006bv}
S.~Ryu and T.~Takayanagi, ``{Holographic derivation of entanglement entropy
  from AdS/CFT},'' \href{http://dx.doi.org/10.1103/PhysRevLett.96.181602}{{\em
  Phys. Rev. Lett.} {\bfseries 96} (2006) 181602},
  \href{http://arxiv.org/abs/hep-th/0603001}{{\ttfamily arXiv:hep-th/0603001}}.

\bibitem{Ryu:2006ef}
S.~Ryu and T.~Takayanagi, ``{Aspects of Holographic Entanglement Entropy},''
  \href{http://dx.doi.org/10.1088/1126-6708/2006/08/045}{{\em JHEP} {\bfseries
  08} (2006) 045}, \href{http://arxiv.org/abs/hep-th/0605073}{{\ttfamily
  arXiv:hep-th/0605073}}.

\bibitem{Hubeny:2007xt}
V.~E. Hubeny, M.~Rangamani, and T.~Takayanagi, ``{A Covariant holographic
  entanglement entropy proposal},''
  \href{http://dx.doi.org/10.1088/1126-6708/2007/07/062}{{\em JHEP} {\bfseries
  07} (2007) 062}, \href{http://arxiv.org/abs/0705.0016}{{\ttfamily
  arXiv:0705.0016 [hep-th]}}.

\bibitem{Lewkowycz:2013nqa}
A.~Lewkowycz and J.~Maldacena, ``{Generalized gravitational entropy},''
  \href{http://dx.doi.org/10.1007/JHEP08(2013)090}{{\em JHEP} {\bfseries 08}
  (2013) 090}, \href{http://arxiv.org/abs/1304.4926}{{\ttfamily arXiv:1304.4926
  [hep-th]}}.

\bibitem{Dong:2016hjy}
X.~Dong, A.~Lewkowycz, and M.~Rangamani, ``{Deriving covariant holographic
  entanglement},'' \href{http://dx.doi.org/10.1007/JHEP11(2016)028}{{\em JHEP}
  {\bfseries 11} (2016) 028}, \href{http://arxiv.org/abs/1607.07506}{{\ttfamily
  arXiv:1607.07506 [hep-th]}}.

\bibitem{Casini:2011kv}
H.~Casini, M.~Huerta, and R.~C. Myers, ``{Towards a derivation of holographic
  entanglement entropy},''
  \href{http://dx.doi.org/10.1007/JHEP05(2011)036}{{\em JHEP} {\bfseries 05}
  (2011) 036}, \href{http://arxiv.org/abs/1102.0440}{{\ttfamily arXiv:1102.0440
  [hep-th]}}.

\bibitem{Vidal:2002zz}
G.~Vidal and R.~F. Werner, ``{Computable measure of entanglement},''
  \href{http://dx.doi.org/10.1103/PhysRevA.65.032314}{{\em Phys. Rev. A}
  {\bfseries 65} (2002) 032314},
  \href{http://arxiv.org/abs/quant-ph/0102117}{{\ttfamily
  arXiv:quant-ph/0102117}}.

\bibitem{Plenio:2005cwa}
M.~B. Plenio, ``{Logarithmic Negativity: A Full Entanglement Monotone That is
  not Convex},'' \href{http://dx.doi.org/10.1103/PhysRevLett.95.090503}{{\em
  Phys. Rev. Lett.} {\bfseries 95} (2005) 090503},
  \href{http://arxiv.org/abs/quant-ph/0505071}{{\ttfamily
  arXiv:quant-ph/0505071}}.

\bibitem{Dutta:2019gen}
S.~Dutta and T.~Faulkner, ``{A canonical purification for the entanglement
  wedge cross-section},'' \href{http://dx.doi.org/10.1007/JHEP03(2021)178}{{\em
  JHEP} {\bfseries 03} (2021) 178},
  \href{http://arxiv.org/abs/1905.00577}{{\ttfamily arXiv:1905.00577
  [hep-th]}}.

\bibitem{Takayanagi:2017knl}
T.~Takayanagi and K.~Umemoto, ``{Entanglement of purification through
  holographic duality},''
  \href{http://dx.doi.org/10.1038/s41567-018-0075-2}{{\em Nature Phys.}
  {\bfseries 14} no.~6, (2018) 573--577},
  \href{http://arxiv.org/abs/1708.09393}{{\ttfamily arXiv:1708.09393
  [hep-th]}}.

\bibitem{Wen:2021qgx}
Q.~Wen, ``{Balanced Partial Entanglement and the Entanglement Wedge Cross
  Section},'' \href{http://dx.doi.org/10.1007/JHEP04(2021)301}{{\em JHEP}
  {\bfseries 04} (2021) 301}, \href{http://arxiv.org/abs/2103.00415}{{\ttfamily
  arXiv:2103.00415 [hep-th]}}.

\bibitem{Zamolodchikov:2004ce}
A.~B. Zamolodchikov, ``{Expectation value of composite field T anti-T in
  two-dimensional quantum field theory},''
  \href{http://arxiv.org/abs/hep-th/0401146}{{\ttfamily arXiv:hep-th/0401146}}.

\bibitem{Cavaglia:2016oda}
A.~Cavagli\`a, S.~Negro, I.~M. Sz\'ecs\'enyi, and R.~Tateo, ``{$T
  \bar{T}$-deformed 2D Quantum Field Theories},''
  \href{http://dx.doi.org/10.1007/JHEP10(2016)112}{{\em JHEP} {\bfseries 10}
  (2016) 112}, \href{http://arxiv.org/abs/1608.05534}{{\ttfamily
  arXiv:1608.05534 [hep-th]}}.

\bibitem{Smirnov:2016lqw}
F.~A. Smirnov and A.~B. Zamolodchikov, ``{On space of integrable quantum field
  theories},'' \href{http://dx.doi.org/10.1016/j.nuclphysb.2016.12.014}{{\em
  Nucl. Phys. B} {\bfseries 915} (2017) 363--383},
  \href{http://arxiv.org/abs/1608.05499}{{\ttfamily arXiv:1608.05499
  [hep-th]}}.

\bibitem{McGough:2016lol}
L.~McGough, M.~Mezei, and H.~Verlinde, ``{Moving the CFT into the bulk with $
  T\overline{T} $},'' \href{http://dx.doi.org/10.1007/JHEP04(2018)010}{{\em
  JHEP} {\bfseries 04} (2018) 010},
  \href{http://arxiv.org/abs/1611.03470}{{\ttfamily arXiv:1611.03470
  [hep-th]}}.

\bibitem{Asrat:2017tzd}
M.~Asrat, A.~Giveon, N.~Itzhaki, and D.~Kutasov, ``{Holography Beyond AdS},''
  \href{http://dx.doi.org/10.1016/j.nuclphysb.2018.05.005}{{\em Nucl. Phys. B}
  {\bfseries 932} (2018) 241--253},
  \href{http://arxiv.org/abs/1711.02690}{{\ttfamily arXiv:1711.02690
  [hep-th]}}.

\bibitem{Shyam:2017znq}
V.~Shyam, ``{Background independent holographic dual to $T\bar{T}$ deformed CFT
  with large central charge in 2 dimensions},''
  \href{http://dx.doi.org/10.1007/JHEP10(2017)108}{{\em JHEP} {\bfseries 10}
  (2017) 108}, \href{http://arxiv.org/abs/1707.08118}{{\ttfamily
  arXiv:1707.08118 [hep-th]}}.

\bibitem{Kraus:2018xrn}
P.~Kraus, J.~Liu, and D.~Marolf, ``{Cutoff AdS$_{3}$ versus the $ T\overline{T}
  $ deformation},'' \href{http://dx.doi.org/10.1007/JHEP07(2018)027}{{\em JHEP}
  {\bfseries 07} (2018) 027}, \href{http://arxiv.org/abs/1801.02714}{{\ttfamily
  arXiv:1801.02714 [hep-th]}}.

\bibitem{Cottrell:2018skz}
W.~Cottrell and A.~Hashimoto, ``{Comments on $T \bar T$ double trace
  deformations and boundary conditions},''
  \href{http://dx.doi.org/10.1016/j.physletb.2018.09.068}{{\em Phys. Lett. B}
  {\bfseries 789} (2019) 251--255},
  \href{http://arxiv.org/abs/1801.09708}{{\ttfamily arXiv:1801.09708
  [hep-th]}}.

\bibitem{Taylor:2018xcy}
M.~Taylor, ``{$T \bar{T}$ deformations in general dimensions},''
  \href{http://dx.doi.org/10.4310/ATMP.2023.v27.n1.a2}{{\em Adv. Theor. Math.
  Phys.} {\bfseries 27} no.~1, (2023) 37--63},
  \href{http://arxiv.org/abs/1805.10287}{{\ttfamily arXiv:1805.10287
  [hep-th]}}.

\bibitem{Hartman:2018tkw}
T.~Hartman, J.~Kruthoff, E.~Shaghoulian, and A.~Tajdini, ``{Holography at
  finite cutoff with a $T^2$ deformation},''
  \href{http://dx.doi.org/10.1007/JHEP03(2019)004}{{\em JHEP} {\bfseries 03}
  (2019) 004}, \href{http://arxiv.org/abs/1807.11401}{{\ttfamily
  arXiv:1807.11401 [hep-th]}}.

\bibitem{Shyam:2018sro}
V.~Shyam, ``{Finite Cutoff AdS$_{5}$ Holography and the Generalized Gradient
  Flow},'' \href{http://dx.doi.org/10.1007/JHEP12(2018)086}{{\em JHEP}
  {\bfseries 12} (2018) 086}, \href{http://arxiv.org/abs/1808.07760}{{\ttfamily
  arXiv:1808.07760 [hep-th]}}.

\bibitem{Caputa:2019pam}
P.~Caputa, S.~Datta, and V.~Shyam, ``{Sphere partition functions
  \textbackslash{}\& cut-off AdS},''
  \href{http://dx.doi.org/10.1007/JHEP05(2019)112}{{\em JHEP} {\bfseries 05}
  (2019) 112}, \href{http://arxiv.org/abs/1902.10893}{{\ttfamily
  arXiv:1902.10893 [hep-th]}}.

\bibitem{Giveon:2017myj}
A.~Giveon, N.~Itzhaki, and D.~Kutasov, ``{A solvable irrelevant deformation of
  AdS$_{3}$/CFT$_{2}$},'' \href{http://dx.doi.org/10.1007/JHEP12(2017)155}{{\em
  JHEP} {\bfseries 12} (2017) 155},
  \href{http://arxiv.org/abs/1707.05800}{{\ttfamily arXiv:1707.05800
  [hep-th]}}.

\bibitem{Tian:2023fgf}
J.~Tian, ``{On-shell action of $\text{T}\bar{\text{T}}$-deformed Holographic
  CFTs},'' \href{http://arxiv.org/abs/2306.01258}{{\ttfamily arXiv:2306.01258
  [hep-th]}}.

\bibitem{Dei:2024sct}
A.~Dei, B.~Knighton, K.~Naderi, and S.~Sethi, ``{Tensionless AdS$_3$/CFT$_2$
  and Single Trace $T\overline{T}$},''
  \href{http://arxiv.org/abs/2408.00823}{{\ttfamily arXiv:2408.00823
  [hep-th]}}.

\bibitem{Du:2024bqk}
Z.~Du, W.-X. Lai, K.~Liu, and W.~Song, ``{Asymptotic Symmetries in the
  TsT/$T\bar{T}$ Correspondence},''
  \href{http://arxiv.org/abs/2407.19495}{{\ttfamily arXiv:2407.19495
  [hep-th]}}.

\bibitem{Tian:2024vln}
J.~Tian, T.~Lai, and F.~Omidi, ``{Modular transformations of on-shell actions
  of (root-)$T\bar{T}$ deformed holographic CFTs},''
  \href{http://dx.doi.org/10.1016/j.nuclphysb.2024.116675}{{\em Nucl. Phys. B}
  {\bfseries 1007} (2024) 116675},
  \href{http://arxiv.org/abs/2404.16354}{{\ttfamily arXiv:2404.16354
  [hep-th]}}.

\bibitem{Chen:2018eqk}
B.~Chen, L.~Chen, and P.-X. Hao, ``{Entanglement entropy in
  $T\overline{T}$-deformed CFT},''
  \href{http://dx.doi.org/10.1103/PhysRevD.98.086025}{{\em Phys. Rev. D}
  {\bfseries 98} no.~8, (2018) 086025},
  \href{http://arxiv.org/abs/1807.08293}{{\ttfamily arXiv:1807.08293
  [hep-th]}}.

\bibitem{Donnelly:2018bef}
W.~Donnelly and V.~Shyam, ``{Entanglement entropy and $T \overline{T}$
  deformation},'' \href{http://dx.doi.org/10.1103/PhysRevLett.121.131602}{{\em
  Phys. Rev. Lett.} {\bfseries 121} no.~13, (2018) 131602},
  \href{http://arxiv.org/abs/1806.07444}{{\ttfamily arXiv:1806.07444
  [hep-th]}}.

\bibitem{Lewkowycz:2019xse}
A.~Lewkowycz, J.~Liu, E.~Silverstein, and G.~Torroba, ``{$ T\overline{T} $ and
  EE, with implications for (A)dS subregion encodings},''
  \href{http://dx.doi.org/10.1007/JHEP04(2020)152}{{\em JHEP} {\bfseries 04}
  (2020) 152}, \href{http://arxiv.org/abs/1909.13808}{{\ttfamily
  arXiv:1909.13808 [hep-th]}}.

\bibitem{Banerjee:2019ewu}
A.~Banerjee, A.~Bhattacharyya, and S.~Chakraborty, ``{Entanglement Entropy for
  $TT$ deformed CFT in general dimensions},''
  \href{http://dx.doi.org/10.1016/j.nuclphysb.2019.114775}{{\em Nucl. Phys. B}
  {\bfseries 948} (2019) 114775},
  \href{http://arxiv.org/abs/1904.00716}{{\ttfamily arXiv:1904.00716
  [hep-th]}}.

\bibitem{Jeong:2019ylz}
H.-S. Jeong, K.-Y. Kim, and M.~Nishida, ``{Entanglement and R\'enyi entropy of
  multiple intervals in $T\overline{T}$-deformed CFT and holography},''
  \href{http://dx.doi.org/10.1103/PhysRevD.100.106015}{{\em Phys. Rev. D}
  {\bfseries 100} no.~10, (2019) 106015},
  \href{http://arxiv.org/abs/1906.03894}{{\ttfamily arXiv:1906.03894
  [hep-th]}}.

\bibitem{Murdia:2019fax}
C.~Murdia, Y.~Nomura, P.~Rath, and N.~Salzetta, ``{Comments on holographic
  entanglement entropy in $TT$ deformed conformal field theories},''
  \href{http://dx.doi.org/10.1103/PhysRevD.100.026011}{{\em Phys. Rev. D}
  {\bfseries 100} no.~2, (2019) 026011},
  \href{http://arxiv.org/abs/1904.04408}{{\ttfamily arXiv:1904.04408
  [hep-th]}}.

\bibitem{He:2019vzf}
S.~He and H.~Shu, ``{Correlation functions, entanglement and chaos in the $
  T\overline{T}/J\overline{T} $-deformed CFTs},''
  \href{http://dx.doi.org/10.1007/JHEP02(2020)088}{{\em JHEP} {\bfseries 02}
  (2020) 088}, \href{http://arxiv.org/abs/1907.12603}{{\ttfamily
  arXiv:1907.12603 [hep-th]}}.

\bibitem{Asrat:2019end}
M.~Asrat, ``{Entropic $c$\textendash{}functions in $T{\bar T}, J{\bar T},
  T{\bar J}$ deformations},''
  \href{http://dx.doi.org/10.1016/j.nuclphysb.2020.115186}{{\em Nucl. Phys. B}
  {\bfseries 960} (2020) 115186},
  \href{http://arxiv.org/abs/1911.04618}{{\ttfamily arXiv:1911.04618
  [hep-th]}}.

\bibitem{Basu:2023bov}
D.~Basu, Lavish, and B.~Paul, ``{Entanglement negativity in
  TT\textasciimacron{}-deformed CFT2s},''
  \href{http://dx.doi.org/10.1103/PhysRevD.107.126026}{{\em Phys. Rev. D}
  {\bfseries 107} no.~12, (2023) 126026},
  \href{http://arxiv.org/abs/2302.11435}{{\ttfamily arXiv:2302.11435
  [hep-th]}}.

\bibitem{Basu:2023aqz}
D.~Basu, S.~Biswas, A.~Dey, B.~Paul, and G.~Sengupta, ``{Odd entanglement
  entropy in TT\textasciimacron{} deformed CFT2s and holography},''
  \href{http://dx.doi.org/10.1103/PhysRevD.108.126013}{{\em Phys. Rev. D}
  {\bfseries 108} no.~12, (2023) 126013},
  \href{http://arxiv.org/abs/2307.04832}{{\ttfamily arXiv:2307.04832
  [hep-th]}}.

\bibitem{Basu:2024bal}
D.~Basu and V.~Raj, ``{Reflected entropy and timelike entanglement in
  TT\textasciimacron{}-deformed CFT2s},''
  \href{http://dx.doi.org/10.1103/PhysRevD.110.046009}{{\em Phys. Rev. D}
  {\bfseries 110} no.~4, (2024) 046009},
  \href{http://arxiv.org/abs/2402.07253}{{\ttfamily arXiv:2402.07253
  [hep-th]}}.

\bibitem{Chang:2024voo}
J.-C. Chang, S.~He, Y.-X. Liu, and L.~Zhao, ``{The holographic $T\bar{T}$
  deformation of the entanglement entropy in (A)dS$_3$/CFT$_2$},''
  \href{http://arxiv.org/abs/2409.08198}{{\ttfamily arXiv:2409.08198
  [hep-th]}}.

\bibitem{Banerjee:2024wtl}
A.~Banerjee and P.~Roy, ``{Bounds on $T\bar {T}$ deformation from
  entanglement},'' \href{http://arxiv.org/abs/2404.16946}{{\ttfamily
  arXiv:2404.16946 [hep-th]}}.

\bibitem{Guica:2019nzm}
M.~Guica and R.~Monten, ``{$T\bar T$ and the mirage of a bulk cutoff},''
  \href{http://dx.doi.org/10.21468/SciPostPhys.10.2.024}{{\em SciPost Phys.}
  {\bfseries 10} no.~2, (2021) 024},
  \href{http://arxiv.org/abs/1906.11251}{{\ttfamily arXiv:1906.11251
  [hep-th]}}.

\bibitem{Doi:2022iyj}
K.~Doi, J.~Harper, A.~Mollabashi, T.~Takayanagi, and Y.~Taki, ``{Pseudoentropy
  in dS/CFT and Timelike Entanglement Entropy},''
  \href{http://dx.doi.org/10.1103/PhysRevLett.130.031601}{{\em Phys. Rev.
  Lett.} {\bfseries 130} no.~3, (2023) 031601},
  \href{http://arxiv.org/abs/2210.09457}{{\ttfamily arXiv:2210.09457
  [hep-th]}}.

\bibitem{Doi:2023zaf}
K.~Doi, J.~Harper, A.~Mollabashi, T.~Takayanagi, and Y.~Taki, ``{Timelike
  entanglement entropy},''
  \href{http://dx.doi.org/10.1007/JHEP05(2023)052}{{\em JHEP} {\bfseries 05}
  (2023) 052}, \href{http://arxiv.org/abs/2302.11695}{{\ttfamily
  arXiv:2302.11695 [hep-th]}}.

\bibitem{Apolo:2023aho}
L.~Apolo, W.~Song, and B.~Yu, ``{On the universal behavior of $ T\overline{T}
  $-deformed CFTs: single and double-trace partition functions at large c},''
  \href{http://dx.doi.org/10.1007/JHEP05(2023)210}{{\em JHEP} {\bfseries 05}
  (2023) 210}, \href{http://arxiv.org/abs/2301.04153}{{\ttfamily
  arXiv:2301.04153 [hep-th]}}.

\bibitem{Apolo:2023vnm}
L.~Apolo, P.-X. Hao, W.-X. Lai, and W.~Song, ``{Glue-on AdS holography for $
  T\overline{T} $-deformed CFTs},''
  \href{http://dx.doi.org/10.1007/JHEP06(2023)117}{{\em JHEP} {\bfseries 06}
  (2023) 117}, \href{http://arxiv.org/abs/2303.04836}{{\ttfamily
  arXiv:2303.04836 [hep-th]}}.

\bibitem{Apolo:2023ckr}
L.~Apolo, P.-X. Hao, W.-X. Lai, and W.~Song, ``{Extremal surfaces in glue-on
  AdS/$ T\overline{T} $ holography},''
  \href{http://dx.doi.org/10.1007/JHEP01(2024)054}{{\em JHEP} {\bfseries 01}
  (2024) 054}, \href{http://arxiv.org/abs/2311.04883}{{\ttfamily
  arXiv:2311.04883 [hep-th]}}.

\bibitem{Brown:1986nw}
J.~D. Brown and M.~Henneaux, ``{Central Charges in the Canonical Realization of
  Asymptotic Symmetries: An Example from Three-Dimensional Gravity},''
  \href{http://dx.doi.org/10.1007/BF01211590}{{\em Commun. Math. Phys.}
  {\bfseries 104} (1986) 207--226}.

\bibitem{Brown:1994gs}
J.~D. Brown, J.~Creighton, and R.~B. Mann, ``{Temperature, energy and heat
  capacity of asymptotically anti-de Sitter black holes},''
  \href{http://dx.doi.org/10.1103/PhysRevD.50.6394}{{\em Phys. Rev. D}
  {\bfseries 50} (1994) 6394--6403},
  \href{http://arxiv.org/abs/gr-qc/9405007}{{\ttfamily arXiv:gr-qc/9405007}}.

\bibitem{Nguyen:2017yqw}
P.~Nguyen, T.~Devakul, M.~G. Halbasch, M.~P. Zaletel, and B.~Swingle,
  ``{Entanglement of purification: from spin chains to holography},''
  \href{http://dx.doi.org/10.1007/JHEP01(2018)098}{{\em JHEP} {\bfseries 01}
  (2018) 098}, \href{http://arxiv.org/abs/1709.07424}{{\ttfamily
  arXiv:1709.07424 [hep-th]}}.

\bibitem{Kusuki:2019evw}
Y.~Kusuki and K.~Tamaoka, ``{Entanglement Wedge Cross Section from CFT:
  Dynamics of Local Operator Quench},''
  \href{http://dx.doi.org/10.1007/JHEP02(2020)017}{{\em JHEP} {\bfseries 02}
  (2020) 017}, \href{http://arxiv.org/abs/1909.06790}{{\ttfamily
  arXiv:1909.06790 [hep-th]}}.

\bibitem{Kusuki:2019rbk}
Y.~Kusuki and K.~Tamaoka, ``{Dynamics of entanglement wedge cross section from
  conformal field theories},''
  \href{http://dx.doi.org/10.1016/j.physletb.2021.136105}{{\em Phys. Lett. B}
  {\bfseries 814} (2021) 136105},
  \href{http://arxiv.org/abs/1907.06646}{{\ttfamily arXiv:1907.06646
  [hep-th]}}.

\bibitem{Basu:2023jtf}
D.~Basu, H.~Chourasiya, V.~Raj, and G.~Sengupta, ``{Reflected entropy in a BCFT
  on a black hole background},''
  \href{http://dx.doi.org/10.1007/JHEP05(2024)054}{{\em JHEP} {\bfseries 05}
  (2024) 054}, \href{http://arxiv.org/abs/2311.17023}{{\ttfamily
  arXiv:2311.17023 [hep-th]}}.

\bibitem{Jeong:2019xdr}
H.-S. Jeong, K.-Y. Kim, and M.~Nishida, ``{Reflected Entropy and Entanglement
  Wedge Cross Section with the First Order Correction},''
  \href{http://dx.doi.org/10.1007/JHEP12(2019)170}{{\em JHEP} {\bfseries 12}
  (2019) 170}, \href{http://arxiv.org/abs/1909.02806}{{\ttfamily
  arXiv:1909.02806 [hep-th]}}.

\bibitem{Asrat:2020uib}
M.~Asrat and J.~Kudler-Flam, ``{$T\bar{T}$, the entanglement wedge cross
  section, and the breakdown of the split property},''
  \href{http://dx.doi.org/10.1103/PhysRevD.102.045009}{{\em Phys. Rev. D}
  {\bfseries 102} no.~4, (2020) 045009},
  \href{http://arxiv.org/abs/2005.08972}{{\ttfamily arXiv:2005.08972
  [hep-th]}}.

\bibitem{Calabrese:2004eu}
P.~Calabrese and J.~L. Cardy, ``{Entanglement entropy and quantum field
  theory},'' \href{http://dx.doi.org/10.1088/1742-5468/2004/06/P06002}{{\em J.
  Stat. Mech.} {\bfseries 0406} (2004) P06002},
  \href{http://arxiv.org/abs/hep-th/0405152}{{\ttfamily arXiv:hep-th/0405152}}.

\bibitem{Calabrese:2009qy}
P.~Calabrese and J.~Cardy, ``{Entanglement entropy and conformal field
  theory},'' \href{http://dx.doi.org/10.1088/1751-8113/42/50/504005}{{\em J.
  Phys. A} {\bfseries 42} (2009) 504005},
  \href{http://arxiv.org/abs/0905.4013}{{\ttfamily arXiv:0905.4013
  [cond-mat.stat-mech]}}.

\bibitem{Sun:2019ijq}
Y.~Sun and J.-R. Sun, ``{Note on the R\'enyi entropy of 2D perturbed
  fermions},'' \href{http://dx.doi.org/10.1103/PhysRevD.99.106008}{{\em Phys.
  Rev. D} {\bfseries 99} no.~10, (2019) 106008},
  \href{http://arxiv.org/abs/1901.08796}{{\ttfamily arXiv:1901.08796
  [hep-th]}}.

\bibitem{Fitzpatrick:2014vua}
A.~L. Fitzpatrick, J.~Kaplan, and M.~T. Walters, ``{Universality of
  Long-Distance AdS Physics from the CFT Bootstrap},''
  \href{http://dx.doi.org/10.1007/JHEP08(2014)145}{{\em JHEP} {\bfseries 08}
  (2014) 145}, \href{http://arxiv.org/abs/1403.6829}{{\ttfamily arXiv:1403.6829
  [hep-th]}}.

\bibitem{Basu:2022nds}
D.~Basu, H.~Parihar, V.~Raj, and G.~Sengupta, ``{Entanglement negativity,
  reflected entropy, and anomalous gravitation},''
  \href{http://dx.doi.org/10.1103/PhysRevD.105.086013}{{\em Phys. Rev. D}
  {\bfseries 105} no.~8, (2022) 086013},
  \href{http://arxiv.org/abs/2202.00683}{{\ttfamily arXiv:2202.00683
  [hep-th]}}. [Erratum: Phys.Rev.D 105, 129902 (2022)].

\bibitem{Afrasiar:2022wzn}
M.~Afrasiar, H.~Chourasiya, V.~Raj, and G.~Sengupta, ``{Covariant holographic
  reflected entropy in AdS3/CFT2},''
  \href{http://dx.doi.org/10.1016/j.physletb.2022.137590}{{\em Phys. Lett. B}
  {\bfseries 835} (2022) 137590},
  \href{http://arxiv.org/abs/2209.02654}{{\ttfamily arXiv:2209.02654
  [hep-th]}}.

\bibitem{Calabrese:2014yza}
P.~Calabrese, J.~Cardy, and E.~Tonni, ``{Finite temperature entanglement
  negativity in conformal field theory},''
  \href{http://dx.doi.org/10.1088/1751-8113/48/1/015006}{{\em J. Phys. A}
  {\bfseries 48} no.~1, (2015) 015006},
  \href{http://arxiv.org/abs/1408.3043}{{\ttfamily arXiv:1408.3043
  [cond-mat.stat-mech]}}.

\bibitem{Zamolodchikov:1986gt}
A.~B. Zamolodchikov, ``{Irreversibility of the Flux of the Renormalization
  Group in a 2D Field Theory},'' {\em JETP Lett.} {\bfseries 43} (1986)
  730--732.

\bibitem{Nakata:2020luh}
Y.~Nakata, T.~Takayanagi, Y.~Taki, K.~Tamaoka, and Z.~Wei, ``{New holographic
  generalization of entanglement entropy},''
  \href{http://dx.doi.org/10.1103/PhysRevD.103.026005}{{\em Phys. Rev. D}
  {\bfseries 103} no.~2, (2021) 026005},
  \href{http://arxiv.org/abs/2005.13801}{{\ttfamily arXiv:2005.13801
  [hep-th]}}.

\bibitem{Basu:2021awn}
D.~Basu, A.~Chandra, V.~Raj, and G.~Sengupta, ``{Entanglement wedge in flat
  holography and entanglement negativity},''
  \href{http://dx.doi.org/10.21468/SciPostPhysCore.5.1.013}{{\em SciPost Phys.
  Core} {\bfseries 5} (2022) 013},
  \href{http://arxiv.org/abs/2106.14896}{{\ttfamily arXiv:2106.14896
  [hep-th]}}.

\bibitem{Hayden:2021gno}
P.~Hayden, O.~Parrikar, and J.~Sorce, ``{The Markov gap for geometric reflected
  entropy},'' \href{http://dx.doi.org/10.1007/JHEP10(2021)047}{{\em JHEP}
  {\bfseries 10} (2021) 047}, \href{http://arxiv.org/abs/2107.00009}{{\ttfamily
  arXiv:2107.00009 [hep-th]}}.

\bibitem{Hayden:2011ag}
P.~Hayden, M.~Headrick, and A.~Maloney, ``{Holographic Mutual Information is
  Monogamous},'' \href{http://dx.doi.org/10.1103/PhysRevD.87.046003}{{\em Phys.
  Rev. D} {\bfseries 87} no.~4, (2013) 046003},
  \href{http://arxiv.org/abs/1107.2940}{{\ttfamily arXiv:1107.2940 [hep-th]}}.

\bibitem{Deng:2023pjs}
F.~Deng, Z.~Wang, and Y.~Zhou, ``{End of the world brane meets $ T\overline{T}
  $},'' \href{http://dx.doi.org/10.1007/JHEP07(2024)036}{{\em JHEP} {\bfseries
  07} (2024) 036}, \href{http://arxiv.org/abs/2310.15031}{{\ttfamily
  arXiv:2310.15031 [hep-th]}}.

\bibitem{Chandrasekaran:2020qtn}
V.~Chandrasekaran, M.~Miyaji, and P.~Rath, ``{Including contributions from
  entanglement islands to the reflected entropy},''
  \href{http://dx.doi.org/10.1103/PhysRevD.102.086009}{{\em Phys. Rev. D}
  {\bfseries 102} no.~8, (2020) 086009},
  \href{http://arxiv.org/abs/2006.10754}{{\ttfamily arXiv:2006.10754
  [hep-th]}}.

\bibitem{Li:2020ceg}
T.~Li, J.~Chu, and Y.~Zhou, ``{Reflected Entropy for an Evaporating Black
  Hole},'' \href{http://dx.doi.org/10.1007/JHEP11(2020)155}{{\em JHEP}
  {\bfseries 11} (2020) 155}, \href{http://arxiv.org/abs/2006.10846}{{\ttfamily
  arXiv:2006.10846 [hep-th]}}.

\bibitem{Skenderis:1999nb}
K.~Skenderis and S.~N. Solodukhin, ``{Quantum effective action from the AdS /
  CFT correspondence},''
  \href{http://dx.doi.org/10.1016/S0370-2693(99)01467-7}{{\em Phys. Lett. B}
  {\bfseries 472} (2000) 316--322},
  \href{http://arxiv.org/abs/hep-th/9910023}{{\ttfamily arXiv:hep-th/9910023}}.

\bibitem{deHaro:2000vlm}
S.~de~Haro, S.~N. Solodukhin, and K.~Skenderis, ``{Holographic reconstruction
  of space-time and renormalization in the AdS / CFT correspondence},''
  \href{http://dx.doi.org/10.1007/s002200100381}{{\em Commun. Math. Phys.}
  {\bfseries 217} (2001) 595--622},
  \href{http://arxiv.org/abs/hep-th/0002230}{{\ttfamily arXiv:hep-th/0002230}}.

\bibitem{Balasubramanian:1999re}
V.~Balasubramanian and P.~Kraus, ``{A Stress tensor for Anti-de Sitter
  gravity},'' \href{http://dx.doi.org/10.1007/s002200050764}{{\em Commun. Math.
  Phys.} {\bfseries 208} (1999) 413--428},
  \href{http://arxiv.org/abs/hep-th/9902121}{{\ttfamily arXiv:hep-th/9902121}}.

\end{thebibliography}\endgroup
\end{document}